\documentclass[12pt]{article}
\usepackage{amssymb}
\usepackage{amsmath}
\usepackage{graphicx}
\usepackage{indentfirst}
\usepackage{cite}

\allowdisplaybreaks

\begin{document}

\title{$K^*$-charmonium dissociation cross sections and \\
charmonium dissociation rates in hadronic matter}
\author{Feng-Rong Liu, Shi-Tao Ji, and Xiao-Ming Xu}
\date{}
\maketitle \vspace{-1cm}
\centerline{Department of Physics, Shanghai University, Baoshan, 
Shanghai 200444, China}

\begin{abstract}
$K^*$-charmonium dissociation reactions in hadronic matter are studied 
in the Born approximation, in the quark-interchange mechanism, and with a 
temperature-dependent quark potential. We obtain the temperature dependence of
unpolarized cross sections for the reactions:
$K^*J/\psi\to\bar{D}D^+_s$, $\bar{D}^{*}D^+_s$, $\bar{D}D^{*+}_s$, and 
$\bar{D}^{*}D^{*+}_s$;
$K^*\psi'\to\bar{D}D^+_s$, $\bar{D}^{*}D^+_s$, $\bar{D}D^{*+}_s$, and 
$\bar{D}^{*}D^{*+}_s$;
$K^*\chi_{c}\to\bar{D}D^+_s$, $\bar{D}^{*}D^+_s$, $\bar{D}D^{*+}_s$, and 
$\bar{D}^{*}D^{*+}_s$. 
We use the cross sections for
charmonium dissociation in collisions with pion, $\rho$ meson, kaon, vector
kaon, and $\eta$ meson to calculate dissociation rates of charmonium with
the five types of mesons. Because of the
temperature dependence of the meson masses, dissociation cross sections, and
meson distribution functions, the charmonium dissociation rates generally
increase with the increase of temperature and decrease with the increase of
charmonium momentum from
2.2 GeV/$c$. We find that the first derivative of the dissociation rate with
respect to the charmonium momentum is zero when the charmonium is at rest.
While the $\eta + \psi'$ and $\eta + \chi_c$ dissociation reactions can be 
neglected, the $J/\psi$, $\psi'$, and $\chi_c$ dissociation are caused by the 
collisions with pion, $\rho$ meson, kaon, vector kaon, and $\eta$ meson.
\end{abstract}



\vspace{0.5cm}
\leftline{\bf I. INTRODUCTION}
\vspace{0.5cm}

Hadronic matter is produced in relativistic heavy ion collisions at the 
Relativistic Heavy Ion Collider and at the Large Hadron Collider.
Pions in hadronic matter have a number density smaller than a quark-gluon 
plasma, but the meson species is not limited to the pion and hadronic matter 
has a lifetime longer than the plasma. Hence, meson-charmonium 
dissociation reactions may cause appreciable
suppression of charmonia in hadronic matter.
In order to separate the suppression of charmonia in hadronic matter so that 
the suppression due to the quark-gluon plasma is identified, we need to
study the meson-charmonium dissociation reactions.

Three approaches have been established for the study of charmonium
dissociation in collisions with hadrons. 
In the short-distance approach the operator product 
expansion of perturbative QCD is applied to heavy quarkonia of small sizes
\cite{KS,Peskin}. Cross sections
for nucleon-$J/\psi$ and pion-$J/\psi$ dissociation have been obtained in 
Refs. \cite{KS,Peskin,AGGA} from existing parton distribution functions 
\cite{Lai,MSR,GR}.
In the quark-interchange approach the quark interchange mechanism between
the incident hadron and the charmonium breaks the charmonium, and produces
charmed mesons and/or charmed strange mesons. Charmonia in collisions with
$\pi$, $\rho$, $K$, and $N$ have been studied in Refs. 
\cite{MBQ,WSB,BSWX,HBBS}. In the meson-exchange approach meson exchange 
between the two initial mesons breaks the charmonium, and effective
Lagrangians with meson couplings are constructed to describe the motion of
meson fields. The $J/\psi$ dissociation in collisions with $\pi$, $\rho$, 
$\omega$, $K$, $K^*$, $\eta$, and $\phi$ has been considered in Refs. 
\cite{MatinyanM,LK,Haglin,OSL,NNR,MPPR,BG}.

The studies in Refs. 
\cite{KS,Peskin,AGGA,MBQ,WSB,BSWX,HBBS,MatinyanM,LK,Haglin,OSL,NNR,MPPR,BG}
concentrate on charmonium dissociation in vacuum. In hadronic matter
charmonium dissociation is affected by the medium \cite{Wong}. 
We have obtained the energy and 
temperature dependence of dissociation cross sections of charmonia
in collisions with $\pi$,
$\rho$, $K$, and $\eta$ mesons in hadronic matter \cite{ZX,JSX}. Charmonium 
dissociation reactions may be endothermic in one temperature region and 
exothermic in another. Peak cross sections of endothermic
reactions change with temperature. The $\eta + J/\psi$ dissociation gives
rise to $J/\psi$ suppression comparable to the suppression caused by the
$\pi + J/\psi$ dissociation \cite{JSX}.

The $K^*+J/\psi$ dissociation in vacuum was considered in the 
meson-exchange approach in Ref. \cite{Haglin}, but no 
cross sections were presented. The energy and temperature 
dependence of $K^*$-charmonium dissociation cross sections are unknown.
In hadronic matter the quark interaction, meson masses, and mesonic 
quark-antiquark relative motion depend on temperature. From 
vacuum to medium the $K^*$-charmonium dissociation reactions must change.
Therefore, in this work we calculate the dissociation 
cross sections of $J/\psi$, $\psi'$, and $\chi_{c}$ in collisions with $K^*$ 
in hadronic matter on the basis of the quark-interchange mechanism
\cite{BS}, the Born approximation, and a temperature-dependent quark
potential. Furthermore, from the energy and temperature dependence of the
cross sections we calculate the dissociation rate of charmonium with vector
kaons. Since the dissociation cross sections of charmonia in collisions with 
$\pi$, $\rho$, $K$, and $\eta$ mesons are provided in Refs. \cite{ZX,JSX}, 
we also calculate the dissociation rates of charmonium with pion, $\rho$ meson,
kaon,
and $\eta$ meson. From these dissociation rates we know different contributions
to the charmonium dissociation from different mesons in hadronic matter.

This paper is organized as follows. In Sec. II we introduce cross-section
formulas, a central spin-independent potential,
and a spin-spin interaction. In Sec. III we present numerical unpolarized 
cross sections for twelve $K^*$-charmonium dissociation reactions and relevant
discussions. In Sec. IV  we define the dissociation rate of charmonium 
with meson in hadronic matter, calculate the dissociation rates of charmonium 
with pion, $\rho$ meson, kaon, vector kaon, and $\eta$ meson, and discuss
relevant results. In Sec. V we summarize the present work.

\vspace{0.5cm}
\leftline{\bf II. CROSS-SECTION FORMULAS}
\vspace{0.5cm}

Let $J_i$, $m_i$, and $P_i=(E_{i},\vec{P}_i)$ be the angular momentum, mass,
and four-momentum of meson $i~(i=q\bar{q},c\bar{c},q\bar{c},c\bar{q})$ in the
reaction $q\bar{q}+c\bar{c}\rightarrow q\bar{c}+c\bar{q}$, respectively.
$q$ stands for the
up quark, down quark or strange quark. The flavor of the quark $q$ may be
different from the flavor of the antiquark $\bar q$.
The unpolarized cross section for 
$q\bar{q}+c\bar{c}\rightarrow q\bar{c}+c\bar{q}$ is
\begin{eqnarray}
\sigma^{\rm unpol}(\sqrt {s},T) & = & \frac {(2\pi)^4}
{4\sqrt {(P_{q\bar {q}} \cdot P_{c\bar {c}})^2 
-m_{q\bar {q}}^2m_{c\bar {c}}^2}}
     \int \frac {d^3P_{q\bar {c}}}{(2\pi)^32E_{q\bar {c}}}
     \frac {d^3P_{c\bar {q}}}{(2\pi)^32E_{c\bar {q}}}
\nonumber     \\
& &
\frac {1}{(2J_{q\bar q}+1)(2J_{c\bar c}+1)}
\sum\limits_{J_{q\bar{q}z}J_{c\bar{c}z}J_{q\bar{c}z}J_{c\bar{q}z}}
\mid {\cal M}_{\rm fi} \mid^2     \delta (E_{\rm f} - E_{\rm i})
\delta (\vec {P}_{\rm f} - \vec {P}_{\rm i}),     
\end{eqnarray}
where $s$ is the Mandelstam variable given by
$s=(E_{q\bar{q}}+E_{c\bar{c}})^2-(\vec{P}_{q\bar{q}}+\vec{P}_{c\bar{c}})^2$,
$T$ is the temperature, $\mathcal{M}_{\rm fi}$ is the transition amplitude,
$J_{iz}$ is the magnetic projection quantum number of
$J_i~(i=q\bar{q},c\bar{c},q\bar{c},c\bar{q})$, 
$E_{\rm i}=E_{q\bar{q}}+E_{c\bar{c}}$,
$E_{\rm f}=E_{q\bar{c}}+E_{c\bar{q}}$, 
$\vec {P}_{\rm i}=\vec {P}_{q\bar{q}}+\vec {P}_{c\bar{c}}$, and
$\vec {P}_{\rm f}=\vec {P}_{q\bar{c}}+\vec {P}_{c\bar{q}}$. If the orbital
angular momenta of the four mesons are zero, the unpolarized cross section is
\begin{eqnarray}
\sigma^{\rm unpol}(\sqrt {s},T) & = & 
\frac {(2\pi)^4} {4\sqrt {(P_{q\bar {q}} \cdot P_{c\bar {c}})^2 
-m_{q\bar {q}}^2m_{c\bar {c}}^2}}
     \int \frac {d^3P_{q\bar {c}}}{(2\pi)^32E_{q\bar {c}}}
     \frac {d^3P_{c\bar {q}}}{(2\pi)^32E_{c\bar {q}}}
\nonumber   \\
& &
\frac {1}{(2S_{q\bar {q}}+1)(2S_{c\bar {c}}+1)} \sum\limits_S (2S+1) 
\mid {\cal M}_{\rm fi} \mid^2  \delta (E_{\rm f} - E_{\rm i})
\delta (\vec {P}_{\rm f} - \vec {P}_{\rm i}),      
\end{eqnarray}
where $S_{q\bar{q}}$ and $S_{c\bar{c}}$ are the spins of
$q\bar{q}$ and $c\bar{c}$, respectively, and $S$ is the total spin of the two
initial mesons. If the orbital angular momenta of $q\bar{q}$, $c\bar{c}$, 
$q\bar{c}$, and $c\bar{q}$ are 0, $L_{c\bar c}$, 0, and 0, respectively, 
the unpolarized cross section is
\begin{eqnarray}
\sigma^{\rm unpol}(\sqrt {s},T) & = & \frac {(2\pi)^4}
{4\sqrt {(P_{q\bar {q}} \cdot P_{c\bar {c}})^2 
-m_{q\bar {q}}^2m_{c\bar {c}}^2}}
     \int \frac {d^3P_{q\bar {c}}}{(2\pi)^32E_{q\bar {c}}}
     \frac {d^3P_{c\bar {q}}}{(2\pi)^32E_{c\bar {q}}}
\nonumber   \\
& &
\sum\limits_{JSL_{c\bar{c}z}} (2J+1)(2S+1)
\left\{
\begin{array}{ccc}
0     &   S_{q\bar q}    &   S_{q\bar q}   \\
L_{c\bar c}  &   S_{c\bar c}   &   J_{c\bar c}   \\
L_{c\bar c}  &   S   &  J
\end{array}
\right\}^2
\nonumber   \\
& &
\mid {\cal M}_{\rm fi} \mid^2     \delta (E_{\rm f} - E_{\rm i})
\delta (\vec {P}_{\rm f} - \vec {P}_{\rm i}),    
\end{eqnarray}
where the braces give the $9j$ coefficient,
$J$ is the total angular momentum of the two
initial mesons, and $L_{c\bar{c}z}$ is the magnetic projection quantum number
of $L_{c\bar{c}}$. Furthermore, if $L_{c\bar c}=1$ and $S_{c\bar c}=0$ or 1,
the unpolarized cross section is
\begin{eqnarray}
\sigma^{\rm unpol}(\sqrt {s},T) & = & 
\frac {(2\pi)^4} {4\sqrt {(P_{q\bar {q}} \cdot P_{c\bar {c}})^2 
-m_{q\bar {q}}^2m_{c\bar {c}}^2}}
     \int \frac {d^3P_{q\bar {c}}}{(2\pi)^32E_{q\bar {c}}}
     \frac {d^3P_{c\bar {q}}}{(2\pi)^32E_{c\bar {q}}}
\nonumber   \\
& &
\frac {1}{(2S_{q\bar {q}}+1)(2S_{c\bar {c}}+1)(2L_{c\bar c}+1)}
\nonumber   \\
& &
\sum\limits_{SL_{c\bar {c}z}} (2S+1) 
\mid {\cal M}_{\rm fi} \mid^2  \delta (E_{\rm f} - E_{\rm i})
\delta (\vec {P}_{\rm f} - \vec {P}_{\rm i}).     
\end{eqnarray}
Define
\begin{eqnarray}
\sigma (S,m_S,\sqrt {s},T) & = & \frac {(2\pi)^4}
{4\sqrt {(P_{q\bar {q}} \cdot P_{c\bar {c}})^2 
-m_{q\bar {q}}^2m_{c\bar {c}}^2}}    \nonumber   \\
& & \int \frac {d^3P_{q\bar {c}}}{(2\pi)^32E_{q\bar {c}}}
     \frac {d^3P_{c\bar {q}}}{(2\pi)^32E_{c\bar {q}}}
\mid {\cal M}_{\rm fi} \mid^2     \delta (E_{\rm f} - E_{\rm i})
\delta (\vec {P}_{\rm f} - \vec {P}_{\rm i}),      
\end{eqnarray}
where $m_S$ is the magnetic projection quantum number of $S$.
Then, Eqs. (2) and (4) are written as
\begin{equation}
\sigma ^{\rm unpol} (\sqrt {s}, T) =
\frac{1}{(2S_{q\bar q} +1) (2S_ {c\bar c}+1) (2L_ {c\bar c}+1)}
\sum \limits_ {SL_{c\bar{c}z}} (2S+1)
\sigma (S,m_S,\sqrt {s},T). 
\end{equation}
In the center-of-mass frame of $q\bar q$ and $c\bar c$ \cite{LX},
\begin{equation}
\sigma(S,m_S,\sqrt {s},T) =\frac{1}{32\pi s}\frac{|\vec{P}^{\prime }(\sqrt{s})|
}{|\vec{P}(\sqrt{s})|}\int_{0}^{\pi }d\theta
|\mathcal{M}_{\rm fi}|^{2}\sin \theta,
\end{equation}
where $\vec {P}_{q\bar q}=\vec P$, $\vec {P}_{q\bar c}=\vec {P}^\prime$,
and $\theta$ is the angle between $\vec{P}$ and $\vec{P}'$.
Either the quark interchange between $q\bar q$ and $c\bar c$ 
or the antiquark interchange
leads to the reaction $q\bar{q}+c\bar{c}\rightarrow q\bar{c}+c\bar{q}$. In
addition to the interchange, an interaction takes place between the quark or
the antiquark of meson $q\bar q$ ($q\bar c$) 
and the quark or the antiquark of meson 
$c\bar c$ ($c\bar q$). Diagrams for the reaction are shown in Fig. 1 for the
prior form and Fig. 2 for the post form \cite{JSX}.
The scattering in the prior form means that gluon exchange occurs before quark
interchange. The corresponding transition amplitude is
\begin{equation}
{\cal M}_{\rm fi}^{\rm prior} = 
4\sqrt {E_{q\bar{q}} E_{c\bar{c}}E_{q\bar{c}}E_{c\bar{q}}}
\langle\psi_{q\bar {c}}|\langle\psi_{c\bar{q}}|
(V_{q\bar {c}} +V_ {c \bar {q}} +V_ {qc}+V_{\bar {q}\bar {c}})
|\psi_{q\bar {q}}\rangle|\psi_{c\bar {c}}\rangle,
\end{equation}
where
$\psi_{q\bar {q}}$ ($\psi_{c\bar {c}}$, $\psi_{q\bar {c}}$,
$\psi_{c\bar{q}}$) represents the product of color, spin, flavor, and
relative-motion wave functions of
$q\bar {q}$ ($c\bar {c}$, $q\bar {c}$, $c\bar{q}$), and
$V_{q\bar c}$ ($V_{c\bar q}$, $V_{qc}$, $V_{\bar {q}\bar c}$)
is the potential of $q$ and $\bar c$ ($c$ and $\bar q$, $q$ and $c$,
$\bar q$ and $\bar c$).
From the transition amplitude in the prior form we get
\begin{equation}
\sigma^{\rm prior}(S,m_S,\sqrt {s},T) 
=\frac{1}{32\pi s}\frac{|\vec{P}^{\prime }(\sqrt{s})|
}{|\vec{P}(\sqrt{s})|}\int_{0}^{\pi }d\theta
|\mathcal{M}_{\rm fi}^{\rm prior}|^{2}\sin \theta.
\end{equation}
The scattering in the post form means that gluon exchange occurs after quark 
interchange. The corresponding transition amplitude is
\begin{equation}
{\cal M}_{\rm fi}^{\rm post}  = 
4\sqrt {E_{q\bar{q}} E_{c\bar{c}}E_{q\bar{c}}E_{c\bar{q}}}
\langle\psi_{q\bar {c}}|\langle\psi_{c\bar{q}}|
(V_{q\bar {q}} +V_ {c \bar {c}} +V_ {qc}+V_{\bar {q}\bar {c}})
|\psi_{q\bar {q}}\rangle|\psi_{c\bar {c}}\rangle,
\end{equation}
which gives
\begin{equation}
\sigma^{\rm post}(S,m_S,\sqrt {s},T) 
=\frac{1}{32\pi s}\frac{|\vec{P}^{\prime }(\sqrt{s})|
}{|\vec{P}(\sqrt{s})|}\int_{0}^{\pi }d\theta
|\mathcal{M}_{\rm fi}^{\rm post}|^{2}\sin \theta.
\end{equation}
Since $\sigma^{\rm prior}$ may differ from $\sigma^{\rm post}$ 
\cite{MottM,BBS,WC}, the unpolarized cross section is
\begin{eqnarray}
\sigma ^{\rm unpol} (\sqrt {s}, T) & = &
\frac{1}{(2S_{q\bar q} +1) (2S_ {c\bar c}+1) (2L_ {c\bar c}+1)}
\sum \limits_ {SL_{c\bar{c}z}} (2S+1)
\nonumber\\
&&\times\frac{\sigma^{\rm prior}(S,m_S,\sqrt {s},T)+\sigma^{\rm post}
(S,m_S,\sqrt {s},T)}{2}.
\end{eqnarray}

The potential used in Eqs. (8) and (10) includes a central spin-independent
potential denoted by $V_{\rm si}$ and a spin-spin interaction denoted by
$V_{\rm ss}$:
\begin{equation}
V_{ab}(\vec {r}) = V_{\rm si}(\vec {r}) + V_{\rm ss}(\vec {r}),
\end{equation}
where $ab$ represents $q\bar c$, $c\bar q$, $qc$, $\bar {q}\bar {c}$, 
$q\bar q$, or $c\bar c$, and $\vec r$ is the relative coordinate of $a$ and 
$b$. The central spin-independent potential is
\begin{equation}
V_{\rm {si}}(\vec {r}) =
-\frac {\vec {\lambda}_a}{2} \cdot \frac {\vec{\lambda}_b}{2}
\frac{3}{4} D \left[ 1.3- \left( \frac {T}{T_{\rm c}} \right)^4 \right]
\tanh (Ar)
+ \frac {\vec {\lambda}_a}{2} \cdot \frac {\vec {\lambda}_b}{2}
\frac {6\pi}{25} \frac {v(\lambda r)}{r} \exp (-Er),
\end{equation}
where $D=0. 7$ GeV, $T_{\rm c}=0.175$ GeV, 
$A=1.5[0.75+0.25 (T/{T_{\rm c}})^{10}]^6$ GeV, $E=0. 6$ GeV,
$\lambda=\sqrt{25/16\pi^2 \alpha'}$ with $\alpha'=1.04$ GeV$^{-2}$, and
$\vec {\lambda}_a$ are the Gell-Mann matrices 
for the color generators of constituent $a$. The dimensionless function
$v(x)$ \cite{BT} is
\begin{eqnarray}
v(x)=\frac
{100}{3\pi} \int^\infty_0 \frac {dQ}{Q} \left[\rho (\vec {Q}^2) -\frac {K}{\vec
{Q}^2}\right] \sin \left(\frac {Q}{\lambda}x\right),
\end{eqnarray}
where $K=3/16\pi^2\alpha'$
and $\rho (\vec {Q} ^2)$ is the physical running coupling constant 
at the gluon momentum $\vec {Q}$. 
At short distances the quark interaction is described by 
perturbative QCD in vacuum, and the expression 
$\frac {\vec {\lambda}_a}{2} \cdot \frac {\vec {\lambda}_b}{2}
\frac {6\pi}{25} \frac {v(\lambda r)}{r}$ in the second term of Eq. (14)
is given by one-gluon exchange plus perturbative one- and two-loop corrections
\cite{BT}. Lattice QCD calculations have provided numerical quark-antiquark
free energies at intermediate and large distances \cite{KLP}.
The potential $V_{\rm si}(\vec {r})$ well fits 
$\frac {\vec {\lambda}_a}{2} \cdot \frac {\vec {\lambda}_b}{2}
\frac {6\pi}{25} \frac {v(\lambda r)}{r}$ at short distances and
the free energies at $T/T_{\rm c}>0.55$.

The spin-spin interaction arises from
perturbative one-gluon exchange plus one- and two-loop corrections \cite{Xu},
and includes relativistic effects \cite{BS,BSWX,GI}:
\begin{eqnarray}
V_{\rm ss}(\vec {r})=
- \frac {\vec {\lambda}_a}{2} \cdot \frac {\vec {\lambda}_b}{2}
\frac {16\pi^2}{25}\frac{d^3}{\pi^{3/2}}\exp(-d^2r^2) \frac {\vec {s}_a \cdot 
\vec {s} _b} {m_am_b}
+ \frac {\vec {\lambda}_a}{2} \cdot \frac {\vec {\lambda}_b}{2}
  \frac {4\pi}{25} \frac {1} {r}
\frac {d^2v(\lambda r)}{dr^2} \frac {\vec {s}_a \cdot \vec {s}_b}{m_am_b} ,
\end{eqnarray}
where $\vec {s}_a$ and $m_a$ are the spin and mass of 
constituent $a$, respectively, and the quantity $d$ is given by
\begin{eqnarray}
d^2=\sigma_{0}^2\left[\frac{1}{2}+\frac{1}{2}\left(\frac{4m_a m_b}{(m_a+m_b)^2}
\right)^4\right]+\sigma_{1}^2\left(\frac{2m_am_b}{m_a+m_b}\right)^2,
\end{eqnarray}
where $\sigma_0=0.15$ GeV and $\sigma_1=0.705$.

In Eqs. (8) and (10) the colour part of the state $\mid \psi_{q\bar q}>$
is the familiar colour-singlet state,
\begin{equation}
\mid q\bar{q}, \rm{colour}>=\frac {1}{\sqrt {3}}\sum\limits_{n=1}^{3}
\sum\limits_{\bar{n}=1}^{3}\delta_{n\bar{n}}\mid n\bar{n}>,
\end{equation}
where $n$ and $\bar n$ denote the quark colour and the antiquark colour, 
respectively. The spin part of $\mid \psi_{q\bar q}>$ is
\begin{equation}
\mid S_{q\bar{q}}S_{q\bar{q}z} >=
\sum\limits_{S_{qz}}\sum\limits_{S_{\bar{q}z}}
<\frac {1}{2}\frac {1}{2}S_{qz}S_{\bar{q}z} \mid S_{q\bar{q}}S_{q\bar{q}z}>
\mid \frac {1}{2}S_{qz}> \mid \frac {1}{2} S_{\bar{q}z} >,
\end{equation}
where $S_{q\bar{q}z}$ is the magnetic projection quantum number of 
$S_{q\bar{q}}$, $<\frac {1}{2}\frac {1}{2}S_{qz}S_{\bar{q}z} \mid 
S_{q\bar{q}}S_{q\bar{q}z}>$ are the Clebsch-Gordan coefficients, 
$\mid \frac {1}{2}S_{qz}>$ is the spin-$\frac{1}{2}$ state with the $z$
component $S_{qz}$, and $\mid \frac {1}{2} S_{\bar{q}z} >$ is the 
spin-$\frac{1}{2}$ state with the $z$ component $S_{\bar{q}z}$. 
Denote the flavour part of $\mid \psi_{q\bar q}>$ by $\mid q\bar{q}, 
\rm{flavour}>$, and the flavour wave functions of mesons in the ground-state
pseudoscalar nonet and the ground-state vector nonet can be found in Refs.
\cite{Close,Griffiths}. For instance, $\mid q\bar{q}, {\rm flavour}>$ is 
$\mid u\bar{s}>$ for $K^{*+}$, $\mid d\bar{s}>$ for $K^{*0}$, 
$-\mid s\bar{d}>$ for $\bar{K}^{*0}$, or $\mid s\bar{u}>$ for $K^{*-}$.
We calculate the transition amplitudes, ${\cal M}_{\rm fi}^{\rm prior}$ and 
${\cal M}_{\rm fi}^{\rm post}$, in momentum space, and thus need 
the relative-motion wave function of $q\bar q$ in momentum space. The 
relative-motion wave function is 
denoted by $\phi_{q\bar{q}{\rm rel}}$ and is the product of the spherical
harmonics and the Fourier transform of the radial wave function obtained from
the Schr\"odinger equation with the potential given in Eq. (13). 
$\phi_{q\bar{q}{\rm rel}}$ is normalized according to $\int \frac 
{d^3p_{q\bar{q}}}{(2\pi)^3} \phi_{q\bar{q}{\rm rel}}^* \phi_{q\bar{q}{\rm rel}}
=1$, where $\vec {p}_{q\bar{q}}$ is the relative momentum of $q$ and $\bar q$.
As the product of the colour, spin, flavour, and relative-motion wave 
functions of $q\bar{q}$, $\mid \psi_{q\bar q}>$ is written as
\begin{equation}
\mid \psi_{q\bar q}> = \phi_{q\bar{q}{\rm rel}} \mid q\bar{q}, {\rm colour}>
\mid q\bar{q}, {\rm flavour}> \mid S_{q\bar{q}}S_{q\bar{q}z} >.
\end{equation}
The other wave functions ($\mid \psi_{c\bar c}>$, $\mid \psi_{q\bar c}>$, and 
$\mid \psi_{c\bar q}>$) used in Eqs. (8) and (10) are written similarly as
\begin{equation}
\mid \psi_{c\bar c}> = \phi_{c\bar{c}{\rm rel}} \mid c\bar{c}, {\rm colour}>
\mid c\bar{c}, {\rm flavour}> \mid S_{c\bar{c}}S_{c\bar{c}z} >,
\end{equation}
\begin{equation}
\mid \psi_{q\bar c}> = \phi_{q\bar{c}{\rm rel}} \mid q\bar{c}, {\rm colour}>
\mid q\bar{c}, {\rm flavour}> \mid S_{q\bar{c}}S_{q\bar{c}z} >,
\end{equation}
\begin{equation}
\mid \psi_{c\bar q}> = \phi_{c\bar{q}\rm{rel}} \mid c\bar{q}, {\rm colour}>
\mid c\bar{q}, {\rm flavour}> \mid S_{c\bar{q}}S_{c\bar{q}z} >,
\end{equation}
where $\phi_{c\bar{c}{\rm rel}}$ ($\phi_{q\bar{c}{\rm rel}}$, 
$\phi_{c\bar{q}\rm{rel}}$), $\mid c\bar{c}, {\rm colour}>$ 
($\mid q\bar{c}, {\rm colour}>$, $\mid c\bar{q}, {\rm colour}>$),
$\mid c\bar{c}, {\rm flavour}>$ ($\mid q\bar{c}, {\rm flavour}>$, 
$\mid c\bar{q}, {\rm flavour}>$), and $\mid S_{c\bar{c}}S_{c\bar{c}z} >$ 
($\mid S_{q\bar{c}}S_{q\bar{c}z} >$, $\mid S_{c\bar{q}}S_{c\bar{q}z} >$) are
the relative-motion, colour, flavour, and spin parts of $c\bar c$ ($q\bar c$,
$c\bar q$), respectively.

In the prior form of the reaction $A(q\bar{q})+B(c\bar{c}) \to 
C(q\bar{c})+D(c\bar{q})$ the colour interaction between a constituent of meson 
$A(q\bar{q})$ and a constituent of meson $B(c\bar{c})$ 
turns the colour-singlet states $A$ and $B$ into 
colour-octet states of $q\bar q$ and $c\bar c$, respectively. During 
propagation of quarks and antiquarks quark interchange, i.e., exchange of $q$
and $c$ or of $\bar q$ and $\bar c$ causes $q$ ($c$) to find $\bar c$ 
($\bar q$) to get the colour-singlet state $C$ ($D$). In the post 
form of $A(q\bar{q})+B(c\bar{c}) \to C(q\bar{c})+D(c\bar{q})$ quark 
interchange, i.e., exchange of a constituent of meson $A$ and a constituent of 
meson $B$ produces a colour-octet state of $q\bar c$ and another
colour-octet state of $c\bar q$. The colour interaction between a constituent 
of $q\bar c$ and a constituent of $c\bar q$
makes $q\bar c$ ($c\bar q$) colourless so that the bound state $C$ ($D$) is
formed.

The first term given in Eq. (14) stands for the confining potential. In the
confinement regime the mesonic quark-antiquark relative-motion wave functions
mainly determined by the confining potential are nonperturbative.
At low energies
near threshold of $A(q\bar{q})+B(c\bar{c}) \to C(q\bar{c})+D(c\bar{q})$ the
nonperturbative part of the
$q\bar q$ and $c\bar c$ wave functions must overlap during
the transition from mesons $A$ and $B$ to the colour-octet states of $q\bar q$
and $c\bar c$. Even though the distance of $q$ and $\bar c$ ($c$ and $\bar q$)
is large, the nonperturbative correlation corresponding to the nonperturbative
part leads to formation of a bound state of $q$ and $\bar c$ 
($c$ and $\bar q$) \cite{MBQ,BR} during the hadronization of $q$, $\bar q$, 
$c$, and $\bar c$ to form mesons $C$ and $D$.

The Schr\"odinger equation with the potential given in Eq. (13) at $T=0$  is 
solved to reproduce the experimental masses of $\pi$, $\rho$, $K$, $K^*$, 
$J/\psi$, $\psi'$, $\chi_{c}$, $D$, $D^*$, $D_s$, and $D^*_s$ mesons 
\cite{PDG}. In the
Schr\"odinger equation and the potential the masses of the up quark, the down
quark, the strange quark, and the charm quark are 0.32 GeV, 0.32 GeV, 0.5 GeV,
and 1.51 GeV, respectively. With the pionic quark-antiquark relative-motion 
wave functions obtained in solving the Schr\"odinger equation,
the experimental data of $S$-wave $I=2$ elastic phase shifts for $\pi\pi$ 
scattering in vacuum for $0 < \sqrt {s} < 2.4$ GeV 
\cite{Colton,Durusoy,Hoogland,Losty} are reproduced in the Born approximation
and in the quark-interchange mechanism.

\vspace{0.5cm}
\leftline{\bf III. NUMERICAL CROSS SECTIONS AND DISCUSSIONS }
\vspace{0.5cm}

We establish the notation 
$ K= \left( \begin{array}{c} K^+ \\ K^0 \end{array} \right) $,
$\bar{K}= \left( \begin{array}{c} \bar{K}^0 \\ K^- \end{array} \right)$,
$ K^*= \left( \begin{array}{c} K^{*+} \\ K^{*0} \end{array} \right) $,
$\bar{K}^*= \left( \begin{array}{c}\bar{K}^{*0} \\ K^{*-} \end{array} \right)$,
$ D= \left( \begin{array}{c} D^+ \\ D^0 \end{array} \right) $,
$\bar{D}= \left( \begin{array}{c} \bar{D}^0 \\ D^- \end{array} \right)$,
$ D^*= \left( \begin{array}{c} D^{*+} \\ D^{*0} \end{array} \right)$, and
$\bar{D}^*= \left( \begin{array}{c} \bar{D}^{*0} \\ D^{*-} \end{array} 
\right)$. 
We consider the following reactions:
\begin{eqnarray}
\nonumber
K^*+J/\psi\to\bar{D}+D^+_s,\quad K^*+J/\psi\to\bar{D}^{*}+D^+_s,
\\ \nonumber
K^*+J/\psi\to\bar{D}+D^{*+}_s,\quad K^*+J/\psi\to\bar{D}^{*}+D^{*+}_s;
\\ \nonumber
K^*+\psi'\to\bar{D}+D^+_s,\quad  K^*+\psi'\to\bar{D}^{*}+D^+_s,
\\ \nonumber
K^*+\psi'\to\bar{D}+D^{*+}_s, \quad K^*+\psi'\to\bar{D}^{*}+D^{*+}_s;
\\ \nonumber
K^*+\chi_{c}\to\bar{D}+D^+_s, \quad K^*+\chi_{c}\to\bar{D}^{*}+D^+_s,
\\ \nonumber
K^*+\chi_{c}\to\bar{D}+D^{*+}_s, \quad K^*+\chi_{c}\to\bar{D}^{*}+D^{*+}_s.
\end{eqnarray}
These reactions are governed by quark interchange.

The transition amplitudes given in Eqs. (8) and (10) are further written as
\begin{eqnarray}
{\cal M}_{\rm fi}^{\rm prior} & = &
4\sqrt {E_{q\bar {q}}E_{c\bar {c}}E_{q\bar {c}}E_{c\bar {q}}}
<q\bar{c},{\rm flavour} \mid <c\bar{q},{\rm flavour} 
\mid q\bar{q},{\rm flavour}> \mid c\bar{c},{\rm flavour}>      \nonumber    \\
& & 
<q\bar {c}, {\rm colour} \mid <c\bar {q}, {\rm colour} \mid
<S_{q\bar{c}}S_{q\bar{c}z}\mid <S_{c\bar{q}}S_{c\bar{q}z}\mid
     \int \frac {d^3 p_{q\bar {c}}}{(2\pi)^3} 
     \frac {d^3 p_{c\bar {q}}}{(2\pi)^3}      \nonumber    \\
& & 
[ \phi^*_{q\bar {c}{\rm rel}} (\vec {p}_{q\bar {c}}) 
\phi^*_{c\bar {q}{\rm rel}} (\vec {p}_{c\bar {q}})
V_{q\bar {c}}
\phi_{q\bar {q}{\rm rel}} (\vec {p}_{c\bar{q}}
+\frac {m_{\bar q}}{m_q+m_{\bar q}}\vec{P}
+\frac {m_{\bar q}}{m_c+m_{\bar q}}\vec{P}^\prime )     \nonumber    \\
& &
\phi_{c\bar {c}{\rm rel}} (\vec {p}_{c\bar {q}}
+\frac {m_c}{m_c+m_{\bar c}}\vec{P}
-\frac {m_c}{m_c+m_{\bar q}}\vec{P}^\prime )    \nonumber   \\
& & 
+\phi^*_{q\bar{c}{\rm rel}} (\vec {p}_{q\bar {c}}) 
\phi^*_{c\bar {q}{\rm rel}} (\vec {p}_{c\bar {q}})
V_{c\bar {q}}
\phi_{q\bar {q}{\rm rel}} (\vec {p}_{q\bar {c}}
-\frac {m_q}{m_q+m_{\bar q}}\vec{P}
+\frac {m_q}{m_q+m_{\bar c}}\vec{P}^\prime )   \nonumber   \\
& &
\phi_{c\bar {c}{\rm rel}} (\vec {p}_{q\bar {c}}
-\frac {m_{\bar c}}{m_c+m_{\bar c}}\vec{P}
-\frac {m_{\bar c}}{m_q+m_{\bar c}}\vec{P}^\prime )    \nonumber   \\
& & 
+\phi^*_{q\bar {c}{\rm rel}} (\vec {p}_{q\bar {c}}) 
\phi^*_{c\bar {q}{\rm rel}} (\vec {p}_{c\bar {q}})
V_{qc}
\phi_{q\bar {q}{\rm rel}} (\vec {p}_{c\bar {q}}
+\frac {m_{\bar q}}{m_q+m_{\bar q}}\vec{P}
+\frac {m_{\bar q}}{m_c+m_{\bar q}}\vec{P}^\prime )    \nonumber   \\
& &
\phi_{c\bar {c}{\rm rel}} (\vec {p}_{q\bar {c}}
-\frac {m_{\bar c}}{m_c+m_{\bar c}}\vec{P}
-\frac {m_{\bar c}}{m_q+m_{\bar c}}\vec{P}^\prime )    \nonumber   \\
& & 
+\phi^*_{q\bar {c}{\rm rel}} (\vec {p}_{q\bar {c}}) 
\phi^*_{c\bar {q}{\rm rel}} (\vec {p}_{c\bar {q}})
V_{\bar{q}\bar{c}}
\phi_{q\bar {q}{\rm rel}} (\vec {p}_{q\bar {c}}
-\frac {m_q}{m_q+m_{\bar q}}\vec{P}
+\frac {m_q}{m_q+m_{\bar c}}\vec{P}^\prime )    \nonumber   \\
& &
\phi_{c\bar {c}{\rm rel}} (\vec {p}_{c\bar {q}}
+\frac {m_c}{m_c+m_{\bar c}}\vec{P}
-\frac {m_c}{m_c+m_{\bar q}}\vec{P}^\prime ) ]    \nonumber   \\
& & 
\mid S_{q\bar{q}}S_{q\bar{q}z}> \mid S_{c\bar{c}}S_{c\bar{c}z}>
\mid q\bar {q}, {\rm colour}> \mid c\bar {c}, {\rm colour}>,
\nonumber   \\
\end{eqnarray}
\begin{eqnarray}
{\cal M}_{\rm fi}^{\rm post} & = & 
4\sqrt {E_{q\bar {q}}E_{c\bar {c}}E_{q\bar {c}}E_{c\bar {q}}}
<q\bar{c},{\rm flavour} \mid <c\bar{q},{\rm flavour} 
\mid q\bar{q},{\rm flavour}> \mid c\bar{c},{\rm flavour}>      \nonumber   \\
& & 
<q\bar {c}, {\rm colour} \mid <c\bar {q}, {\rm colour} \mid
<S_{q\bar{c}}S_{q\bar{c}z}\mid <S_{c\bar{q}}S_{c\bar{q}z}\mid
     \int \frac {d^3 p_{q\bar {q}}}{(2\pi)^3} 
     \frac {d^3 p_{c\bar {c}}}{(2\pi)^3}    \nonumber   \\
& & 
[ \phi^*_{q\bar {c}{\rm rel}} (\vec {p}_{c\bar {c}}
+\frac {m_{\bar c}}{m_c+m_{\bar c}}\vec{P}
+\frac {m_{\bar c}}{m_q+m_{\bar c}}\vec{P}^\prime ) 
\phi^*_{c\bar {q}{\rm rel}} (\vec {p}_{c\bar {c}}
-\frac {m_c}{m_c+m_{\bar c}}\vec{P}
+\frac {m_c}{m_c+m_{\bar q}}\vec{P}^\prime )    \nonumber   \\
& &
V_{q\bar{q}}
\phi_{q\bar {q}{\rm rel}} (\vec {p}_{q\bar {q}}) 
\phi_{c\bar {c}{\rm rel}} (\vec {p}_{c\bar {c}})       \nonumber   \\
& & 
+\phi^*_{q\bar {c}{\rm rel}} (\vec {p}_{q\bar{q}}
+\frac {m_q}{m_q+m_{\bar q}}\vec{P}
-\frac {m_q}{m_q+m_{\bar c}}\vec{P}^\prime ) 
\phi^*_{c\bar {q}{\rm rel}} (\vec {p}_{q\bar {q}}
-\frac {m_{\bar q}}{m_q+m_{\bar q}}\vec{P}
-\frac {m_{\bar q}}{m_c+m_{\bar q}}\vec{P}^\prime )    \nonumber   \\
& &
V_{c\bar{c}}
\phi_{q\bar {q}{\rm rel}} (\vec {p}_{q\bar {q}}) 
\phi_{c\bar {c}{\rm rel}} (\vec {p}_{c\bar {c}})       \nonumber   \\
& & 
+\phi^*_{q\bar {c}{\rm rel}} (\vec {p}_{c\bar {c}}
+\frac {m_{\bar c}}{m_c+m_{\bar c}}\vec{P}
+\frac {m_{\bar c}}{m_q+m_{\bar c}}\vec{P}^\prime ) 
\phi^*_{c\bar {q}{\rm rel}} (\vec {p}_{q\bar {q}}
-\frac {m_{\bar q}}{m_q+m_{\bar q}}\vec{P}
-\frac {m_{\bar q}}{m_c+m_{\bar q}}\vec{P}^\prime )    \nonumber   \\
& &
V_{qc}
\phi_{q\bar {q}{\rm rel}} (\vec {p}_{q\bar {q}}) 
\phi_{c\bar {c}{\rm rel}} (\vec {p}_{c\bar {c}})       \nonumber   \\
& & 
+\phi^*_{q\bar {c}{\rm rel}} (\vec {p}_{q\bar {q}}
+\frac {m_q}{m_q+m_{\bar q}}\vec{P}
-\frac {m_q}{m_q+m_{\bar c}}\vec{P}^\prime ) 
\phi^*_{c\bar {q}{\rm rel}} (\vec {p}_{c\bar {c}}
-\frac {m_c}{m_c+m_{\bar c}}\vec{P}
+\frac {m_c}{m_c+m_{\bar q}}\vec{P}^\prime )    \nonumber   \\
& &
V_{\bar{q}\bar{c}}
\phi_{q\bar {q}{\rm rel}} (\vec {p}_{q\bar {q}}) 
\phi_{c\bar {c}{\rm rel}} (\vec {p}_{c\bar {c}}) ]     \nonumber   \\
& & 
\mid S_{q\bar{q}}S_{q\bar{q}z}> \mid S_{c\bar{c}}S_{c\bar{c}z}>
\mid q\bar {q}, {\rm colour}> \mid c\bar {c}, {\rm colour}>,
\nonumber   \\
\end{eqnarray}
where $\vec{p}_{c\bar c}$ ($\vec{p}_{q\bar c}$, $\vec{p}_{c\bar q}$) is the
relative momentum of $c$ and $\bar c$ ($q$ and $\bar c$, $c$ and $\bar q$).
Given the wave functions in Eqs. (20)-(23), we calculate colour matrix 
elements, flavour matrix elements, spin matrix elements, and spatial matrix
elements. The colour matrix elements $<q\bar {c}, {\rm colour} \mid 
<c\bar {q}, {\rm colour} \mid \frac {\vec{\lambda}_a}{2} \cdot \frac
{\vec{\lambda}_b}{2} \mid q\bar {q}, {\rm colour}>\mid c\bar{c}, {\rm colour}>$
are -4/9 for $V_{q\bar c}$, $V_{c\bar{q}}$, $V_{q\bar q}$, and $V_{c\bar c}$
and 4/9 for $V_{qc}$ and $V_{\bar{q}\bar{c}}$. For antiquarks 
$\frac {\vec{\lambda}_a}{2}$ ($\frac {\vec{\lambda}_b}{2}$) is as usual
replaced by
$-\frac {\vec{\lambda}_a^T}{2}$ ($-\frac {\vec{\lambda}_b^T}{2}$). For the
twelve $K^*$-charmonium dissociation reactions the flavour matrix elements
$<q\bar{c},{\rm flavour} \mid <c\bar{q},{\rm flavour} 
\mid q\bar{q},{\rm flavour}>
\mid c\bar{c},{\rm flavour}>$ equal 1. Since the spin operator $\vec{s}_q +
\vec{s}_{\bar q} + \vec{s}_c +\vec{s}_{\bar c}$ commutes with the potential
$V_{ab}$, the total spin of the two initial mesons equals the total spin of the
two final mesons. The spin matrix elements 
$<S_{q\bar{c}}S_{q\bar{c}z}\mid <S_{c\bar{q}}S_{c\bar{q}z}\mid 
\vec{s}_a \cdot \vec{s}_b
\mid S_{q\bar{q}}S_{q\bar{q}z}> \mid S_{c\bar{c}}S_{c\bar{c}z}>$
are independent of the magnetic projection quantum number $m_S$, and
depend on the total spin and the spins of the four mesons. The values
of the spin matrix elements can be found in Ref. \cite{BS}. 

The spatial matrix elements are expressed as the two integrals in Eqs. (24) and
(25). The two integrals involve the quark-antiquark relative-motion wave 
functions. Diagrams ``C1 prior'' and ``C2 prior'' in Fig. 1 are ``capture''
diagrams \cite{BS} because the interacting quark-antiquark pair scatter into
the same final meson. Diagrams ``C1 post'' and ``C2 post'' in Fig. 2 are also
``capture'' diagrams because the interacting quark-antiquark pair come from 
the same initial meson. The interacting quark-antiquark pair are in the
colour-singlet state, so are the other quark-antiquark pair. Diagrams ``T1 
prior'', ``T2 prior'', ``T1 post'', and ``T2 post'' are ``transfer'' diagrams
\cite{BS} because the interacting quark-quark or antiquark-antiquark pair
scatter (transfer momentum) into different final mesons in the prior form or
come (bring momentum) from different initial mesons in the post form. The
mesonic quark-antiquark relative-motion wave functions are functions of the
quark-antiquark relative momenta. The interaction between two constituents
and the quark interchange result in that the quark-antiquark relative momenta
are related between an initial meson and a final meson
through the linear combination of $\vec P$ and $\vec{P}^\prime$. For example,
from the first term enclosed by brackets in ${\cal M}_{\rm fi}^{\rm prior}$
we get
$\vec{p}_{q\bar {q}} = \vec {p}_{c\bar{q}}
+\frac {m_{\bar q}}{m_q+m_{\bar q}}\vec{P}
+\frac {m_{\bar q}}{m_c+m_{\bar q}}\vec{P}^\prime$ and
$\vec{p}_{c\bar {c}} = \vec {p}_{c\bar {q}}
+\frac {m_c}{m_c+m_{\bar c}}\vec{P}
-\frac {m_c}{m_c+m_{\bar q}}\vec{P}^\prime$,
and from the third term enclosed by brackets in ${\cal M}_{\rm fi}^{\rm post}$
we get
$\vec{p}_{q\bar {c}} = \vec {p}_{c\bar {c}}
+\frac {m_{\bar c}}{m_c+m_{\bar c}}\vec{P}
+\frac {m_{\bar c}}{m_q+m_{\bar c}}\vec{P}^\prime$ and
$\vec{p}_{c\bar {q}} = \vec {p}_{q\bar {q}}
-\frac {m_{\bar q}}{m_q+m_{\bar q}}\vec{P}
-\frac {m_{\bar q}}{m_c+m_{\bar q}}\vec{P}^\prime$.
For ``capture'' diagrams and ``transfer'' diagrams the
relation is different. Corresponding to the ``capture'' diagrams ``C1 prior''
and ``C2 prior'', the quark-antiquark relative momenta of the two initial 
mesons are only related to the quark-antiquark relative momentum of the final 
meson which is not the interacting quark-antiquark pair. Corresponding to the
``capture'' diagrams ``C1 post'' and ``C2 post'', the quark-antiquark relative
momenta of the two final mesons are only related to the quark-antiquark 
relative
momentum of the initial meson which is not the interacting quark-antiquark
pair. As to the ``transfer'' diagrams the quark-antiquark relative momentum of
a final meson (another final meson) is only related to the one of an initial
meson (another initial meson). In low-energy meson-meson collisions 
$\mid \vec{P} \mid$ and $\mid \vec{P}^\prime \mid$ are small. This allows small
values for $\mid \vec{p}_{q\bar q} \mid$ and $\mid \vec{p}_{c\bar c} \mid$ 
as seen from ${\cal M}_{\rm fi}^{\rm prior}$ and $\mid \vec{p}_{q\bar c} \mid$
and $\mid \vec{p}_{c\bar q} \mid$ as seen from ${\cal M}_{\rm fi}^{\rm post}$. 
$\phi_{q\bar{q}{\rm rel}} (\vec{p}_{q\bar{q}})$ 
($\phi_{c\bar{c}{\rm rel}} (\vec{p}_{c\bar{c}})$,
$\phi_{q\bar{c}{\rm rel}} (\vec{p}_{q\bar{c}})$,
$\phi_{c\bar{q}{\rm rel}} (\vec{p}_{c\bar{q}})$) decreases rapidly while
$\mid \vec{p}_{q\bar q} \mid$ ($\mid \vec{p}_{c\bar c} \mid$,
$\mid \vec{p}_{q\bar c} \mid$, $\mid \vec{p}_{c\bar q} \mid$) increases.
Then, the nonperturbative
part of the mesonic quark-antiquark relative-motion wave functions makes
dominant contributions to ${\cal M}_{\rm fi}^{\rm prior}$ and
${\cal M}_{\rm fi}^{\rm post}$. In high-energy meson-meson collisions 
$\mid \vec{P} \mid$ and $\mid \vec{P}^\prime \mid$ are large. 
The absolute values of $\vec{p}_{q\bar q}$ and $\vec{p}_{c\bar c}$
in ${\cal M}_{\rm fi}^{\rm prior}$ and of $\vec{p}_{q\bar c}$ and 
$\vec{p}_{c\bar q}$ in ${\cal M}_{\rm fi}^{\rm post}$ are generally large. 
The perturbative part of the mesonic quark-antiquark relative-motion wave
functions, which is mainly determined by the second term in Eq. (14) and the 
spin-spin interaction, significantly affects ${\cal M}_{\rm fi}^{\rm prior}$
and ${\cal M}_{\rm fi}^{\rm post}$.

According to Eq. (12)
we calculate unpolarized cross sections at the six temperatures 
$T/T_{\rm c} =0$, $0.65$, $0.75$, $0.85$, $0.9$, and $0.95$. In Figs. 3-14
we plot the unpolarized cross sections for the twelve $K^*$-charmonium
dissociation reactions.
Depending on temperature, a reaction is either endothermic or exothermic.
The numerical cross sections for endothermic reactions are parametrized as
\begin{eqnarray}
\sigma^{\rm unpol}(\sqrt {s},T)
&=&a_1 \left( \frac {\sqrt {s} -\sqrt {s_0}} {b_1} \right)^{c_1}
\exp \left[ c_1 \left( 1-\frac {\sqrt {s} -\sqrt {s_0}} {b_1} \right) \right] 
\nonumber \\
&&+ a_2 \left( \frac {\sqrt {s} -\sqrt {s_0}} {b_2} \right)^{c_2}
\exp \left[ c_2 \left( 1-\frac {\sqrt {s} -\sqrt {s_0}} {b_2} \right) \right],
\end{eqnarray}
where $\sqrt{s_0}$ is the threshold energy, and $a_1$, $b_1$, $c_1$, $a_2$, 
$b_2$, and $c_2$ are parameters. The numerical cross sections
for exothermic reactions are parametrized as
\begin{eqnarray}
\sigma^{\rm unpol}(\sqrt {s},T)
&=&\frac{\vec{P}^{\prime 2}}{\vec{P}^2}
\left\{a_1 \left( \frac {\sqrt {s} -\sqrt {s_0}} {b_1} \right)^{c_1}
\exp \left[ c_1 \left( 1-\frac {\sqrt {s} -\sqrt {s_0}} {b_1} \right) \right] 
\right.
\nonumber \\
&&+ \left.
a_2 \left( \frac {\sqrt {s} -\sqrt {s_0}} {b_2} \right)^{c_2}
\exp \left[ c_2 \left( 1-\frac {\sqrt {s} -\sqrt {s_0}} {b_2} \right) \right]
\right\}.
\end{eqnarray}
The parameter values are listed in Tables
1-3. We follow a procedure presented in Ref. 
\cite{ZX} to get cross sections at any temperature between $0.65T_{\rm c}$
and $T_{\rm c}$. This needs the quantities $d_0$ and $\sqrt{s_z}$ which are
also listed in Tables 1-3. 
$d_0$ is the separation between the peak's location on 
the $\sqrt s$-axis and the threshold energy, and $\sqrt{s_z}$ is 
the square root of the Mandelstam variable at which the cross section is 
1/100 of the peak cross section. 

The cross sections for the three $K^*+J/\psi$ reactions in Figs. 4-6 show that 
the peak cross section of each reaction decreases when temperature changes
from $T/T_{\rm c}=0.65$ to 0.85 and increases when temperature changes
from $T/T_{\rm c}=0.85$ to 0.95.
While temperature increases, the value of the central spin-independent 
potential at large distances
becomes smaller and smaller (the confinement becomes weaker and weaker), 
and the Schr\"{o}dinger equation
produces increasing meson radii. Decrease in peak cross section is caused by 
weakening confinement. Increase in peak cross section
is caused by increasing radii of initial mesons. When the decrease is faster 
than the increase, the peak cross section of each reaction shown in Figs. 4-6
goes down as temperature changes from $0.65T_{\rm c}$ to $0.85T_{\rm c}$. 
When the decrease is slower than the increase, the peak cross section 
of each reaction goes up as temperature changes
from $0.85T_{\rm c}$ to $0.95T_{\rm c}$. 

As shown by Fig. 1 of Ref. \cite{ZX}, the $\psi'$ mass is very close to the
$\chi_c$ mass for $0.6T_c \leq T < T_c$. While temperature increases, the
sum of the initial-meson masses decreases more rapidly than the sum of the
final-meson masses. Any $K^*+\psi'$ reaction shown in Figs. 7-10 is
endothermic above a temperature and exothermic below the temperature. The
$K^*+\chi_c$ reaction that has the same final states as the $K^*+\psi'$ 
reaction is endothermic above or exothermic below almost the same temperature.
However, since the $\chi_c$ and $\psi'$ mesons 
have different quantum numbers, the cross section for the $K^*+\chi_c$
reaction is not identical to the one for the $K^*+\psi'$ reaction.

The $K^*$ meson and the $K$ meson have the same strangeness, and the difference
of the $K^*$ mass ($m_{K^*}$) and the $K$ mass ($m_K$) becomes smaller and 
smaller with increasing temperature.
Since the experimental mass splitting $m_{K^\ast}-m_K=0.3963$
GeV is large, the largest cross section of an exothermic
$K^* + {\rm charmonium}$ reaction
shown by the curve at $T=0$ is larger than the one of the exothermic
$K + {\rm charmonium}$ reaction with the same final states \cite{JSX}.
At $T \to T_{\rm c}$ the $K^\ast$ and $K$ mesons
become degenerate in mass. As shown by the curves at $T/T_{\rm c}=0.95$, all 
the $K^*$+charmonium reactions become endothermic. The largest
cross sections of the four reactions, 
$K^* + \psi^\prime \to \bar{D} +D^{*+}_s$, 
$K^* + \chi_c \to \bar{D}^* + D^+_s$, $K^* + \chi_c \to \bar{D} +D^{*+}_s$,
and $K^* + \chi_c \to \bar{D}^* + D^{*+}_s$,
become smaller than the ones of the four reactions,
$K + \psi^\prime \to \bar{D} +D^{*+}_s$, 
$K + \chi_c \to \bar{D}^* + D^+_s$, $K + \chi_c \to \bar{D} +D^{*+}_s$,
and $K + \chi_c \to \bar{D}^* + D^{*+}_s$,
respectively. This is owed to the difference in the mesonic space
wave function, in the spin matrix element of the spin-spin
interaction, and in the overlap of the initial and final spin wave functions
related to the central spin-independent potential.

The $\bar {K}^*$-charmonium dissociation includes
\begin{eqnarray}
\nonumber
\bar{K}^* + J/\psi \to D_s^- + D,\quad \bar{K}^* + J/\psi \to D_s^{*-} + D,
\\ \nonumber
\bar{K}^* + J/\psi \to D_s^- + D^*,\quad \bar{K}^* + J/\psi \to D_s^{*-} + D^*;
\\ \nonumber
\bar{K}^* + \psi' \to D_s^- + D,\quad \bar{K}^* + \psi' \to D_s^{*-} + D,
\\ \nonumber
\bar{K}^* + \psi' \to D_s^- + D^*,\quad \bar{K}^* + \psi' \to D_s^{*-} + D^*;
\\ \nonumber
\bar{K}^* + \chi_c \to D_s^- + D,\quad \bar{K}^* + \chi_c \to D_s^{*-} + D,
\\ \nonumber
\bar{K}^* + \chi_c \to D_s^- + D^*,\quad \bar{K}^* + \chi_c \to D_s^{*-} + D^*.
\end{eqnarray}
Cross sections for these reactions are obtained from the $K^*$-charmonium
dissociation reactions. For example, the cross section for 
$\bar{K}^* + J/\psi \to D_s^- + D$ equals the cross section for
$K^* + J/\psi \to \bar {D} + D^+_s$.

With the quark potential given in Eq. (13) at $T=0$ we can reproduce the
experimental masses of $\pi$, $\rho$, $K$, $K^\ast$, $J/\psi$, $\psi^\prime$,
$\chi_c$, $D$, $D^*$, $D_s$, and $D_s^*$ mesons. In the Born approximation and
in the quark-interchange mechanism the experimental data of $S$-wave $I=2$
elastic phase shifts for $\pi\pi$ scattering in vacuum can be reproduced. 
Therefore, we calculate $K^*$-charmonium dissociation cross sections with the 
quark potential, in the Born approximation and in the quark-interchange
mechanism. The study of meson-charmonium dissociation reactions is itself of 
theoretical interest. Unfortunately, no direct meson-charmonium collision
experiments are possible. Up to now, attempts to constrain theoretical models
about $\pi J/\psi \to \pi \psi'$ have been made in Refs. \cite{BK,SSZ,CS} 
from the experimental data on the hadronic decay $\psi' \to \pi \pi J/\psi$. 
We can not use the cross sections obtained in Refs. \cite{BK,SSZ,CS} for
$\pi J/\psi \to \pi \psi'$
to constrain our charmonium dissociation cross sections since the reaction can
not be studied in the quark-interchange mechanism. However, we can use the
experimental data on elastic $\pi \pi$ scattering for $I=2$ and elastic $\pi K$
scattering for $I=3/2$ to examine our cross sections for $\pi J/\psi$, 
$\pi \chi_c$, and $\pi \psi'$ dissociation. This is because the five types of
reactions are governed by the quark-interchange mechanism and have the pion as
an incident hadron. The comparison of our theoretical results with the
experimental data is shown in Fig. 30. The solid, dashed, and dotted curves
indicate the cross sections obtained in Ref. \cite{ZX} for 
$\pi J/\psi \to \bar{D}^*D+\bar{D}D^*+\bar{D}^*D^*$, 
for $\pi \psi^\prime \to \bar{D}^*D+\bar{D}D^*+\bar{D}^*D^*$,
and for $\pi \chi_c \to \bar{D}^*D+\bar{D}D^*+\bar{D}^*D^*$, respectively.
The experimental data were obtained in the region $\sqrt {s}<2.8$ GeV, and the
theoretical results are given in the region $\sqrt {s} \geq 3.87583$ GeV.
The two regions are different. To make an easy comparison, 
the three curves have been translated by -3 GeV in $\sqrt s$.
Solving the Schr\"odinger equation with the potential
given in Eq. (13) at $T=0$, we obtain 0.45, 0.534, 0.595, 0.672, and 0.757 fm
as the radii of $J/\psi$, $\pi$, $K$, $\chi_c$, and $\psi'$ mesons, 
respectively. The larger the radii of initial mesons, the larger the cross
section for a reaction. The spin-spin interaction is proportional to the
inverse of the product of the masses of two interacting constituents. From the
elastic $\pi \pi$ scattering or the elastic $\pi K$ scattering to the 
$\pi$-charmonium
dissociation reactions, at least one quark mass changes from the up-quark or
strange-quark mass to the charm-quark mass. Correspondingly, the contribution
of the spin-spin interaction to the cross section gets appreciably smaller.
Therefore, the peak cross section of $\pi J/\psi$ dissociation is much 
smaller than the experimental data on the elastic $\pi K$ scattering for 
$I=3/2$. Even though the $\chi_c$ radius is by 13\% larger than the $K$ radius,
the peak cross section of the $\pi \chi_c$ dissociation is smaller than the 
largest experimental datum
of the elastic $\pi K$ scattering. Since the $\psi'$ radius is largest
among the radii of $J/\psi$, $\pi$, $K$, $\chi_c$, and $\psi^\prime$ mesons,
the peak cross section of the $\pi \psi'$ dissociation is close to the 
experimental data on the elastic $\pi \pi$ scattering for $I=2$ given in Ref.
\cite{CFSW}. Hence, the peak cross
sections of the $\pi$-charmonium dissociation have reasonable orders of
magnitude. We can thus expect that other meson-charmonium dissociation has
reasonable orders of magnitude of peak cross sections.

An indirect way to examine meson-charmonium dissociation cross sections, 
which has been suggested in references, e.g. Ref. \cite{Haglin}, is to use
the cross sections to calculate
nuclear modification factor of $J/\psi$ produced in relativistic heavy-ion
collisions. The collisions may produce charmonia and hadronic matter 
simultaneously, and the charmonia may break up travelling through hadronic 
matter. If the cross sections are unreasonably large, the charmonia dissociated
by hadronic matter cause $J/\psi$  suppression stronger than observed
suppression. The measured nuclear modification factor of $J/\psi$ thus places 
a 
constraint on the charmonium dissociation cross sections. In a future study we 
will calculate the nuclear modification factor using our charmonium 
dissociation cross sections.

\vspace{0.5cm}
\leftline{\bf IV.  DISSOCIATION RATES AND DISCUSSIONS }
\vspace{0.5cm}

In hadronic matter random motion of a kind of hadrons leads to a momentum
distribution.
We use the meson distribution function and the unpolarized cross section to 
define the dissociation rate of charmonium in the interaction with meson. 
The unpolarized
cross sections obtained with the quark potential, in the quark-interchange
mechanism, and in the Born approximation are used to calculate the dissociation
rates in hadronic matter.

\vspace{0.5cm}
\leftline{\bf A. Dissociation rate}
\vspace{0.5cm}

Cross sections for $\pi$-charmonium dissociation reactions and 
$\rho$-charmonium dissociation reactions were obtained in Ref. \cite{ZX}, 
and cross sections for $K$-charmonium dissociation reactions, 
$\bar K$-charmonium dissociation reactions, and 
$\eta$-charmonium dissociation reactions were obtained in Ref. \cite{JSX}. In 
hadronic matter the meson distribution is given by
\begin{equation}
f_i(\vec{k})=\frac{1}{e^{\sqrt{\vec{k}^2+m^2_i}/T}-1},
\end{equation}
where $m_i$ and $\vec{k}$ are the mass and momentum of the meson, respectively.
The number density of meson species $i$ is
\begin{equation}
n_i=g_i\int\frac{d^3k}{(2\pi)^3}f_i(\vec{k}),
\end{equation}
where $g_i$ is the spin-isospin degeneracy factor, and equals 3 for $\pi$, 
9 for $\rho$, 4 for $K$ and $\bar K$, 12 for $K^*$ and $\bar {K}^*$, and 1 
for $\eta$, respectively. The thermal-averaged meson-charmonium 
dissociation cross section is
\begin{equation}
\langle v_{\rm rel}\sigma^{\rm unpol}(\sqrt{s}, T)\rangle
=\frac{g_i\int\frac{d^3k}{(2\pi)^3}v_{\rm rel}\sigma^{\rm unpol}
(\sqrt{s}, T)f_i(\vec{k})}{g_i\int\frac{d^3k}{(2\pi)^3}f_i(\vec{k})},
\end{equation}
where $v_{\rm rel}$ is the relative velocity of the meson and the charmonium. 
Since $v_{\rm rel}$ relies on the meson and charmonium masses, it depends on
temperature.
The dissociation rate of charmonium in the interaction with meson species $i$ 
in hadronic matter is
\begin{equation}
n_i\langle v_{\rm rel}\sigma^{\rm unpol}(\sqrt{s}, T)\rangle
=\frac{g_i}{4\pi^2}\int^\infty_0\int^\pi_0d|\vec{k}|d\theta \sin \theta
\vec{k}^2v_{\rm rel}\sigma^{\rm unpol}(\sqrt{s}, T)f_i(\vec{k}),
\end{equation}
where $\theta$ is the angle between the meson momentum and the charmonium
momentum and
which determines the charmonium suppression due to meson-charmonium 
dissociation in hadronic matter. Since the meson distribution, the unpolarized
cross section, and the relative velocity vary with temperature, the 
dissociation rate depends on temperature.
The dissociation rate of $J/\psi$ with $\pi$
is obtained while $\sigma^{\rm unpol}$ is the sum of the cross sections for
$\pi + J/\psi \to \bar {D}^* + D$, $\pi + J/\psi \to \bar {D} + D^*$, and
$\pi + J/\psi \to \bar {D}^* + D^*$. The dissociation rate of $\psi'$ with
vector kaon is obtained while $\sigma^{\rm unpol}$ is the sum of the cross 
sections for $K^* + \psi' \to \bar {D} + D^+_s$, 
$K^* + \psi' \to \bar {D}^* + D^+_s$, $K^* + \psi' \to \bar {D} + D^{*+}_s$,
and $K^* + \psi' \to \bar {D}^* + D^{*+}_s$. Other dissociation rates for 
charmonium with meson can be similarly calculated according to Table 4.

\vspace{0.5cm}
\leftline{\bf B. Numerical results and discussions}
\vspace{0.5cm}

We calculate dissociation rates of charmonium with pion, $\rho$ meson,
kaon, vector
kaon, and $\eta$ meson in hadronic matter. Numerical results of the rates as 
functions of charmonium momentum are plotted in Figs. 15-29. 
The dissociation rates of charmonium with pion, $\rho$ meson, kaon, vector
kaon, and $\eta$ meson
are shown by the dashed curves, the dotted curves, the dot-dashed curves, 
the dot-dash-dashed curves, and the dot-dot-dashed curves, respectively.
In Figs. 15-19 (Figs. 20-24, Figs. 25-29) every solid curve stands for the
$J/\psi$ ($\psi'$, $\chi_c$) dissociation rate, and is the sum of the 
dissociation rates of $J/\psi$ ($\psi'$, $\chi_c$) with pion, $\rho$ meson,
kaon, vector kaon, and $\eta$ meson. While
temperature increases, the charmonium dissociation rates increase except
the $\chi_c$ dissociation rate at $T/T_{\rm c}=0.9$. This indicates that the
higher the temperature is, the stronger charmonium suppression the mesons
cause. Excluding the $J/\psi$ dissociation rate at $T/T_{\rm c}=0.65$ and 0.75,
the dissociation rates decrease while the charmonium momentum increases. 
At $T/T_{\rm c}=0.65$ in Fig. 15 the dissociation rate
of $J/\psi$ with $\pi$  increases (decreases) when the $J/\psi$ momentum goes
up to (from) 4.7 GeV/$c$, and the one of  $J/\psi$ with kaon
increases (decreases) when the $J/\psi$ momentum goes
up to (from) 3.3 GeV/$c$. Hence, the $J/\psi$ dissociation rate at 
$T/T_{\rm c}=0.65$ has a maximum at $\sqrt {s}=2.2$ GeV. A
similar case at $T/T_{\rm c}=0.75$ is shown in Fig. 16. The decrease of the 
dissociation rates with the increase of charmonium momentum can
be understood with two situations in the following. While the charmonium
momentum increases, the Mandelstam variable $s$ of meson and charmonium
increases, $\sqrt s$ becomes far away from the threshold energy, and the
unpolarized cross section decreases. While the charmonium momentum increases, 
it is also possible that $\sqrt s$ does not become far away from the threshold 
energy. This corresponds to the nearly
collinear motion of meson and charmonium. The angle $\theta$ in Eq. (31) is
thus small, and the contribution to the integration from such $\vec k$ is
negligible. The two situations make the dissociation rate decrease while the 
charmonium momentum increases.

In Eq. (31) the relative velocity is
\begin{equation}
v_{\rm rel}=\frac {\sqrt {(P_i \cdot P_{c\bar c})^2-m_i^2m_{c\bar c}^2}}
{E_iE_{c\bar c}},
\end{equation}
where the meson four-momentum is $P_i=(E_i,\vec {k})$ and from which we get
\begin{equation}
\frac {\partial v_{\rm rel}}{\partial \mid \vec {P}_{c\bar c} \mid} =
\left( \frac {P_i \cdot P_{c\bar c}}{v_{\rm rel} E_i E_{c\bar c}}-v_{\rm rel} 
\right)
\frac {\mid \vec {P}_{c\bar c} \mid}{E_{c\bar c}^2}
-\frac {\mid \vec {k} \mid P_i \cdot P_{c\bar c}}
{v_{\rm rel} E_i^2 E_{c\bar c}^2} \cos \theta .
\end{equation}
The unpolarized cross
sections are functions of temperature and the Mandelstam variable
\begin{equation}
s=(P_i+P_{c\bar c})^2=m_i^2+m_{c\bar c}^2+2P_i \cdot P_{c\bar c},
\end{equation}
from which we have
\begin{equation}
\frac {\partial \sqrt s}{\partial \mid \vec {P}_{c\bar c} \mid} =
\frac {1}{\sqrt s}
\left( \frac {E_i}{E_{c\bar c}} \mid \vec {P}_{c\bar c} \mid
- \mid \vec {k} \mid \cos \theta \right) .
\end{equation}
The first derivative of the dissociation rate with respect to the charmonium 
momentum is
\begin{eqnarray}
& & 
\frac {\partial n_i\langle v_{\rm rel}\sigma^{\rm unpol}(\sqrt{s}, T)\rangle}
{\partial \mid \vec {P}_{c\bar c} \mid}
= \frac{g_i}{4\pi^2}\int^\infty_0\int^\pi_0d|\vec{k}|d\theta \sin \theta
\vec{k}^2
\nonumber     \\
& & \left( \frac {\partial v_{\rm rel}}{\partial \mid \vec {P}_{c\bar c} \mid}
\sigma^{\rm unpol}(\sqrt{s}, T)
+ v_{\rm rel}\frac {\partial \sigma^{\rm unpol}(\sqrt{s}, T)}{\partial 
\sqrt s}\frac {\partial \sqrt s}{\partial \mid \vec {P}_{c\bar c} \mid} \right)
f_i(\vec{k}).
\end{eqnarray}
When the charmonium momentum is 0, the first derivative is 0. 
This result is indicated by all
curves in Figs. 15-29, and means that the dissociation rate of very slowly
moving charmonium equals the dissociation rate of charmonium at rest.

In hadronic matter the pion number density is higher than the number density
of any other meson species. Since the pion is lightest among all mesons, the
relative velocity of pion and charmonium is larger than the one of any other
meson species with the same momentum
and the charmonium. Since the dissociation rate is proportional
to the relative velocity and the meson distribution function, the dissociation
rate of charmonium with $\pi$ is most concerned, and is a benchmark for 
determining the importance of the dissociation rate of charmonium with any 
other meson species.

When the $J/\psi$ meson is at rest, the dissociation rate of $J/\psi$ 
with $\rho$ ($K^*+\bar {K}^*$, $K+\bar K$) is about 
4.37 (3.52, 0.76), 1.23 (1.93, 0.86), 1.16 (1.52, 0.94), 1.78 (1.78, 0.97),
and 4.98 (1.17, 0.3) times the one of $J/\psi$ with $\pi$ at 
$T/T_{\rm c}$ = 0.65, 0.75, 0.85, 0.9, and 0.95, respectively.
The dissociation rate of $J/\psi$ with $\rho$ ($K^\ast +\bar {K}^\ast$, 
$K+\bar K$) is larger (larger, smaller) than the one of 
$J/\psi$ with $\pi$ at low $J/\psi$ momenta. 
At $T/T_{\rm c}$=0.85 the dissociation
rate of $J/\psi$ with $\eta$ is larger than the dissociation rate of $J/\psi$
with $\pi$ when the $J/\psi$ momentum is smaller than 1 GeV/$c$. Only at
$T/T_{\rm c}$=0.95 the dissociation rate of $J/\psi$ with $\eta$ is much 
smaller than the dissociation rate of $J/\psi$ with $\pi$.
The $J/\psi$ dissociation rate at  $T/T_{\rm c}$ 
=0.95 is much larger than the one at $T/T_{\rm c}$ = 0.65, 0.75, 0.85, and 0.9.
For example, the 
$J/\psi$ dissociation rate at $T/T_{\rm c}$ =0.95 is about 7.94 times the one 
at $T/T_{\rm c}$ =0.9. This is because that the endothermic $J/\psi$
dissociation in collisions with pion, $\rho$ meson, kaon, and vector kaon
takes a rapid rise in peak cross sections when temperature 
approaches the critical temperature. 

The dissociation rate of $\psi'$ with $\rho$ is smaller than the one of 
$\psi'$ with $\pi$ at $T/T_{\rm c}$ = 0.65 and 0.75, and larger than the one of
$\psi'$ with $\pi$ at 
$T/T_{\rm c}$ = 0.85, 0.9, and 0.95 when the $\psi'$ momentum is smaller than
3.3 GeV/$c$. At $T/T_{\rm c}$ = 0.85 the dissociation 
rate of $\psi'$ with vector kaon is larger 
than the one of $\psi'$ with $\pi$ when the $\psi'$ momentum is smaller than 
2.6 GeV/$c$. The dissociation rate of $\psi'$ with vector kaon is
smaller than the one of $\psi'$ with $\pi$
at $T/T_{\rm c}$ = 0.9 and 0.95, and is very small
at $T/T_{\rm c}$ = 0.65 and 0.75. At any temperature in this region 
$0.6\leq T/T_{\rm c}<1$, the dissociation rate of $\psi'$ 
with kaon is smaller than the one of $\psi'$ with pion. The dissociation rate
of $\psi'$ with $\eta$ is very small, and the $\eta + \psi'$ reactions can be
neglected.

The dissociation rate of $\chi_c$ with $\rho$ is smaller than the one of 
$\chi_c$ with $\pi$ at $T/T_{\rm c}$ = 0.65, 0.75, and 0.95, and larger than 
the one of $\chi_c$ with $\pi$ at 
$T/T_{\rm c}$ = 0.85 and 0.9 when the $\chi_c$ momentum is smaller than 4 
GeV/$c$. The dissociation rate of $\chi_c$ with vector kaon is much smaller 
than the one of $\chi_c$ with $\pi$ at $T/T_{\rm c}$ = 0.65 and 0.75. 
However, at $T/T_{\rm c}$ = 0.85 and 0.9 the dissociation rate of $\chi_c$ 
with vector kaon is larger than the one of $\chi_c$ with $\pi$ when the 
$\chi_c$ momentum is smaller than about 2 GeV/$c$. 
For $0.6\leq T/T_{\rm c}<1$ the dissociation rate of $\chi_c$ 
with kaon is smaller than the one of $\chi_c$ with pion. The dissociation rate
of $\chi_c$ with $\eta$ is quite small, and the $\eta + \chi_c$ reactions can 
be neglected.

The vector-kaon mass is larger than the kaon mass, and the vector-kaon 
distribution function is smaller than the kaon distribution function. When 
a vector kaon and a kaon have the same
momentum, the relative velocity between the vector kaon
and a charmonium is smaller than the
one between the kaon and the charmonium. However, these do not mean that the 
dissociation rate of charmonium with vector kaon must be smaller than the 
dissociation rate of charmonium with kaon. We account 
for this by the dissociation rates of $J/\psi$ with vector kaon and of
$J/\psi$ with kaon at $T/T_{\rm c}=0.75$. 
The peak cross section of the endothermic
reaction $K^*+J/\psi \to \bar{D}^* + D^+_s$ is larger than the one of the
endothermic reaction $K + J/\psi \to \bar{D}^* + D^+_s$. The peak cross section
of $K^* + J/\psi \to \bar{D}^* + D^{*+}_s$ is larger than the one of 
$K + J/\psi \to \bar{D}^* + D^{*+}_s$. Even though the peak cross section of
$K^* + J/\psi \to \bar{D} + D^{*+}_s$ is smaller than the one of 
$K + J/\psi \to \bar{D} + D^{*+}_s$, the sum of the peak cross sections of 
$K^* + J/\psi \to \bar{D}^* + D^+_s$, $K^* + J/\psi \to \bar{D} + D^{*+}_s$,
and $K^* + J/\psi \to \bar{D}^* + D^{*+}_s$ is larger than the sum of the peak
cross sections of $K + J/\psi \to \bar{D}^* + D^+_s$, 
$K + J/\psi \to \bar{D} + D^{*+}_s$, and 
$K + J/\psi \to \bar{D}^* + D^{*+}_s$. The 
$K^*$-induced $J/\psi$ dissociation has the exothermic reaction
$K^* + J/\psi \to \bar{D} + D^+_s$, but the $K$-induced $J/\psi$ dissociation
does not produce the final states, $\bar{D}$ and $D^+_s$. In addition, while
$\sqrt s$ increases from the threshold energy, the cross
sections for the $K^* + J/\psi$ dissociation reactions
increase more rapidly than the ones for the $K + J/\psi$ dissociation 
reactions. Therefore, at $T/T_{\rm c}=0.75$ the dissociation rate of 
$J/\psi$ with vector kaon is larger than the one of $J/\psi$ 
with kaon when the $J/\psi$ momentum is smaller than 2.1 GeV/$c$. 
In the same way we can understand that the dissociation rate of $\chi_c$ with
vector kaon at $T/T_{\rm c}=0.95$ is larger than the one of $\chi_c$ with kaon
even though the largest cross sections of the three reactions, 
$K^* + \chi_c \to \bar{D}^* + D^+_s$, $K^* + \chi_c \to \bar{D} +D^{*+}_s$,
and $K^* + \chi_c \to \bar{D}^* + D^{*+}_s$,
are smaller than the ones of the three reactions,
$K + \chi_c \to \bar{D}^* + D^+_s$, $K + \chi_c \to \bar{D} +D^{*+}_s$, and
$K + \chi_c \to \bar{D}^* + D^{*+}_s$, respectively, as mentioned in Sec. III.

\vspace{0.5cm}
\leftline{\bf V. SUMMARY }
\vspace{0.5cm}

The unpolarized cross sections for twelve $K^*$-charmonium dissociation 
reactions have been obtained in the quark-interchange mechanism, in the Born
approximation and with the quark potential. The temperature dependence of the
quark potential, the meson masses, and the mesonic quark-antiquark
relative-motion wave functions leads to the temperature dependence of the
unpolarized cross sections. Even though $\psi'$ and $\chi_c$ have very similar
masses, their different quantum numbers cause different cross sections for
$K^* + \psi'$ and $K^* + \chi_c$ reactions. Even though the $K^\ast$
mass is larger than the $K$ mass, the cross sections for some 
$K^*$-charmonium dissociation reactions may be smaller than the ones for the
$K$-charmonium dissociation reactions with the same final states at 
$T \to T_c$.

Using the dissociation cross sections of charmonia in collisions with pion,
$\rho$ meson, kaon, vector kaon, and $\eta$ meson, we have calculated the
dissociation rates of charmonium with pion, $\rho$ meson, kaon, vector kaon, 
and $\eta$ meson. The first derivative of the dissociation rates with respect
to the charmonium momentum is zero at zero charmonium momentum.
The temperature dependence of the cross section, the 
relative velocity, and the distribution function brings about the temperature
dependence of the dissociation rates. 
When the temperature increases, charmonium dissociation
rates generally increase. With charmonium momentum increasing from 2.2 
GeV/$c$, the dissociation rates decrease. The dissociation rates of $J/\psi$
with the five species of mesons are comparable at low $J/\psi$ momenta, and 
the five species of mesons contribute to the $J/\psi$ suppression in hadronic 
matter. The dissociation rates of $\psi'$ with $\rho$ meson and vector
kaon are larger than the one of $\psi'$ with pion in some momentum and 
temperature regions. To study the $\psi'$ suppression, 
the $\psi'$ dissociation in collisions with pion, $\rho$ meson, kaon, 
and vector kaon needs to be considered, but the $\eta + \psi'$
reactions can be neglected. This also holds true for the $\chi_c$ case.

\vspace{0.5cm}
\leftline{\bf ACKNOWLEDGEMENTS}
\vspace{0.5cm}
We thank Prof. H. J. Weber for helpful discussions.
This work was supported by the National Natural Science Foundation of China 
under Grant No. 11175111.

\newpage
\begin{figure}[htbp]
  \centering
    \includegraphics[width=65mm,height=60mm,angle=0]{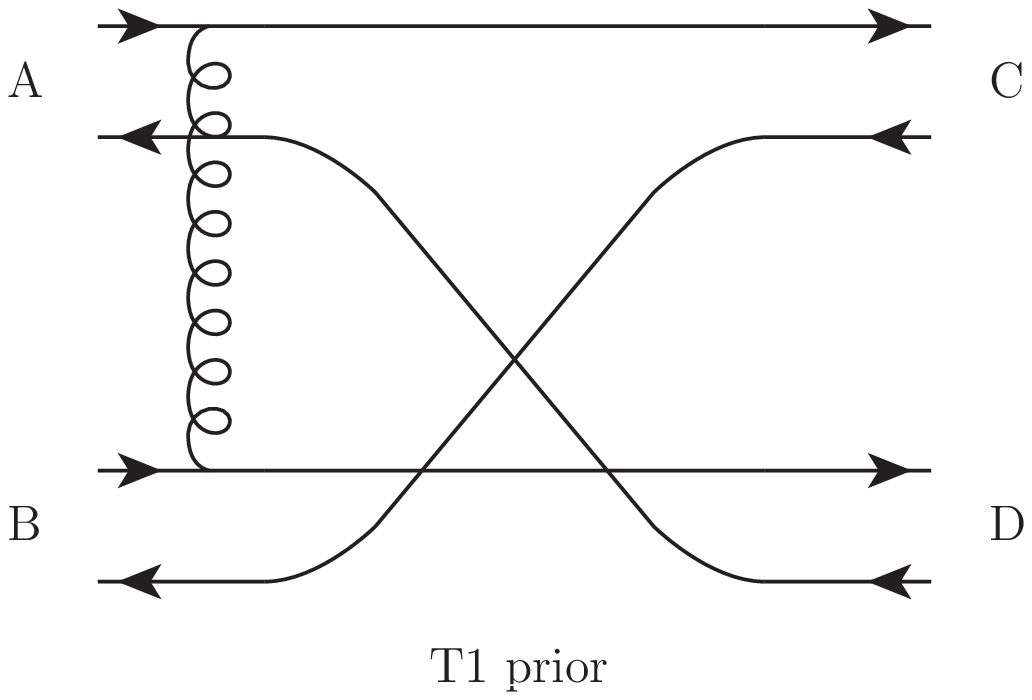}
      \hspace{1.5cm}
    \includegraphics[width=65mm,height=60mm,angle=0]{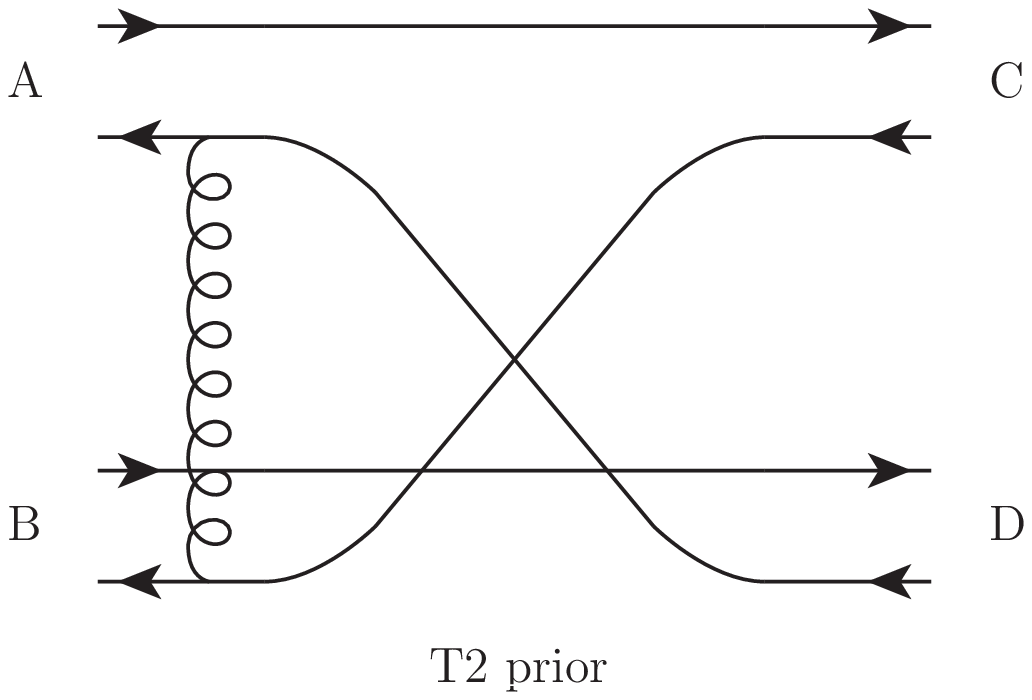}
      \vskip 36pt
    \includegraphics[width=65mm,height=60mm,angle=0]{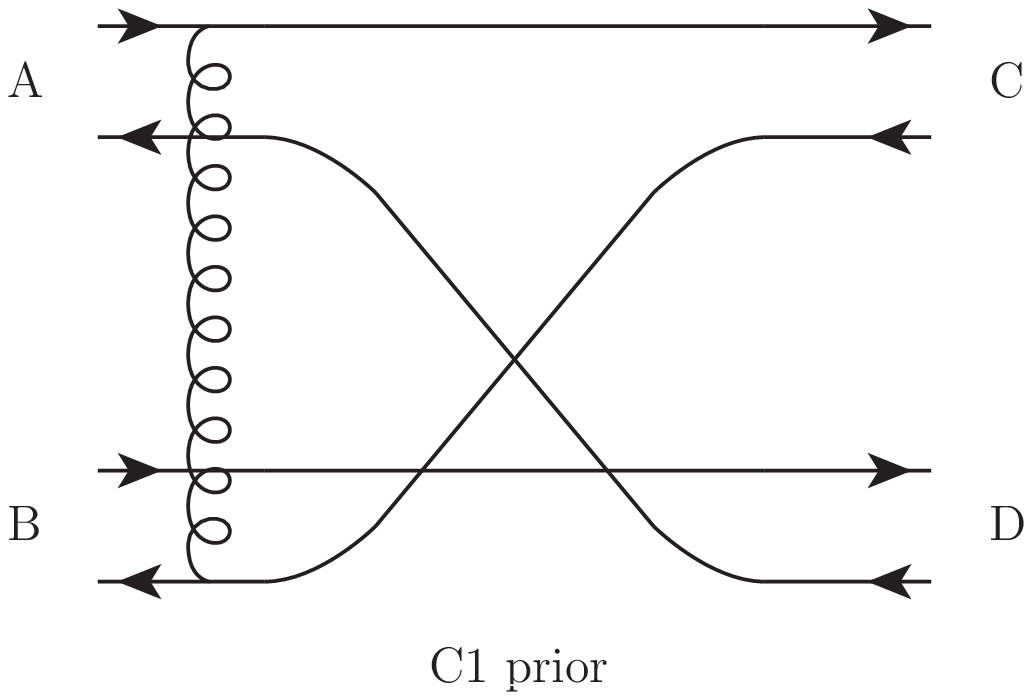}
      \hspace{1.5cm}
    \includegraphics[width=65mm,height=60mm,angle=0]{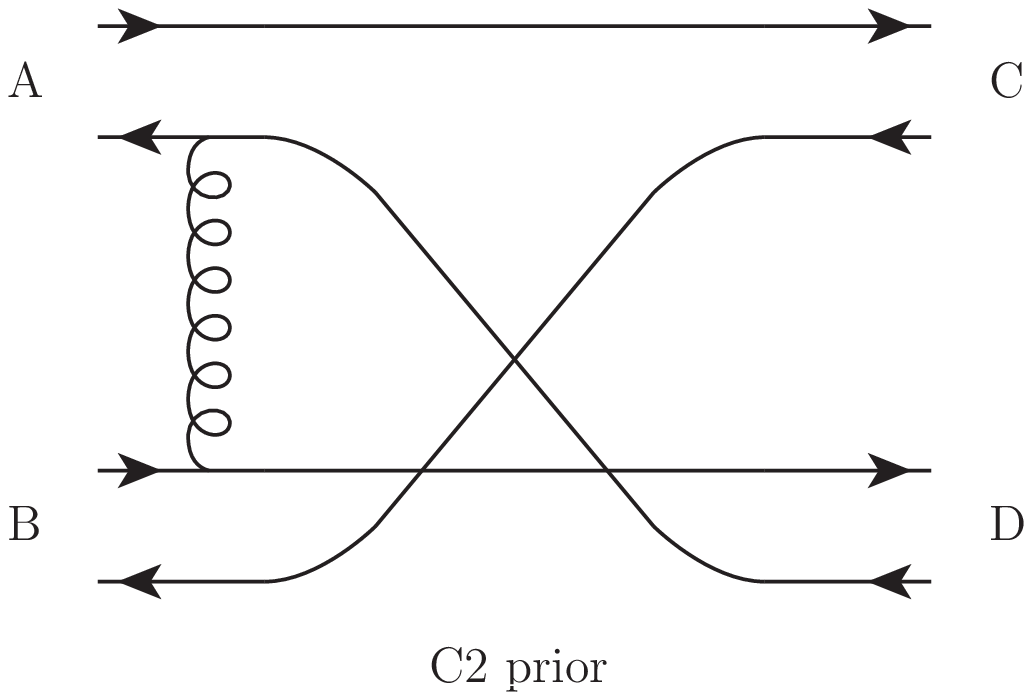}
\caption{'Prior' diagrams. Solid (wavy) lines represent quarks or antiquarks
(interaction). This figure is reproduced under permission from the article with
doi:10.1088/0954-3899/42/9/095110.}
\label{fig1}
\end{figure}

\newpage
\begin{figure}[htbp]
  \centering
    \includegraphics[width=65mm,height=60mm,angle=0]{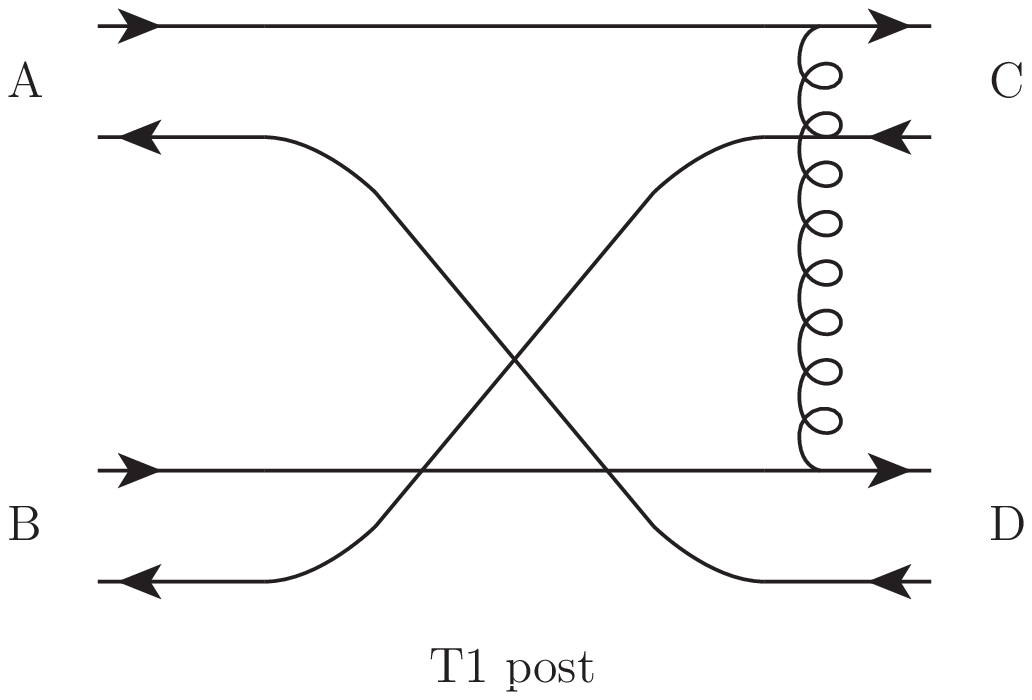}
      \hspace{1.5cm}
    \includegraphics[width=65mm,height=60mm,angle=0]{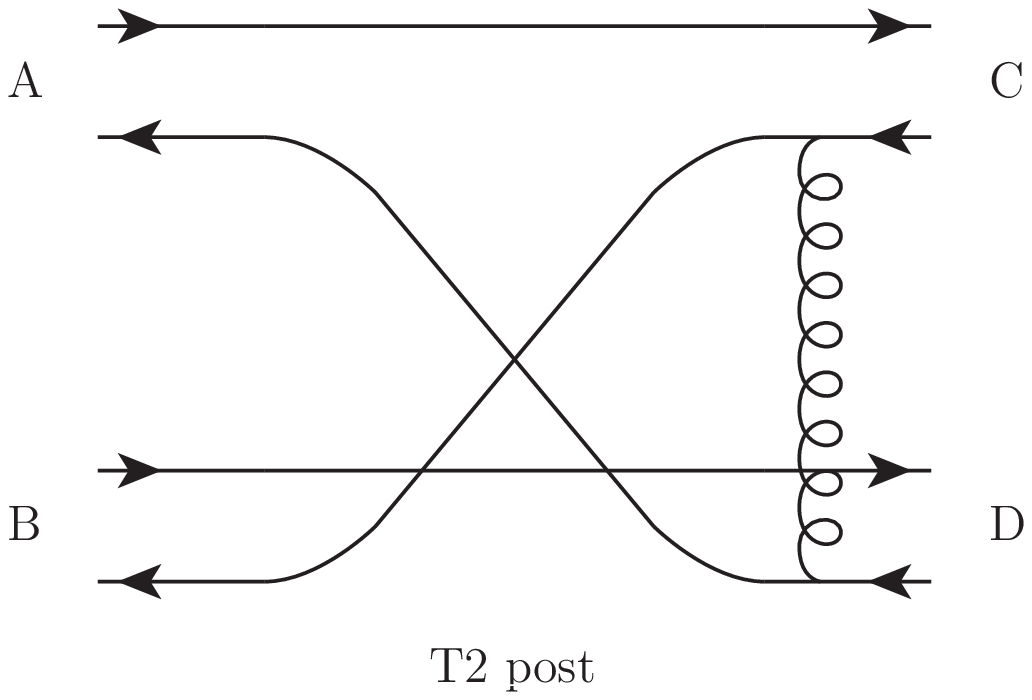}
      \vskip 36pt
    \includegraphics[width=65mm,height=60mm,angle=0]{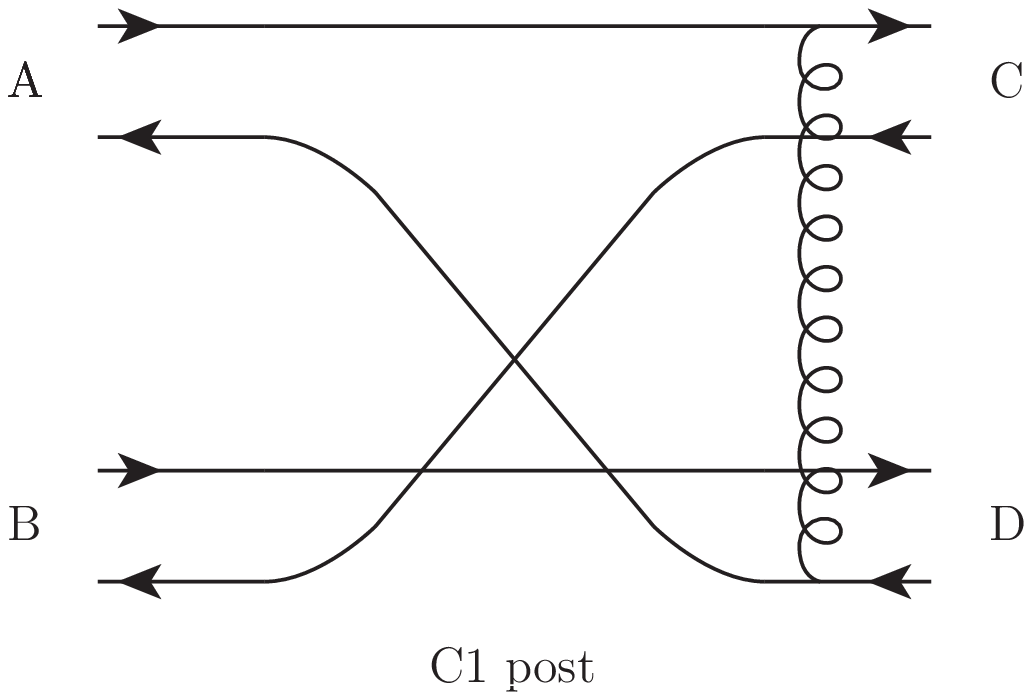}
      \hspace{1.5cm}
    \includegraphics[width=65mm,height=60mm,angle=0]{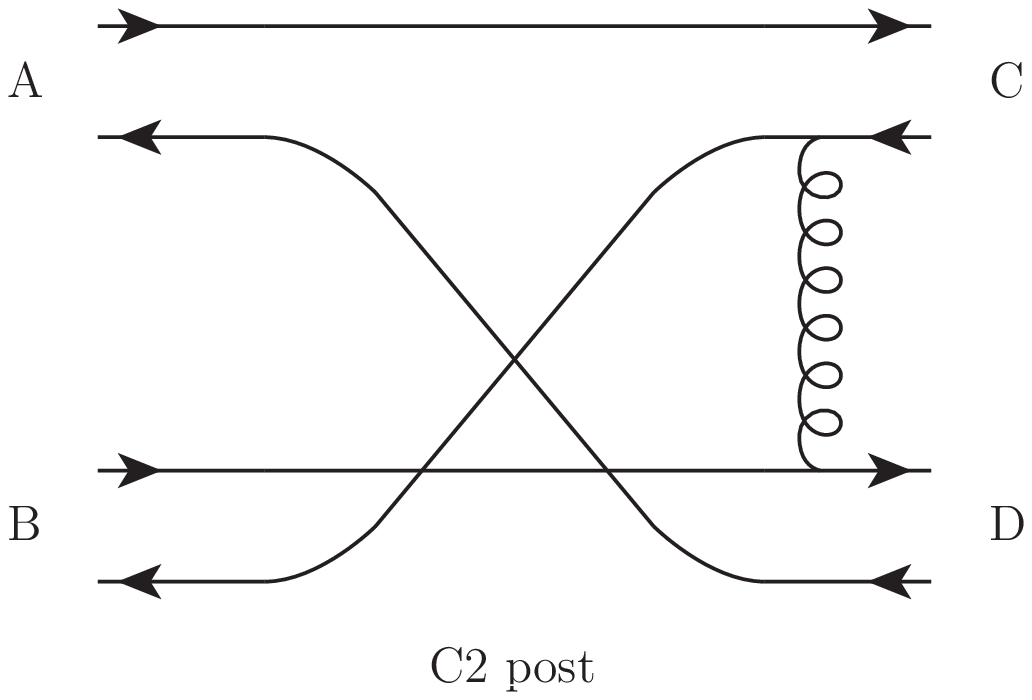}
\caption{'Post' diagrams. Solid (wavy) lines represent quarks or antiquarks
(interaction). This figure is reproduced under permission from the article with
doi:10.1088/0954-3899/42/9/095110.}
\label{fig2}
\end{figure}

\newpage
\begin{figure}[htbp]
  \centering
    \includegraphics[scale=0.6]{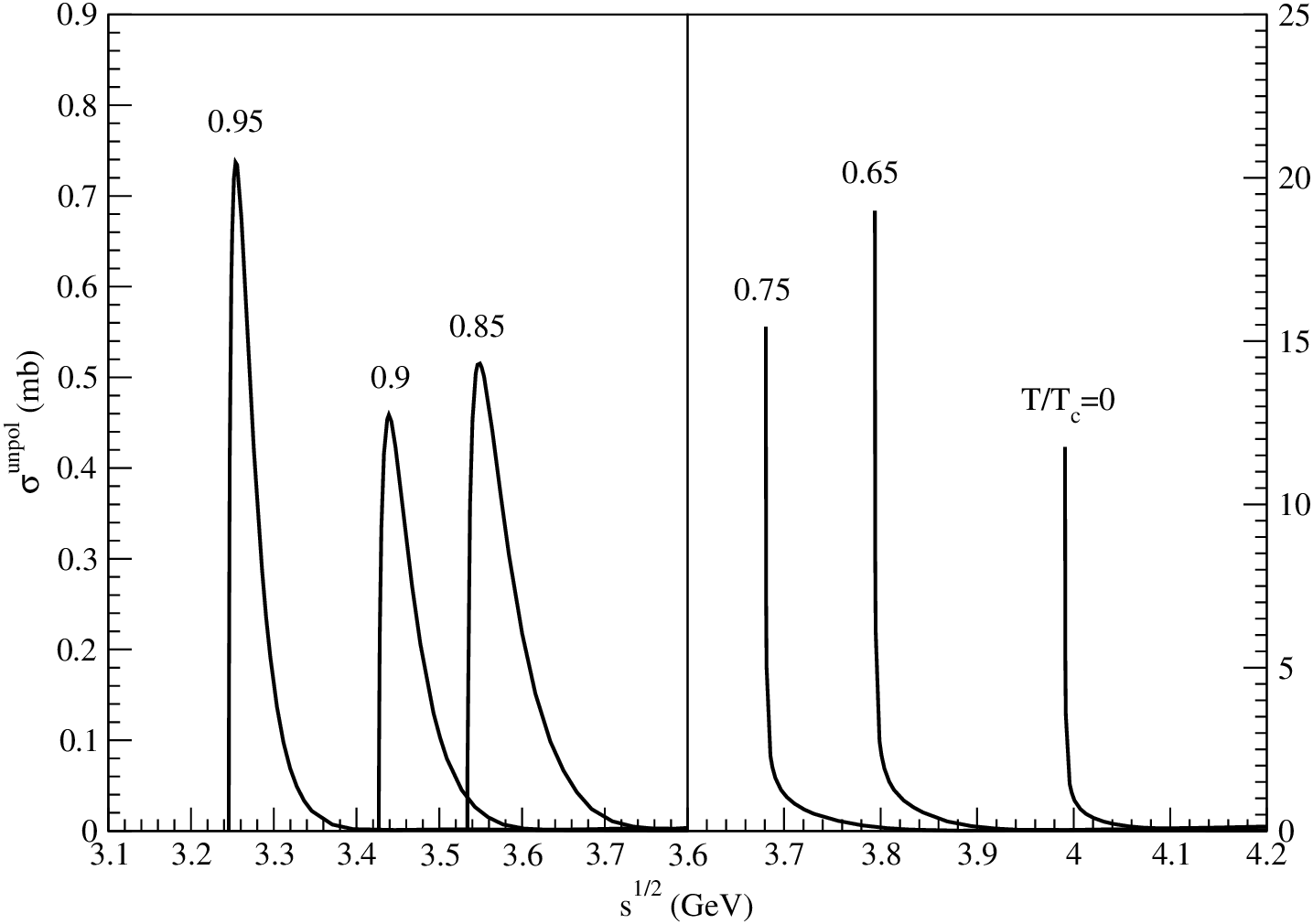}%
\caption{Cross sections for $K^*+J/\psi\to\bar{D}+D^+_s$
at various temperatures.}
\label{fig3}
\end{figure}

\newpage
\begin{figure}[htbp]
  \centering
    \includegraphics[scale=0.6]{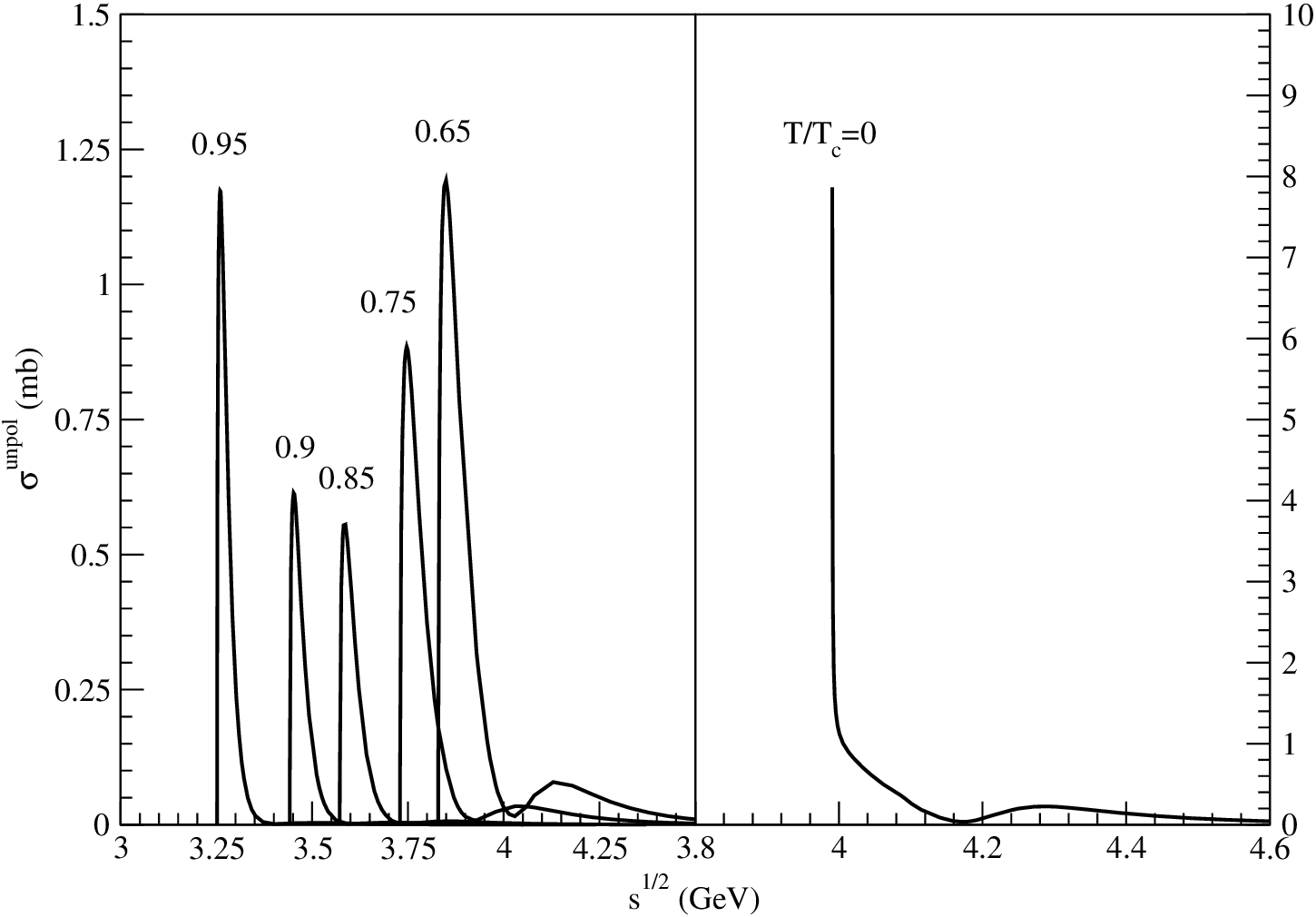}%
\caption{Cross sections for $K^*+J/\psi\to\bar{D}^*+D^+_s$
at various temperatures.}
\label{fig4}
\end{figure}

\newpage
\begin{figure}[htbp]
  \centering
    \includegraphics[scale=0.6]{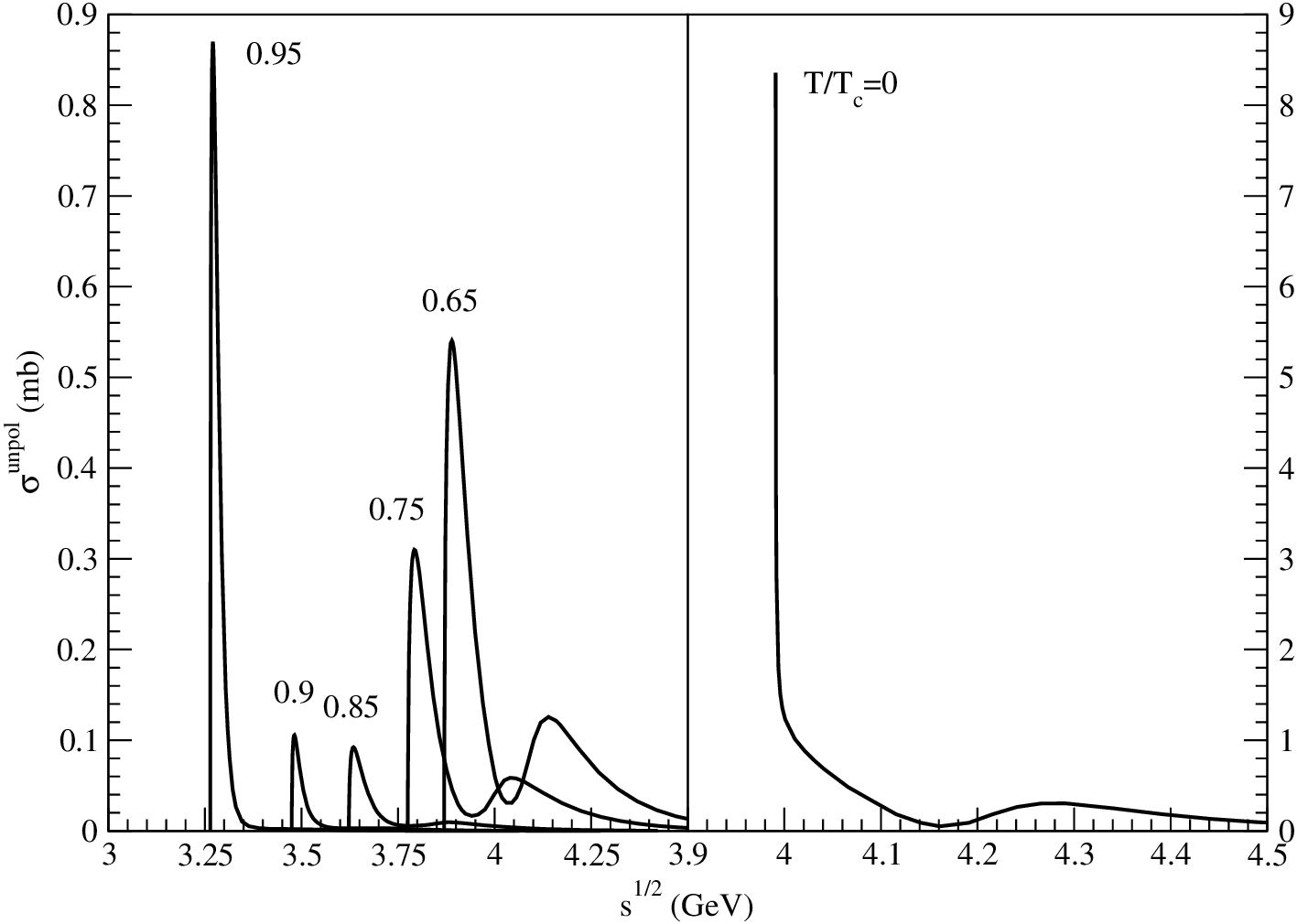}%
\caption{Cross sections for $K^*+J/\psi\to\bar{D}+D^{*+}_s$
at various temperatures.}
\label{fig5}
\end{figure}

\newpage
\begin{figure}[htbp]
  \centering
    \includegraphics[scale=0.6]{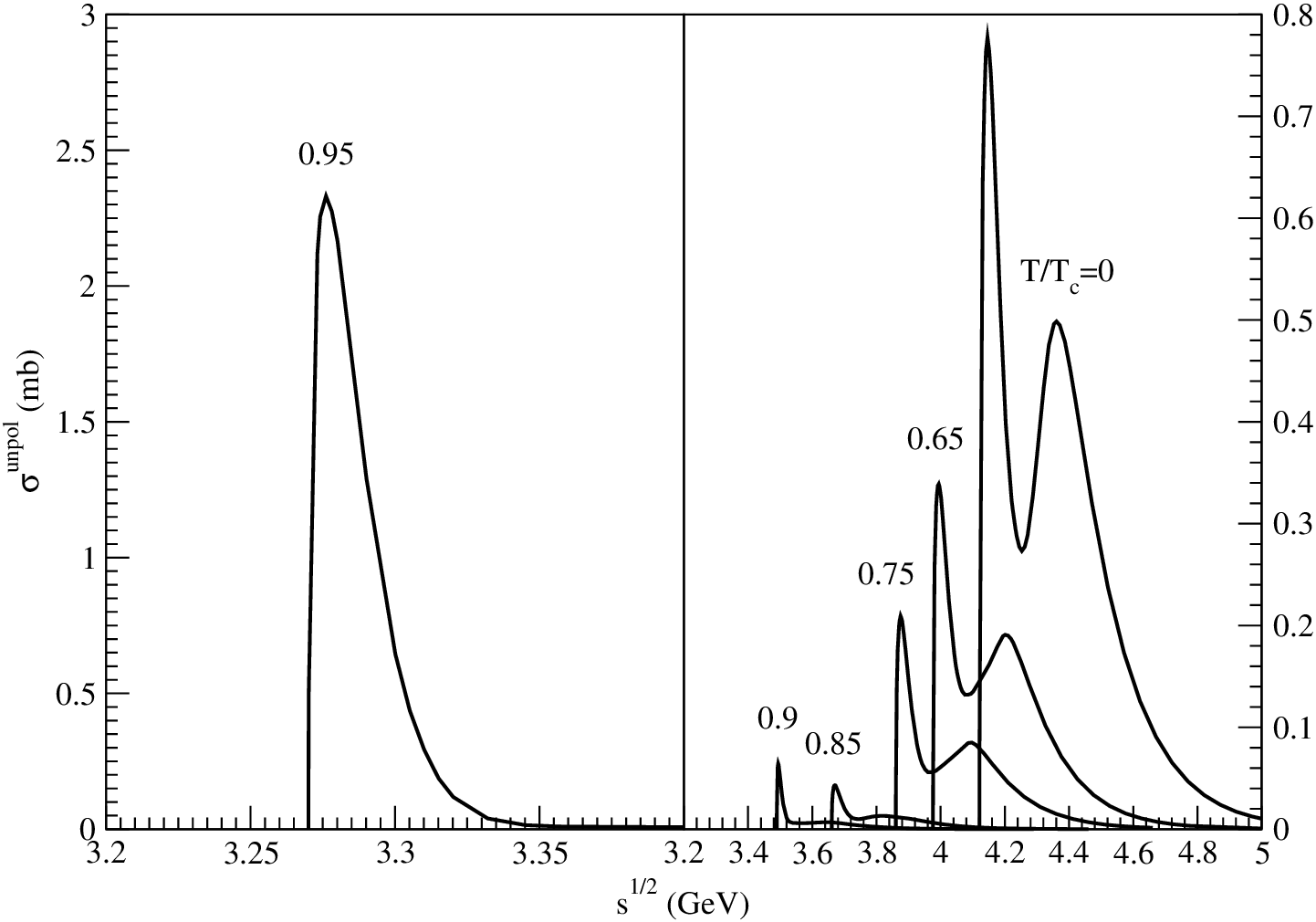}%
\caption{Cross sections for $K^*+J/\psi\to\bar{D}^*+D^{*+}_s$
at various temperatures.}
\label{fig6}
\end{figure}

\newpage
\begin{figure}[htbp]
  \centering
    \includegraphics[scale=0.6]{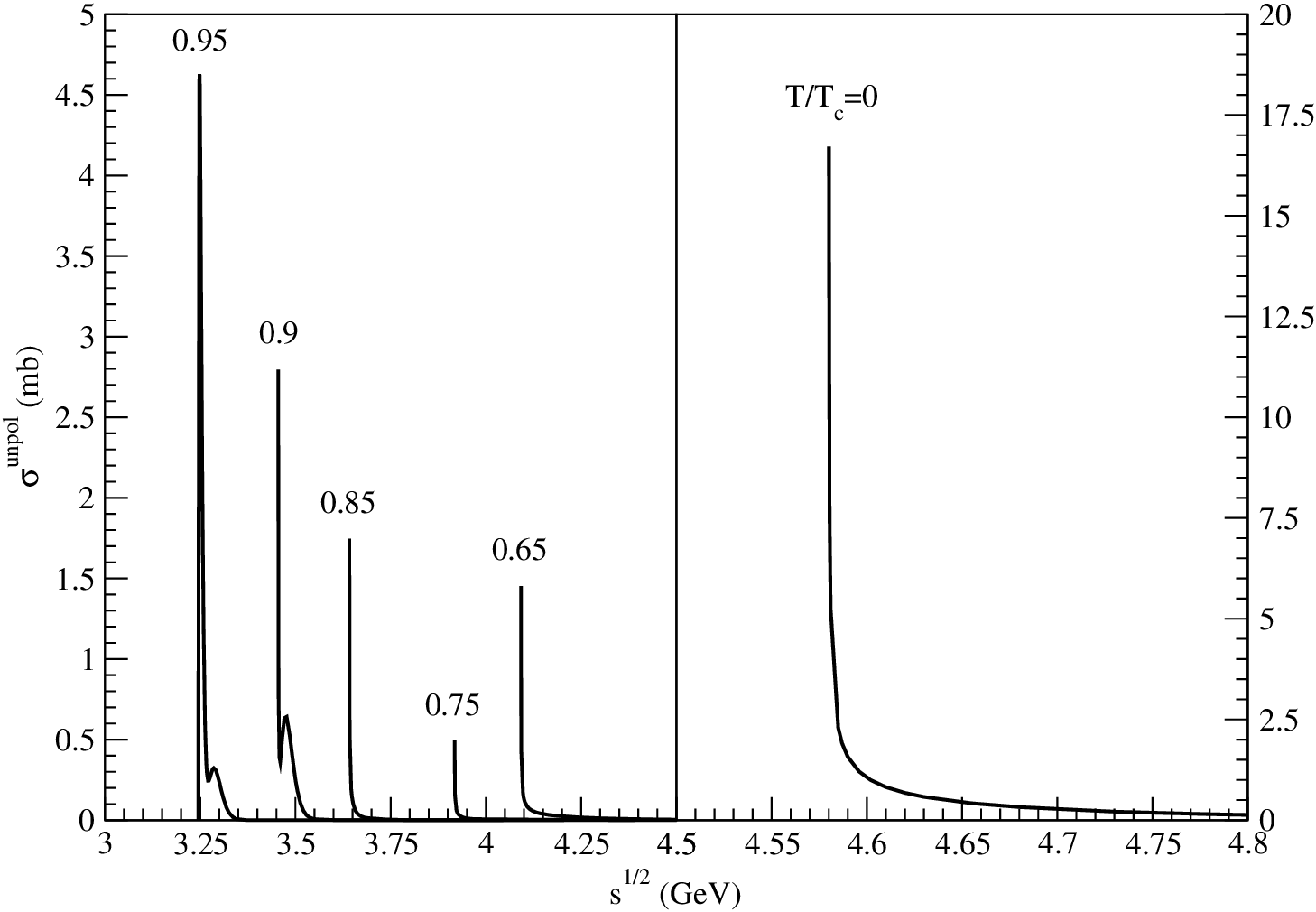}%
\caption{Cross sections for $K^*+\psi'\to\bar{D}+D^+_s$
at various temperatures.}
\label{fig7}
\end{figure}

\newpage
\begin{figure}[htbp]
  \centering
    \includegraphics[scale=0.6]{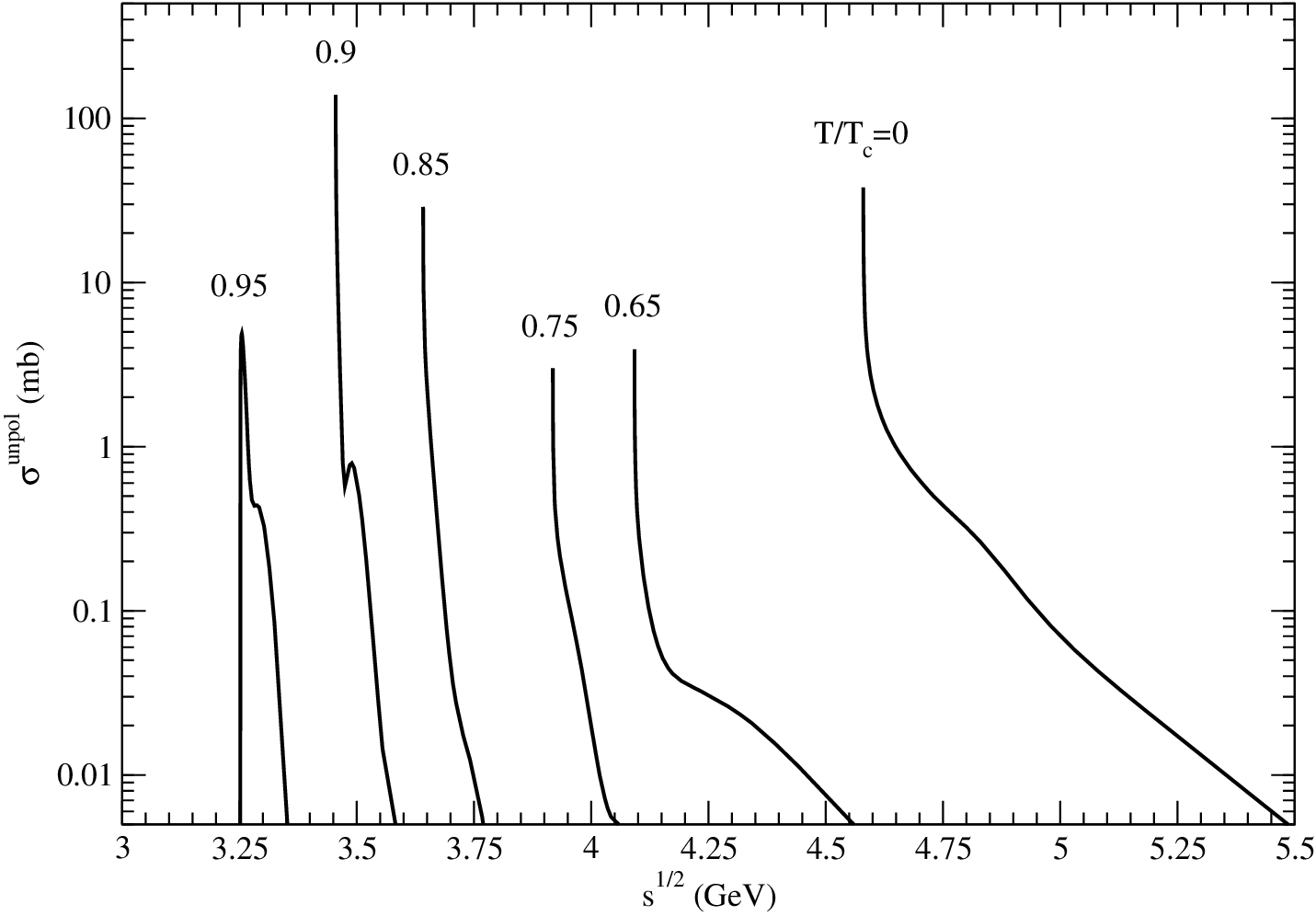}%
\caption{Cross sections for $K^*+\psi'\to\bar{D}^*+D^+_s$
at various temperatures.}
\label{fig8}
\end{figure}

\newpage
\begin{figure}[htbp]
  \centering
    \includegraphics[scale=0.6]{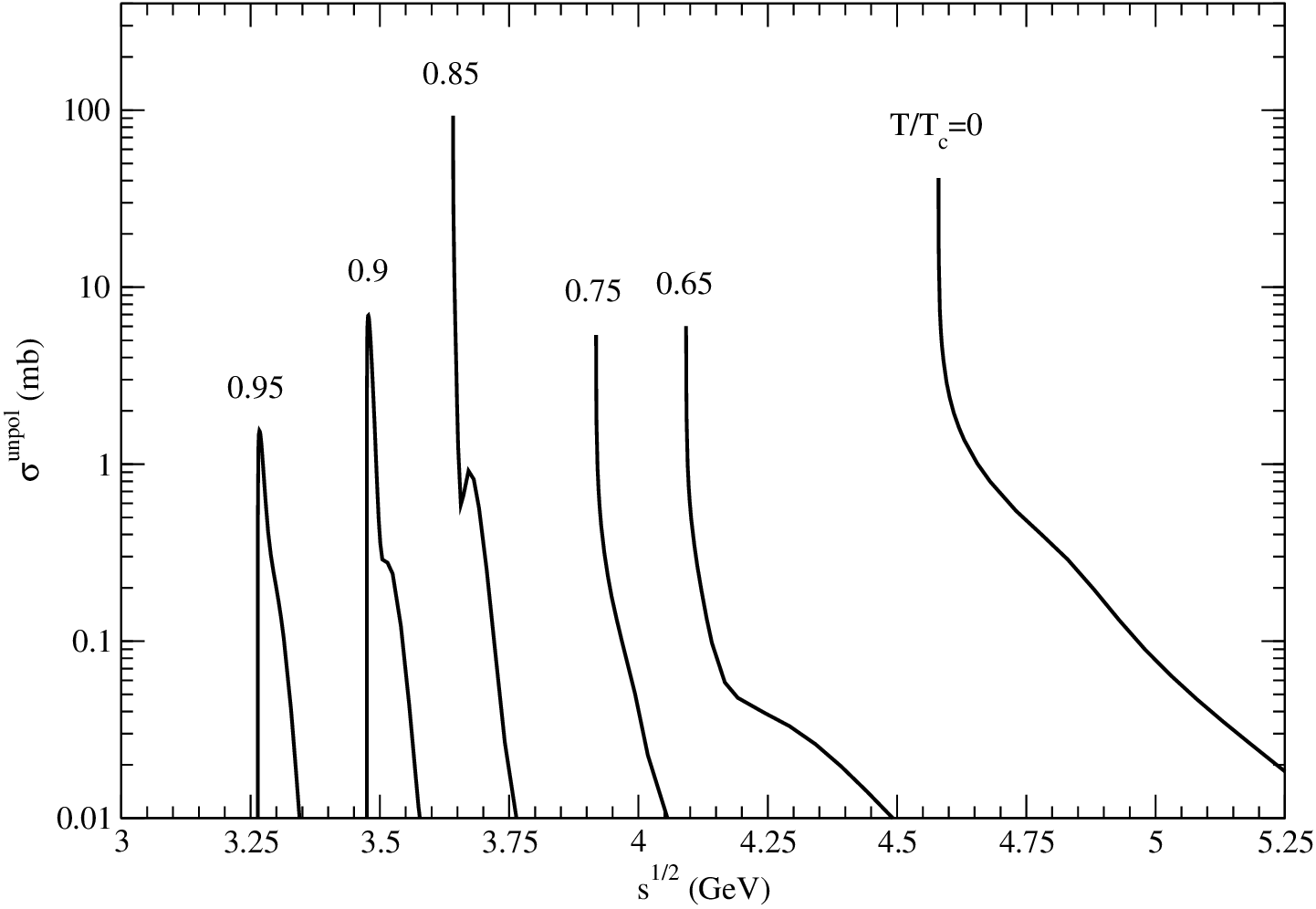}%
\caption{Cross sections for $K^*+\psi'\to\bar{D}+D^{*+}_s$
at various temperatures.}
\label{fig9}
\end{figure}

\newpage
\begin{figure}[htbp]
  \centering
    \includegraphics[scale=0.6]{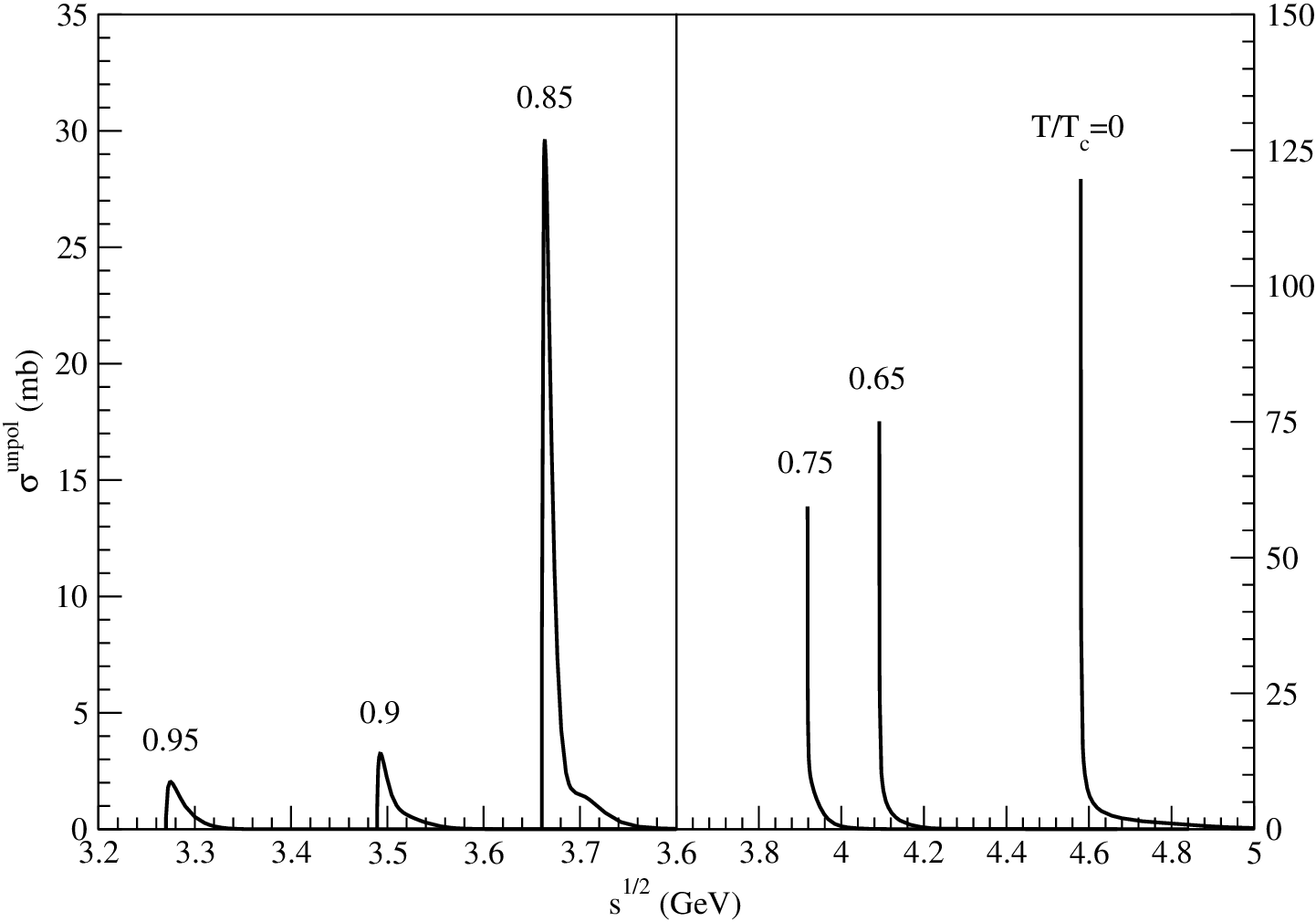}%
\caption{Cross sections for $K^*+\psi'\to\bar{D}^*+D^{*+}_s$
at various temperatures.}
\label{fig10}
\end{figure}

\newpage
\begin{figure}[htbp]
  \centering
    \includegraphics[scale=0.6]{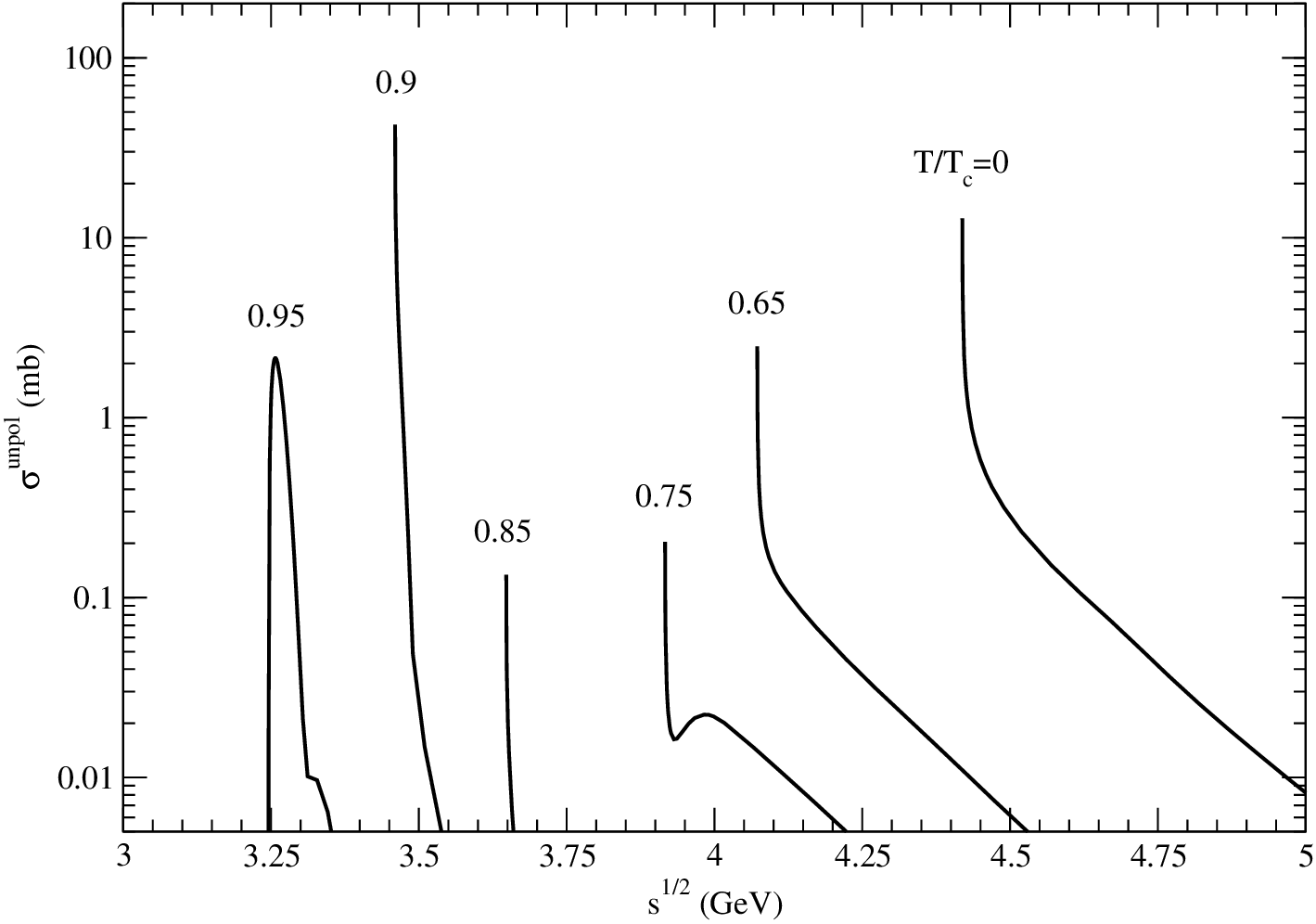}%
\caption{Cross sections for $K^*+\chi_{c}\to\bar{D}+D^+_s$
at various temperatures.}
\label{fig11}
\end{figure}

\newpage
\begin{figure}[htbp]
  \centering
    \includegraphics[scale=0.6]{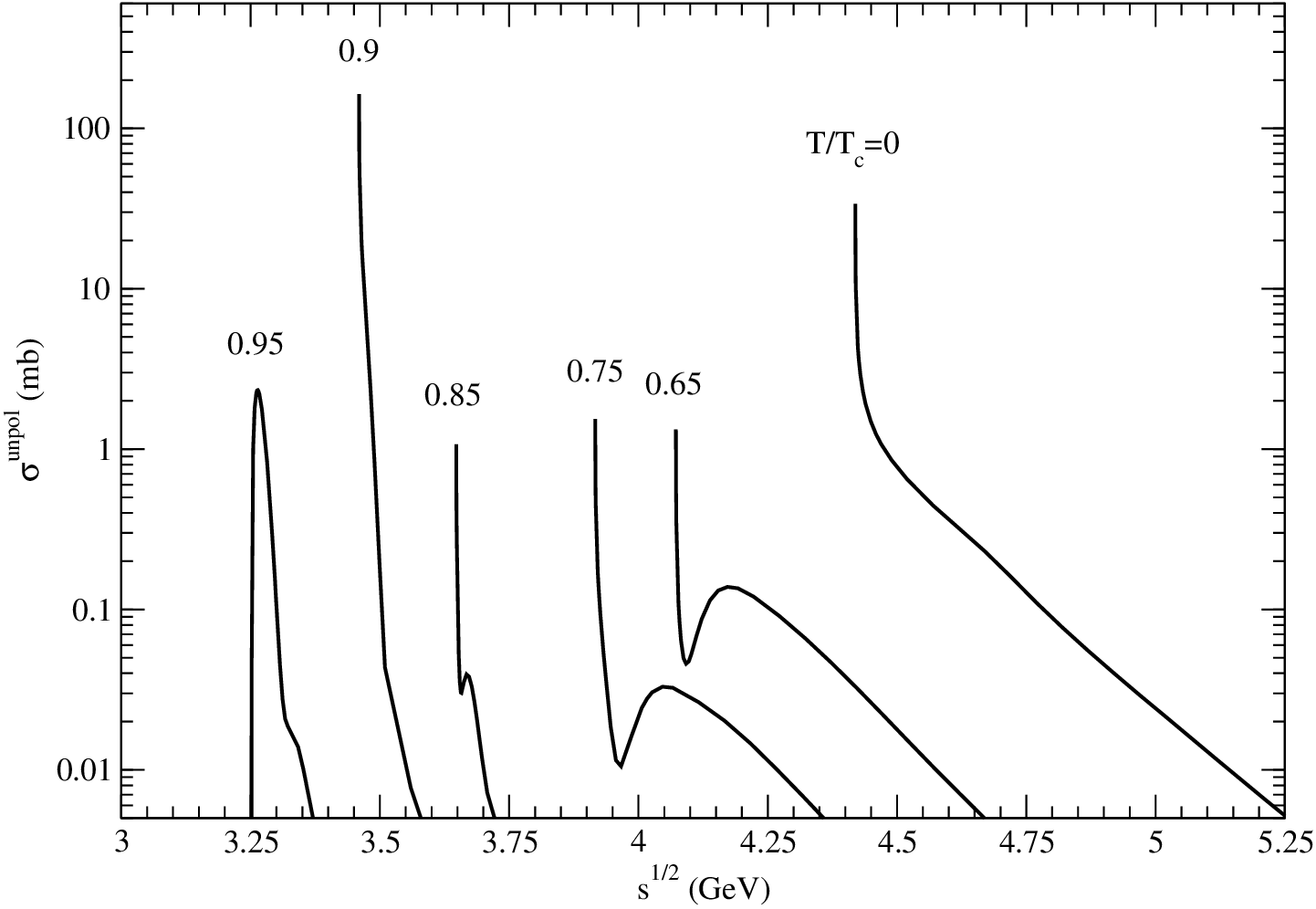}%
\caption{Cross sections for $K^*+\chi_{c}\to\bar{D}^*+D^+_s$
at various temperatures.}
\label{fig12}
\end{figure}

\newpage
\begin{figure}[htbp]
  \centering
    \includegraphics[scale=0.6]{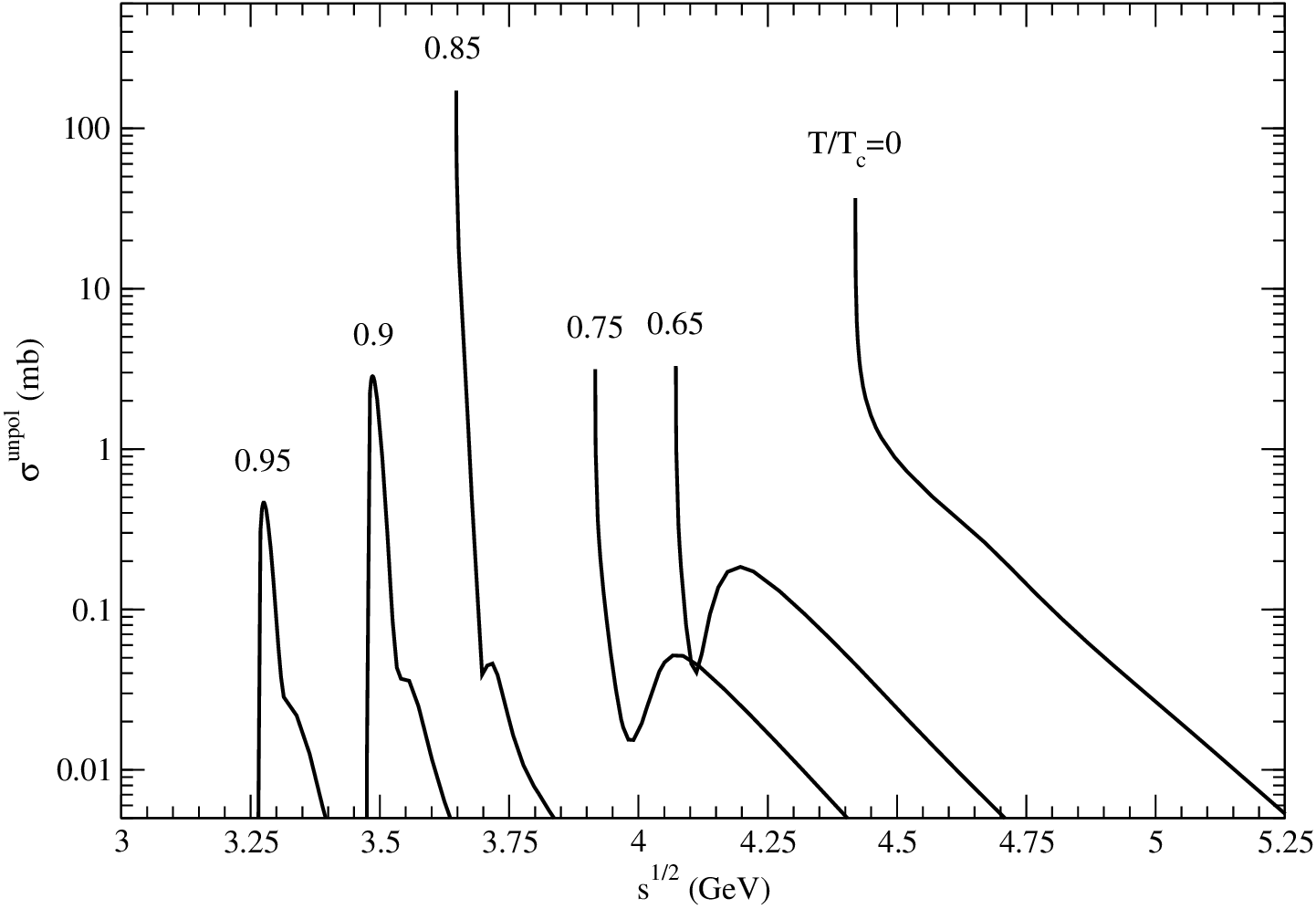}%
\caption{Cross sections for $K^*+\chi_{c}\to\bar{D}+D^{*+}_s$
at various temperatures.}
\label{fig13}
\end{figure}

\newpage
\begin{figure}[htbp]
  \centering
    \includegraphics[scale=0.6]{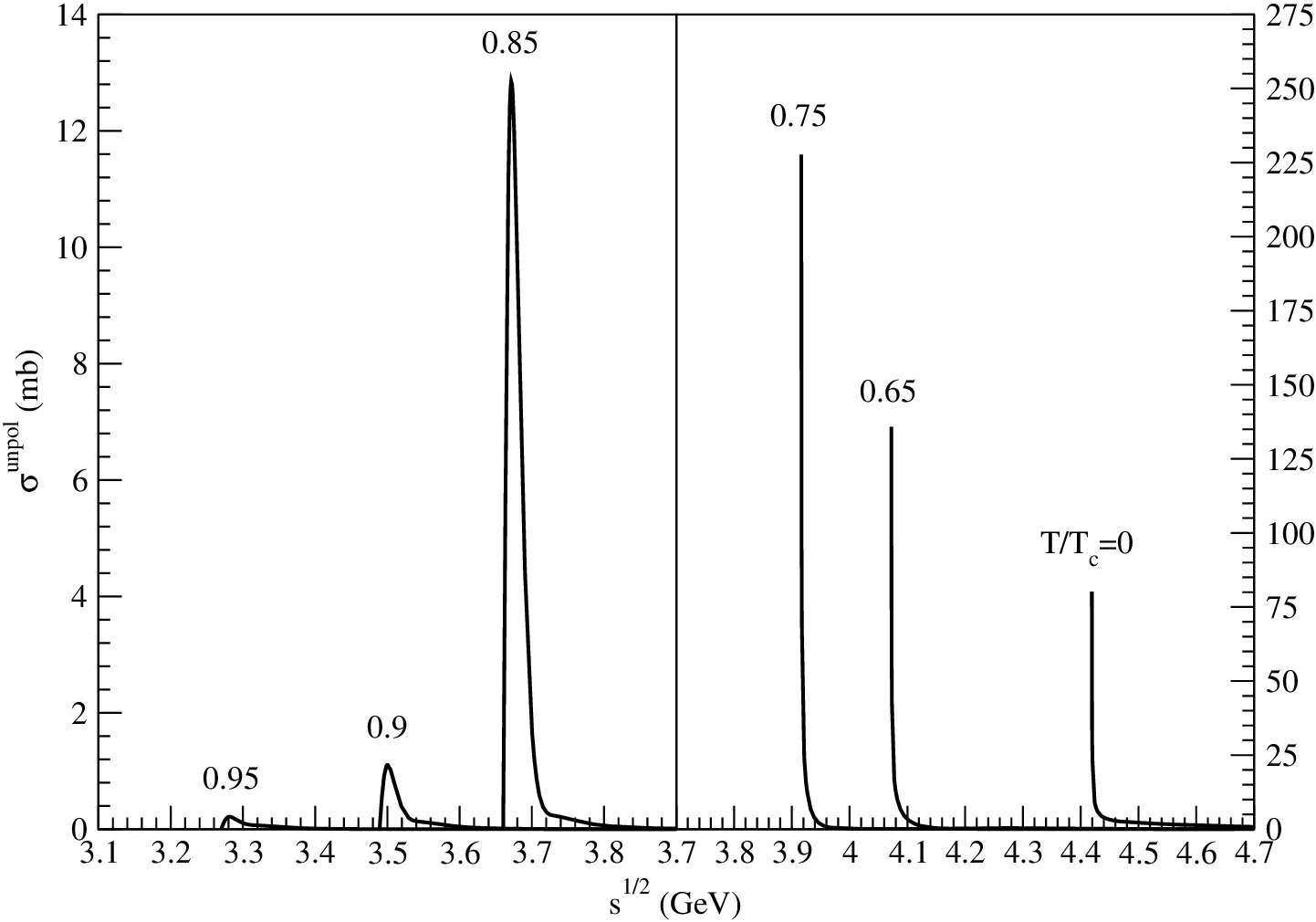}%
\caption{Cross sections for $K^*+\chi_{c}\to\bar{D}^*+D^{*+}_s$
at various temperatures.}
\label{fig14}
\end{figure}

\newpage
\begin{figure}[htbp]
  \centering
    \includegraphics[scale=0.6]{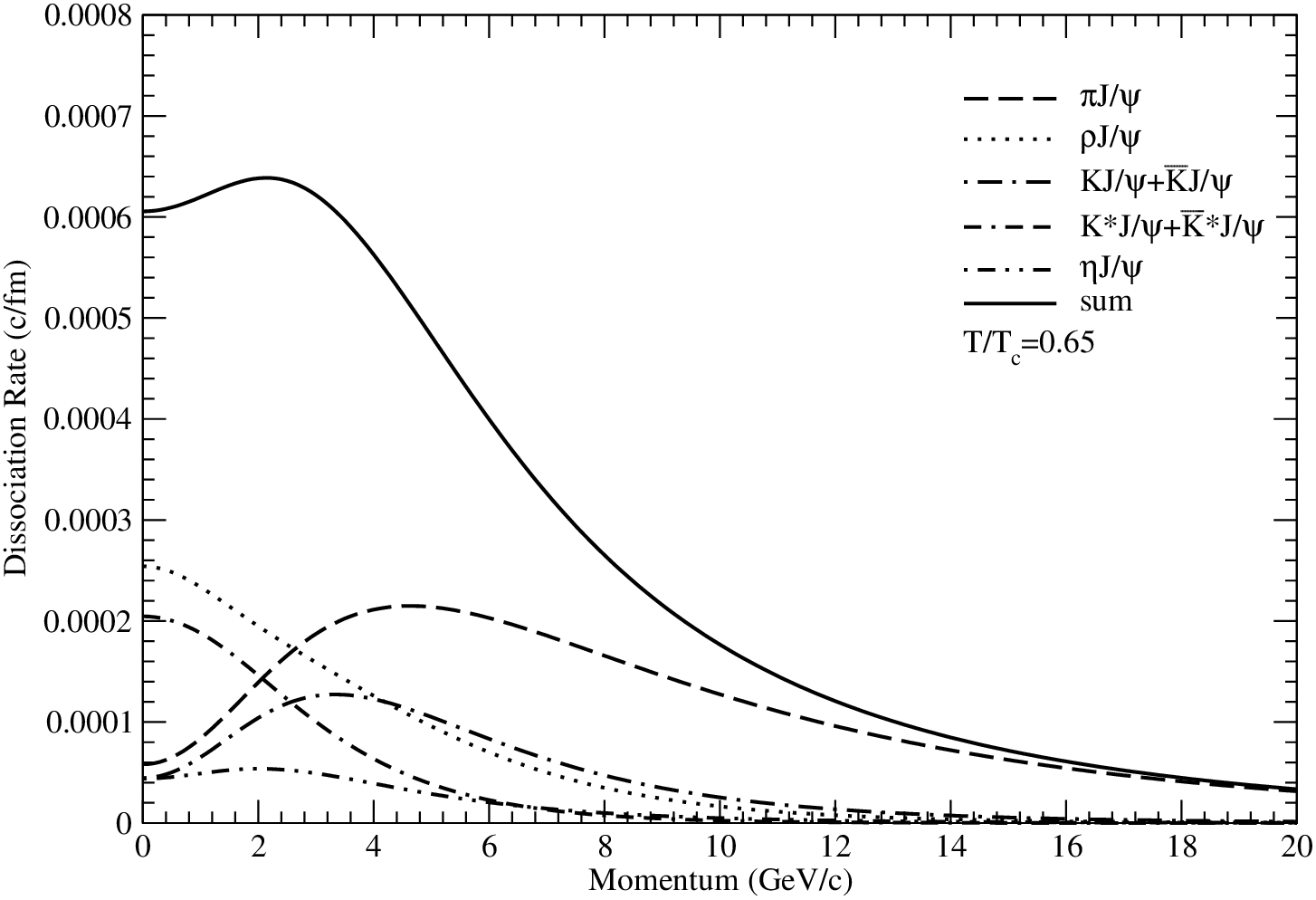}%
\caption{Versus the $J/\psi$ momentum
the dissociation rate of $J/\psi$ with $\pi$ (dashed curve), the one 
of $J/\psi$ with $\rho$ (dotted curve), the sum (dot-dashed curve) of the 
ones of $J/\psi$ with $K$ and with $\bar K$, the sum (dot-dash-dashed curve) 
of the ones of $J/\psi$ with $K^*$ and with $\bar {K}^*$, and the one of 
$J/\psi$
with $\eta$ (dot-dot-dashed curve) at $T=0.65T_{\rm c}$. The solid curve 
represents the sum of these dissociation rates.}
\label{fig15}
\end{figure}

\newpage
\begin{figure}[htbp]
  \centering
    \includegraphics[scale=0.6]{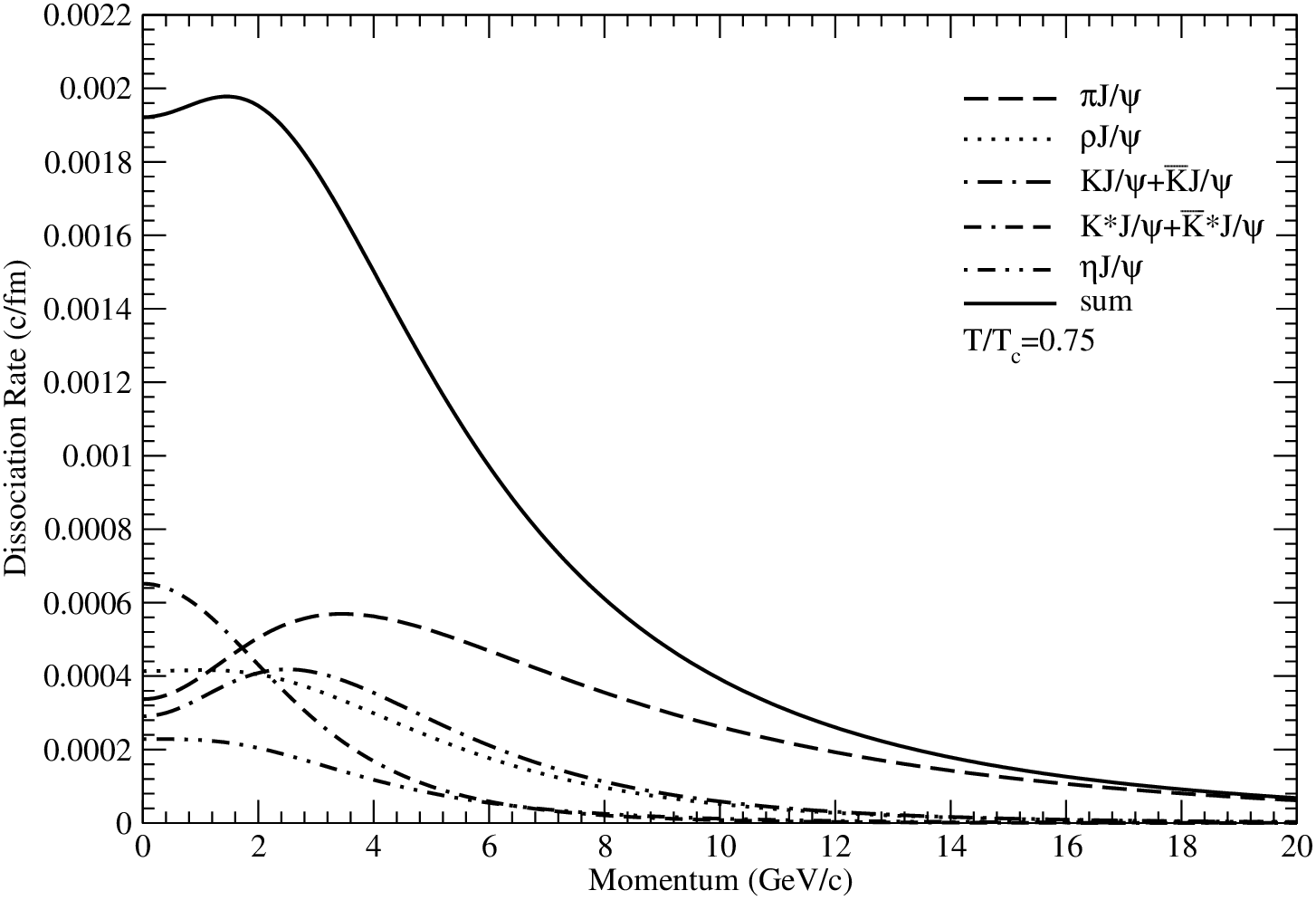}%
\caption{The same as Fig. 15 except for the temperature $0.75T_{\rm c}$ .}
\label{fig16}
\end{figure}

\newpage
\begin{figure}[htbp]
  \centering
    \includegraphics[scale=0.6]{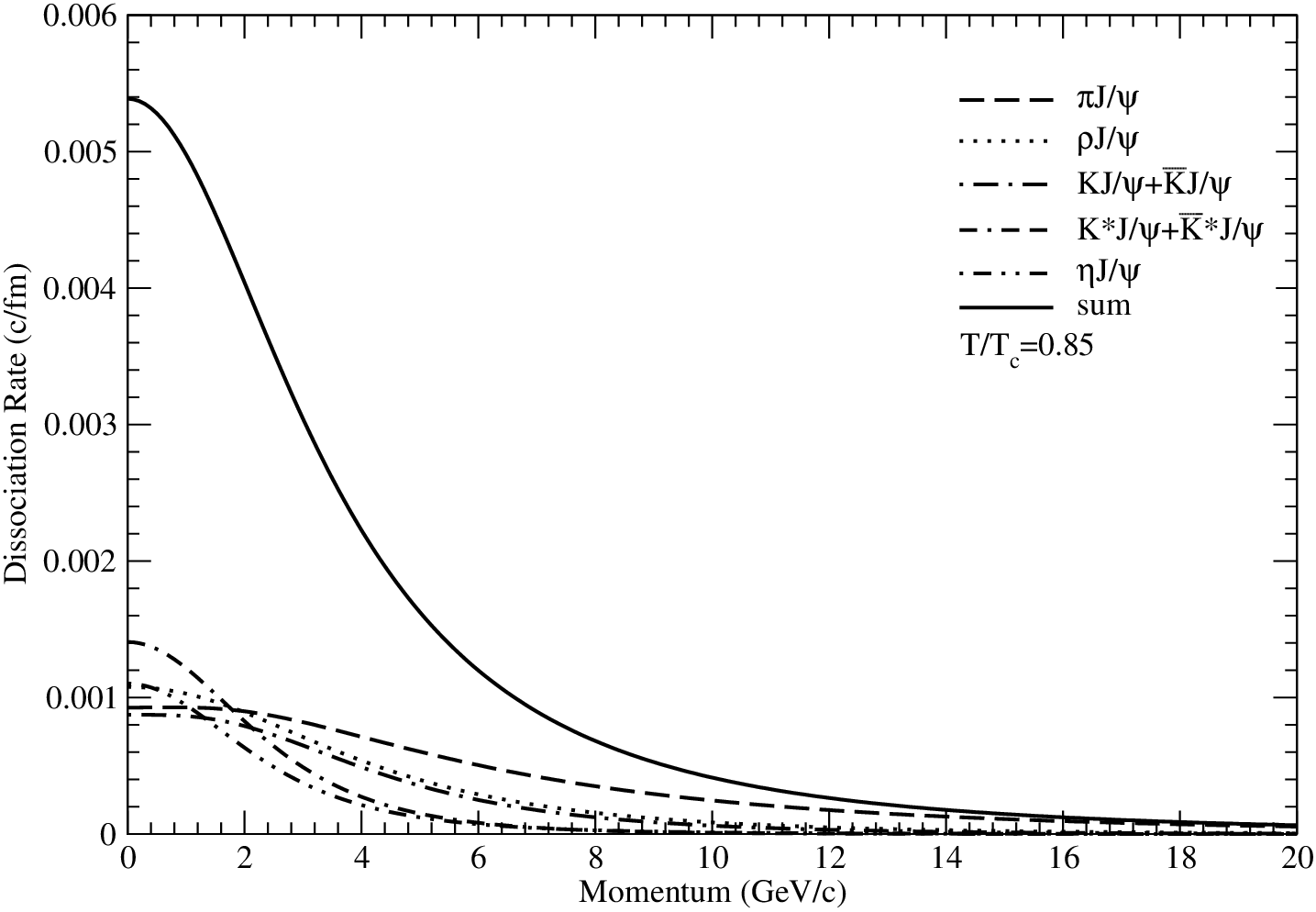}%
\caption{The same as Fig. 15 except for the temperature $0.85T_{\rm c}$ .}
\label{fig17}
\end{figure}

\newpage
\begin{figure}[htbp]
  \centering
    \includegraphics[scale=0.6]{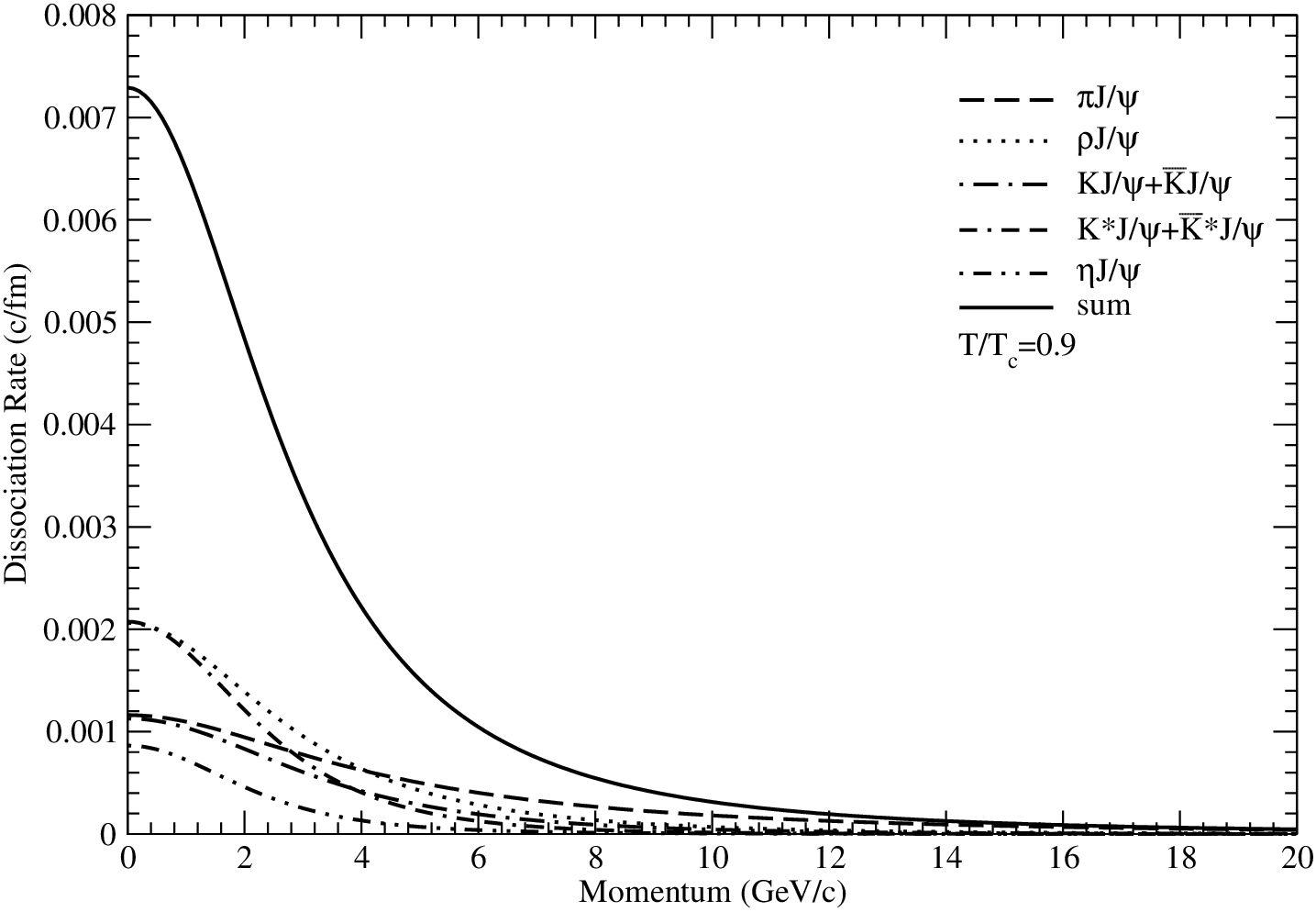}%
\caption{The same as Fig. 15 except for the temperature $0.9T_{\rm c}$ .}
\label{fig18}
\end{figure}

\newpage
\begin{figure}[htbp]
  \centering
    \includegraphics[scale=0.6]{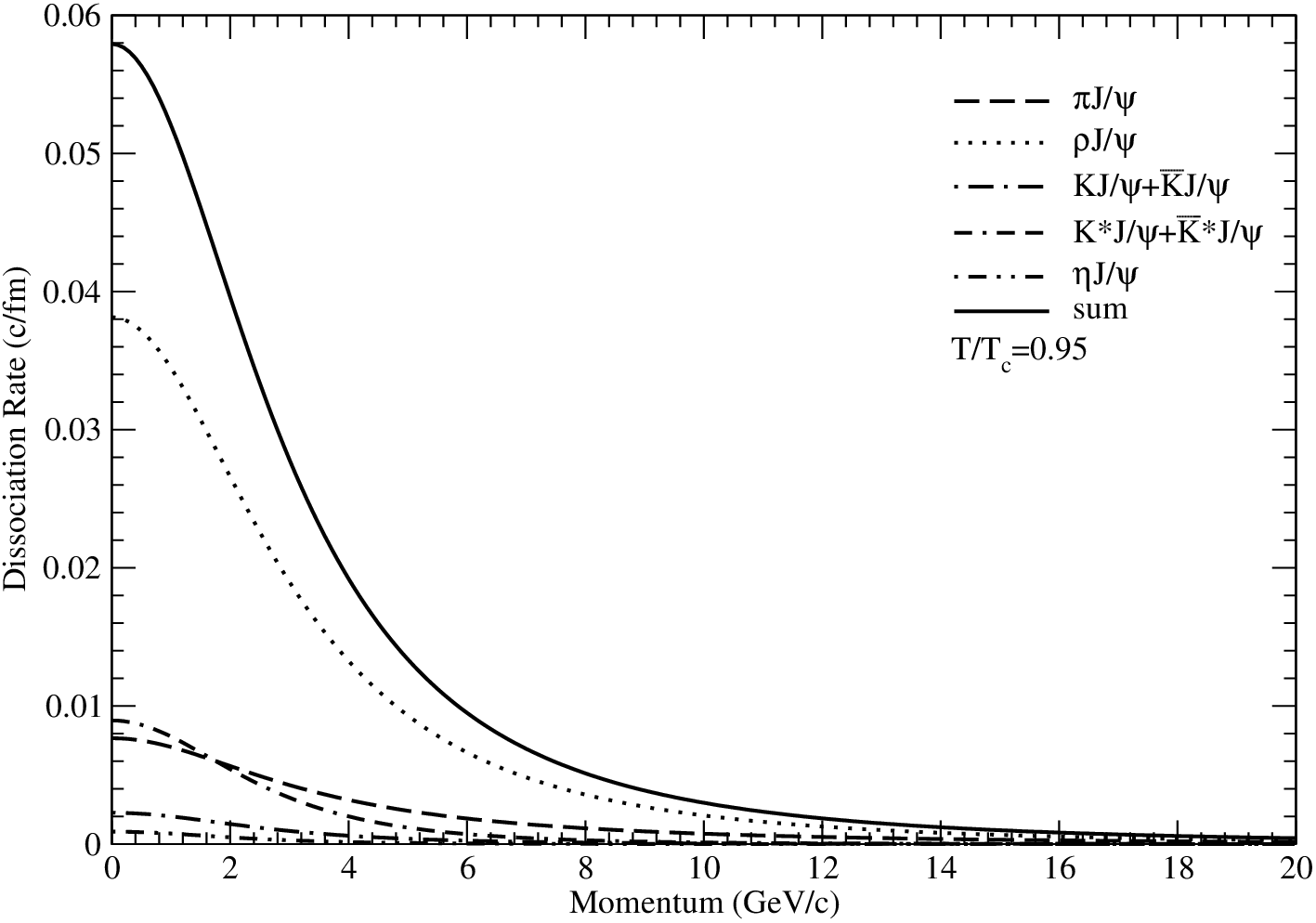}%
\caption{The same as Fig. 15 except for the temperature $0.95T_{\rm c}$ .}
\label{fig19}
\end{figure}

\newpage
\begin{figure}[htbp]
  \centering
    \includegraphics[scale=0.6]{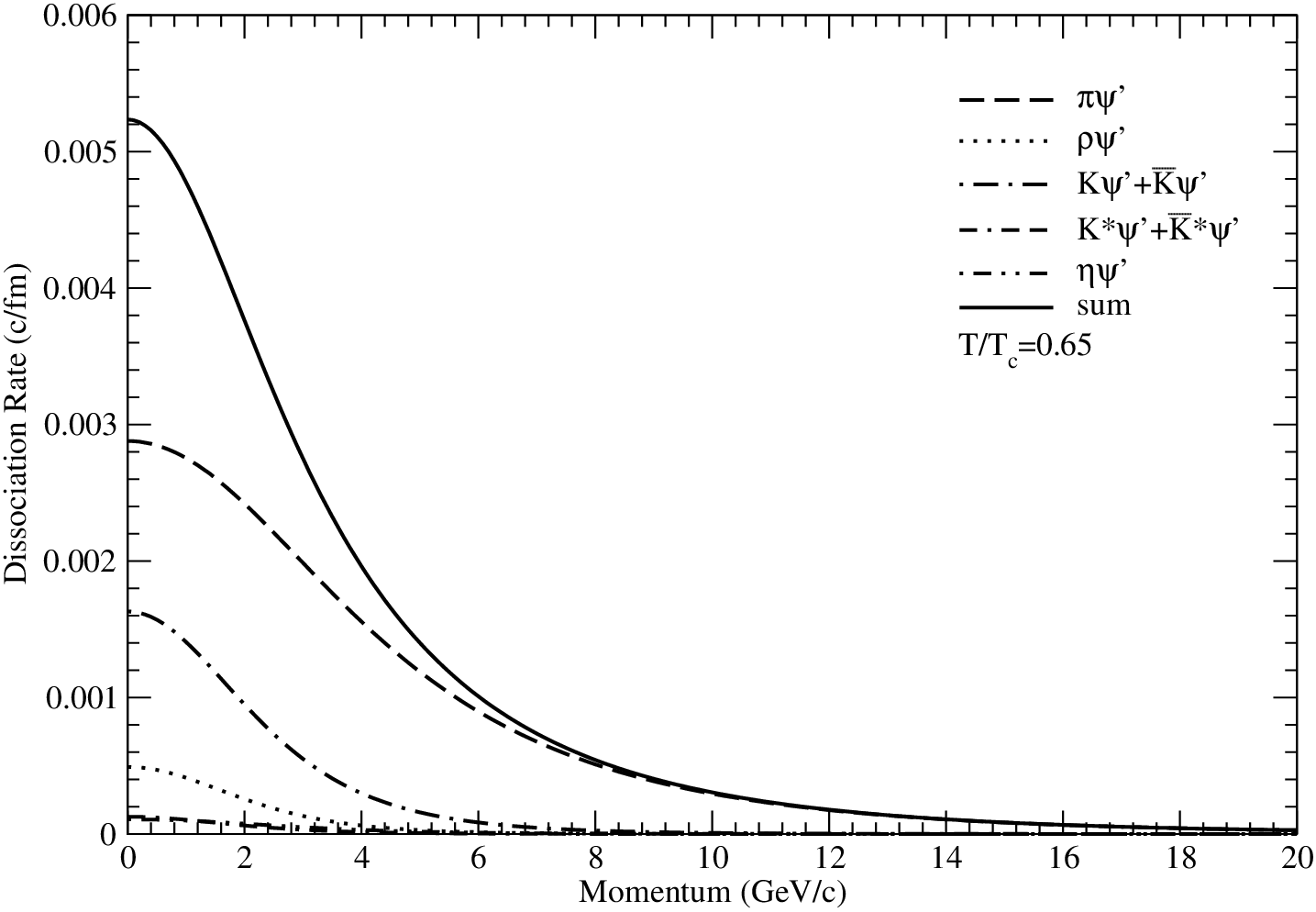}%
\caption{Versus the $\psi'$ momentum
the dissociation rate of $\psi'$ with $\pi$ (dashed curve), the one 
of $\psi'$ with $\rho$ (dotted curve), the sum (dot-dashed curve) of the 
ones of $\psi'$ with $K$ and with $\bar K$, the sum (dot-dash-dashed curve) 
of the ones of $\psi'$ with $K^*$ and with $\bar {K}^*$, and the one of $\psi'$
with $\eta$ (dot-dot-dashed curve) at $T=0.65T_{\rm c}$. The solid curve 
represents the sum of these dissociation rates.}
\label{fig20}
\end{figure}

\newpage
\begin{figure}[htbp]
  \centering
    \includegraphics[scale=0.6]{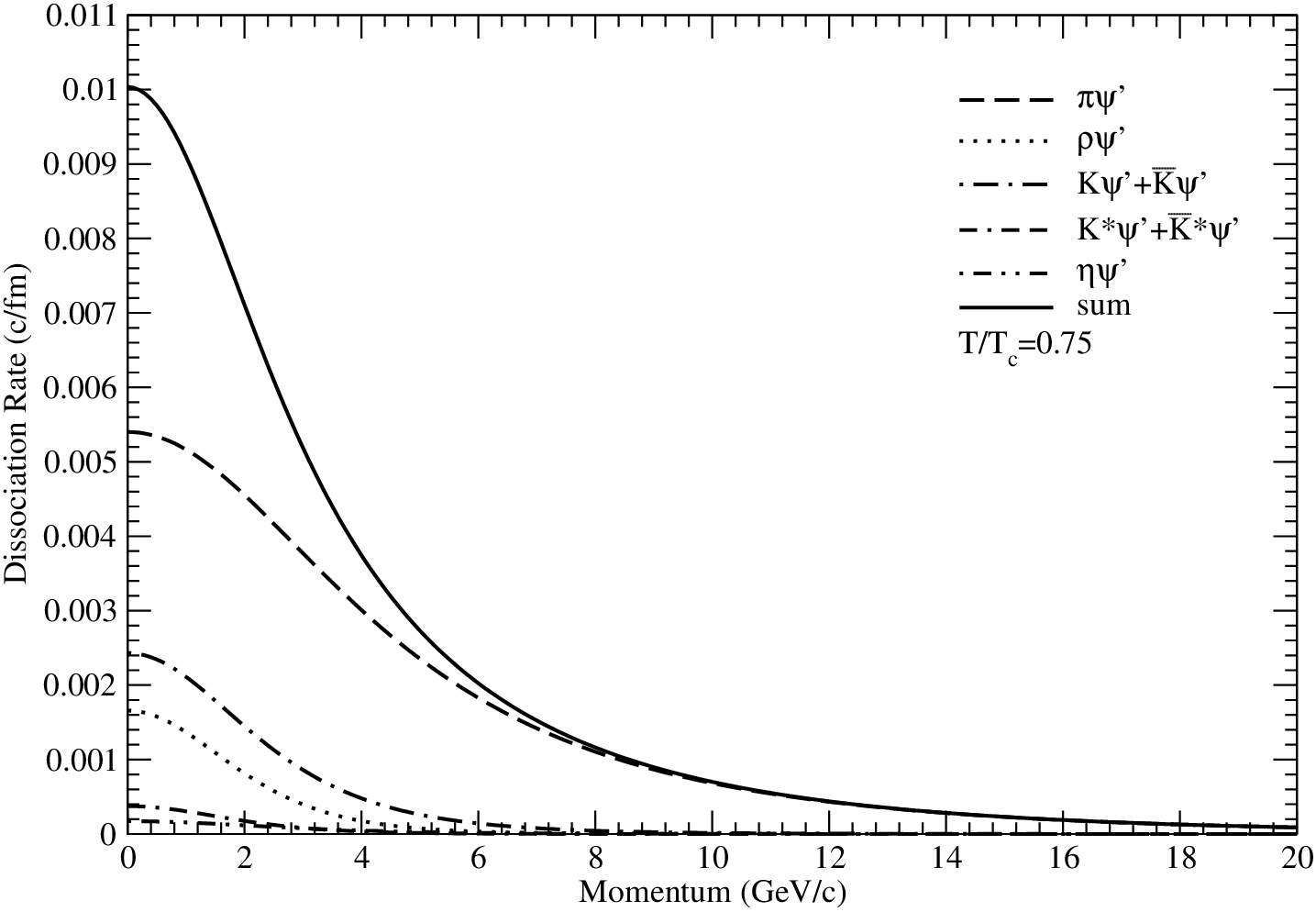}%
\caption{The same as Fig. 20 except for the temperature $0.75T_{\rm c}$.}
\label{fig21}
\end{figure}

\newpage
\begin{figure}[htbp]
  \centering
    \includegraphics[scale=0.6]{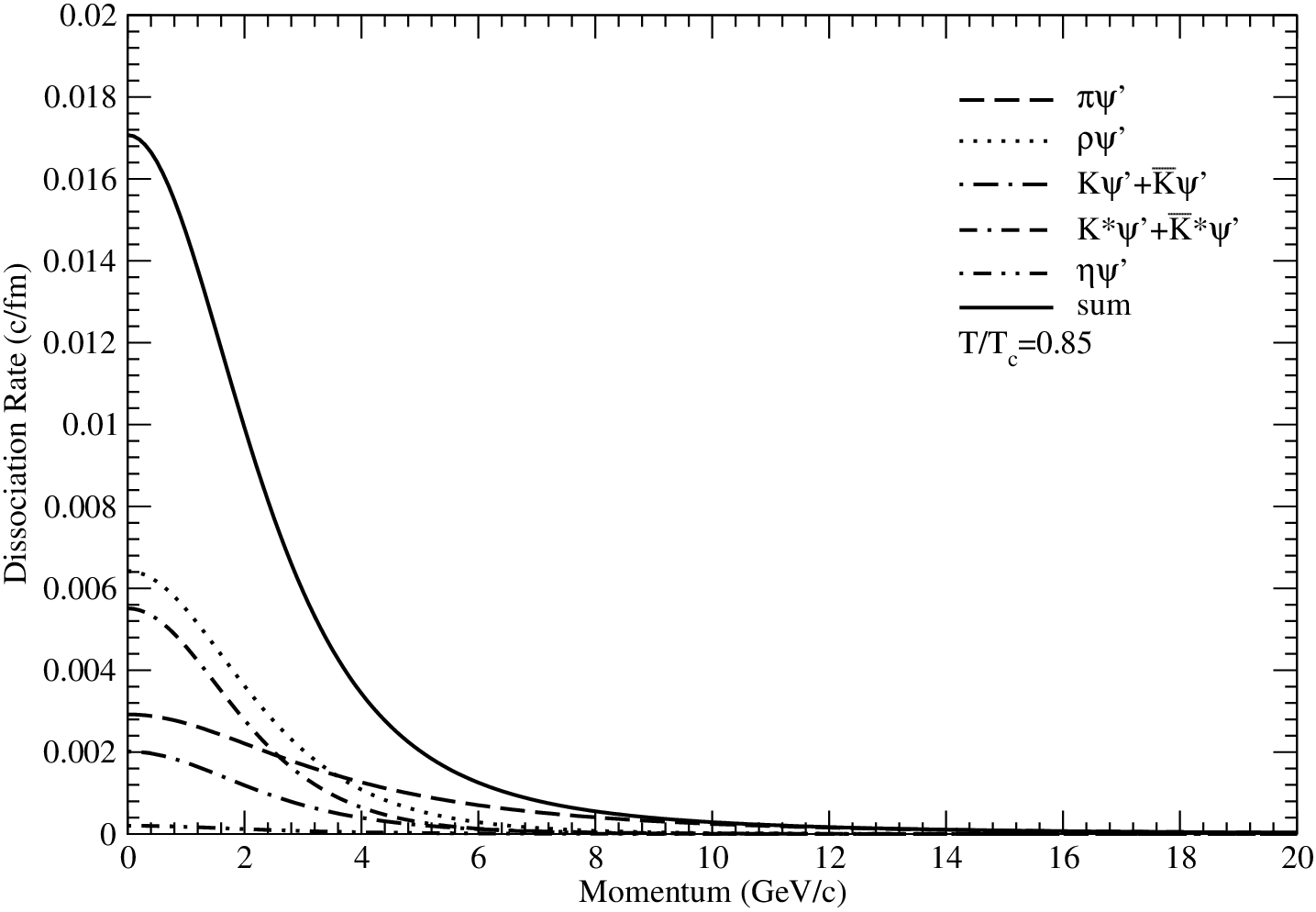}%
\caption{The same as Fig. 20 except for the temperature $0.85T_{\rm c}$.}
\label{fig22}
\end{figure}

\newpage
\begin{figure}[htbp]
  \centering
    \includegraphics[scale=0.6]{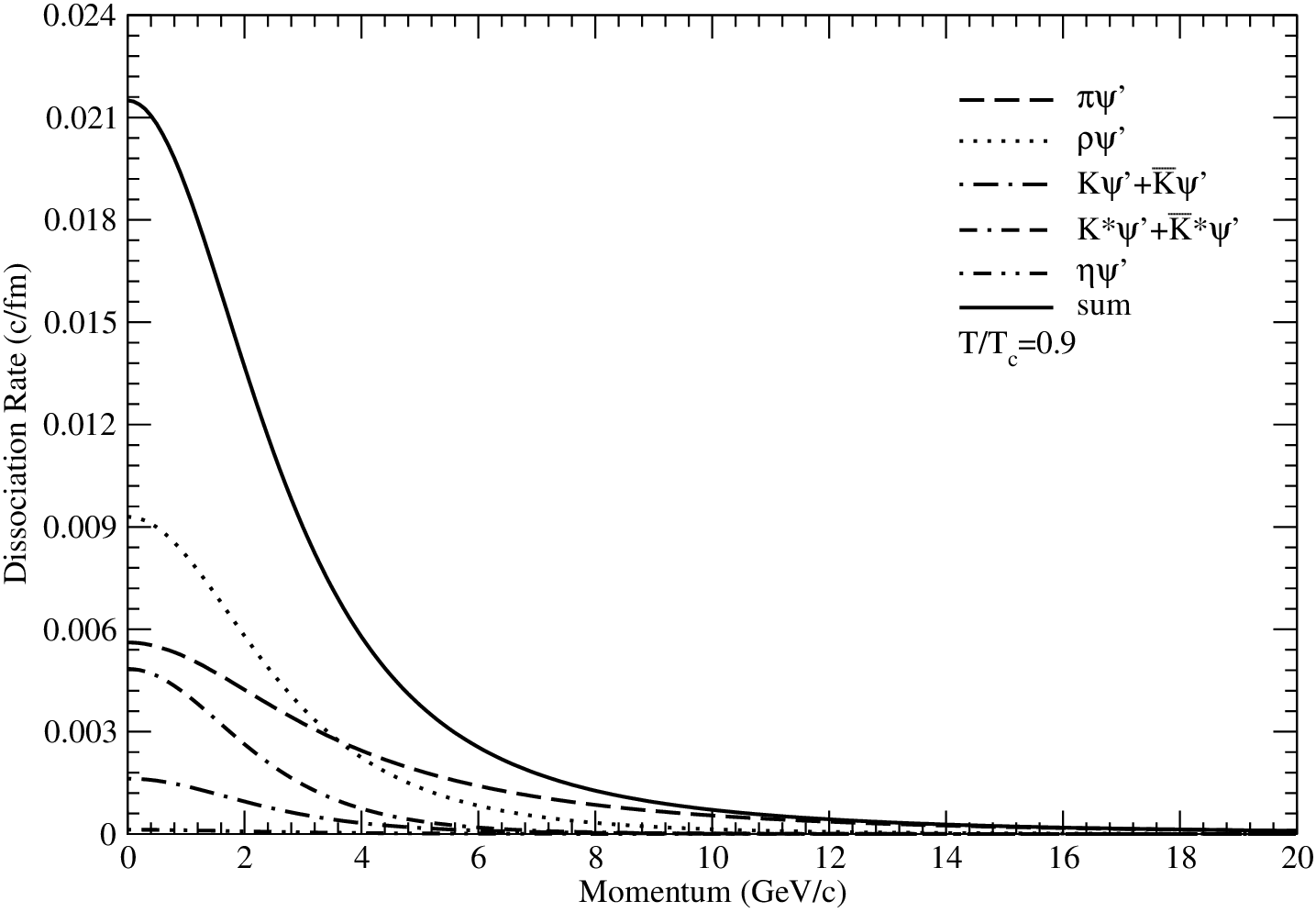}%
\caption{The same as Fig. 20 except for the temperature $0.9T_{\rm c}$.}
\label{fig23}
\end{figure}

\newpage
\begin{figure}[htbp]
  \centering
    \includegraphics[scale=0.6]{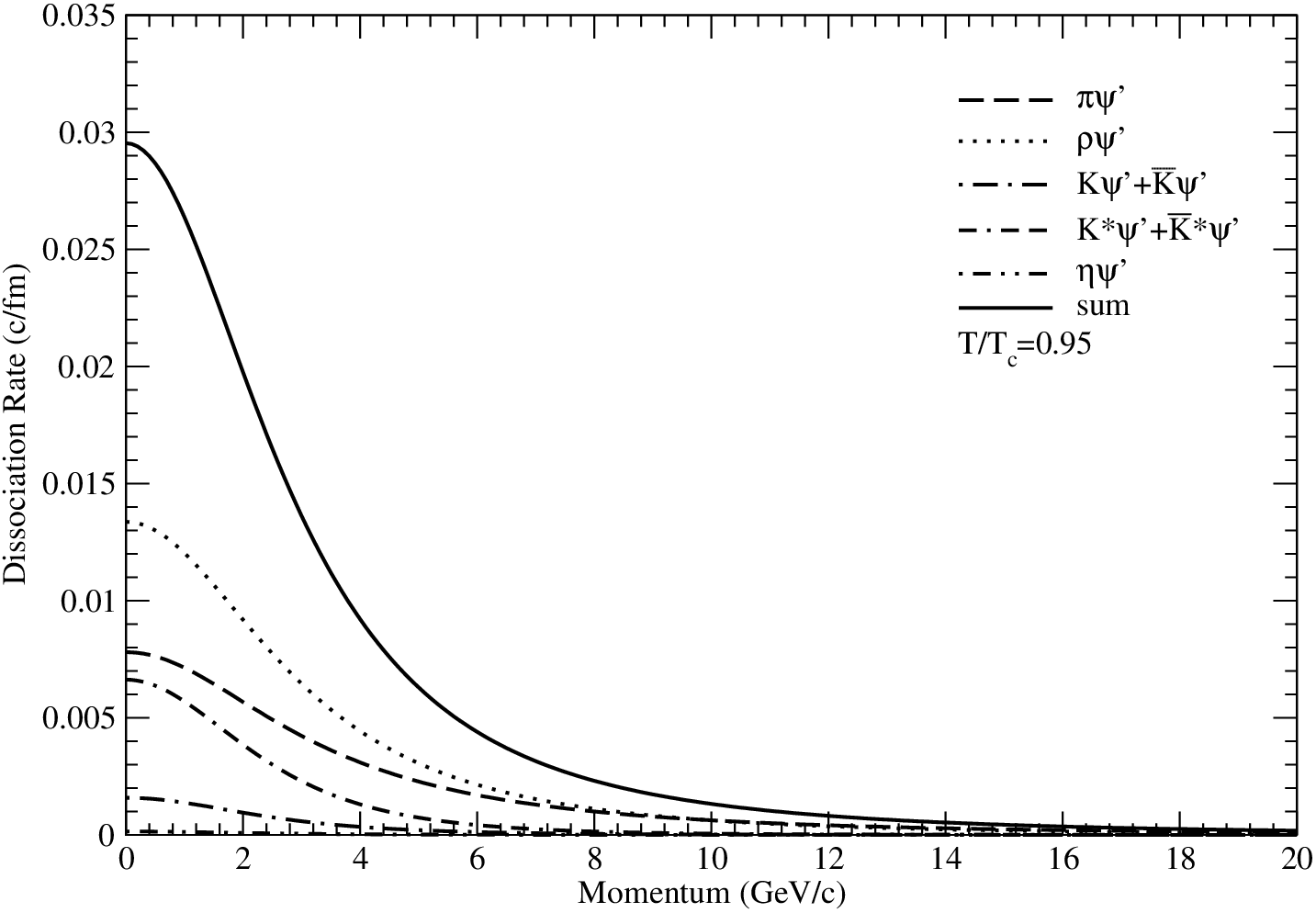}%
\caption{The same as Fig. 20 except for the temperature $0.95T_{\rm c}$.}
\label{fig24}
\end{figure}

\newpage
\begin{figure}[htbp]
  \centering
    \includegraphics[scale=0.6]{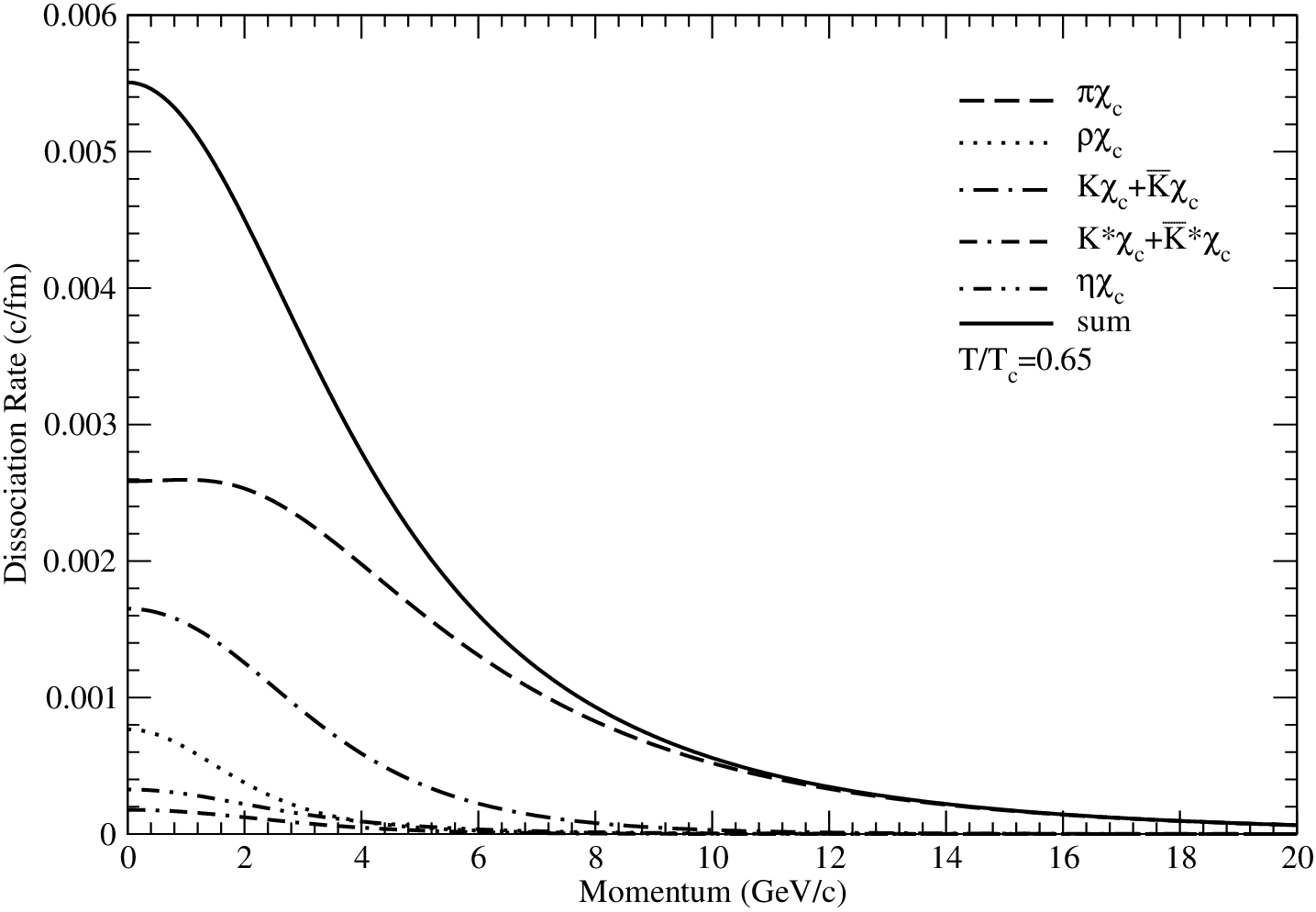}%
\caption{Versus the $\chi_c$ momentum
the dissociation rate of $\chi_c$ with $\pi$ (dashed curve), the one 
of $\chi_c$ with $\rho$ (dotted curve), the sum (dot-dashed curve) of the 
ones of $\chi_c$ with $K$ and with $\bar K$, the sum (dot-dash-dashed curve) 
of the ones of $\chi_c$ with $K^*$ and with $\bar {K}^*$, and the one of 
$\chi_c$
with $\eta$ (dot-dot-dashed curve) at $T=0.65T_{\rm c}$. The solid curve 
represents the sum of these dissociation rates.}
\label{fig25}
\end{figure}

\newpage
\begin{figure}[htbp]
  \centering
    \includegraphics[scale=0.6]{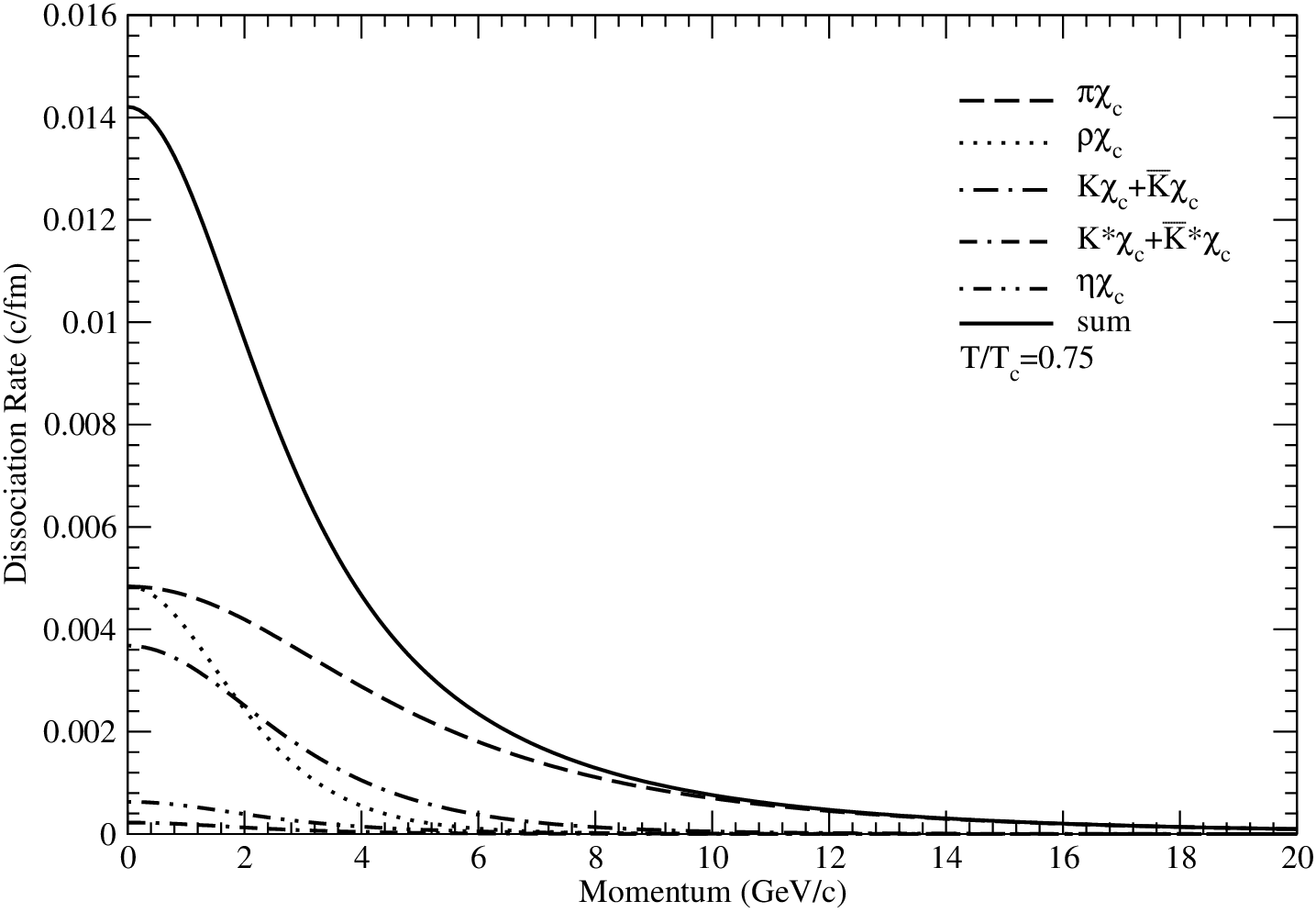}%
\caption{The same as Fig. 25 except for the temperature $0.75T_{\rm c}$.}
\label{fig26}
\end{figure}

\newpage
\begin{figure}[htbp]
  \centering
    \includegraphics[scale=0.6]{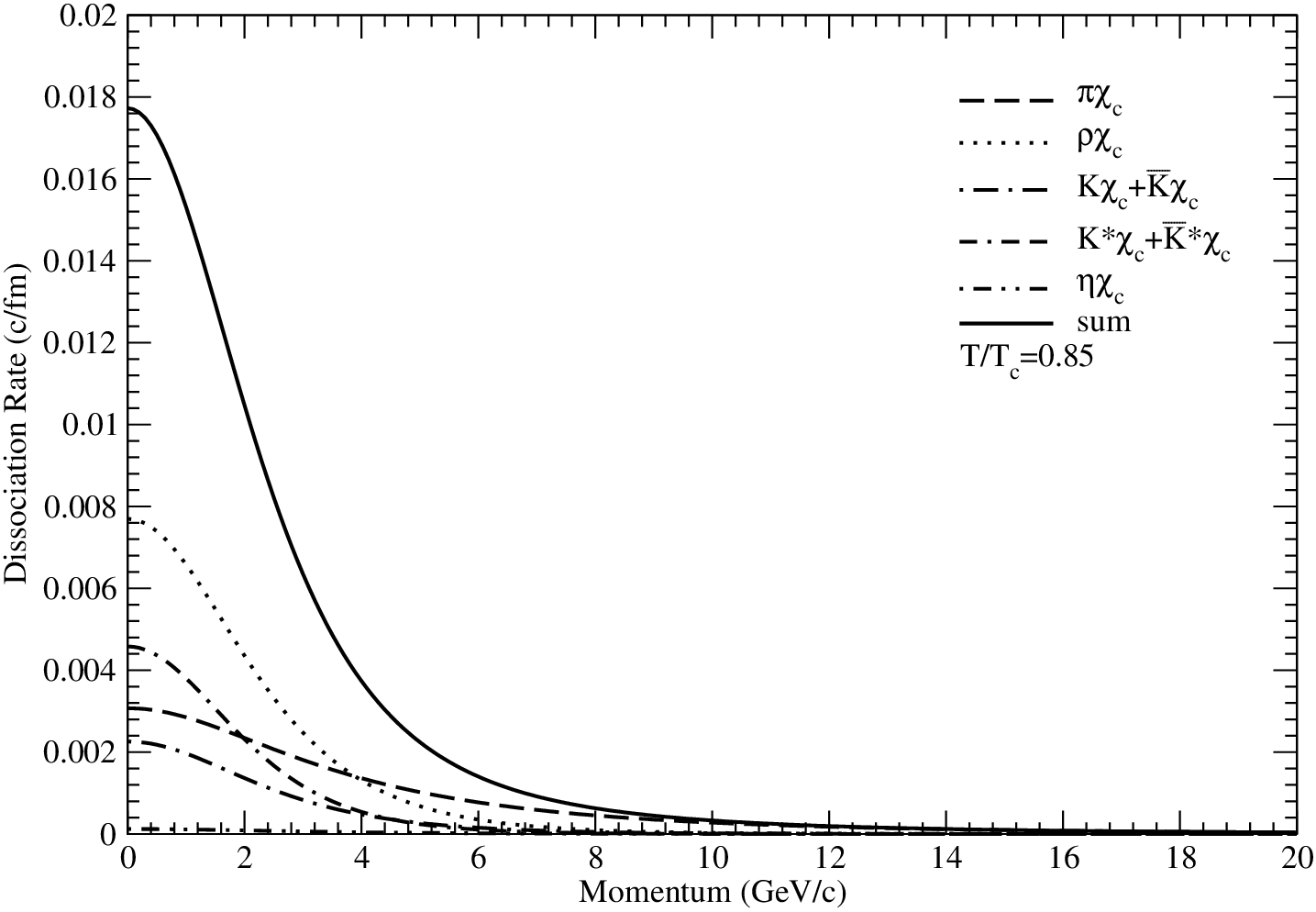}%
\caption{The same as Fig. 25 except for the temperature $0.85T_{\rm c}$.}
\label{fig27}
\end{figure}

\newpage
\begin{figure}[htbp]
  \centering
    \includegraphics[scale=0.6]{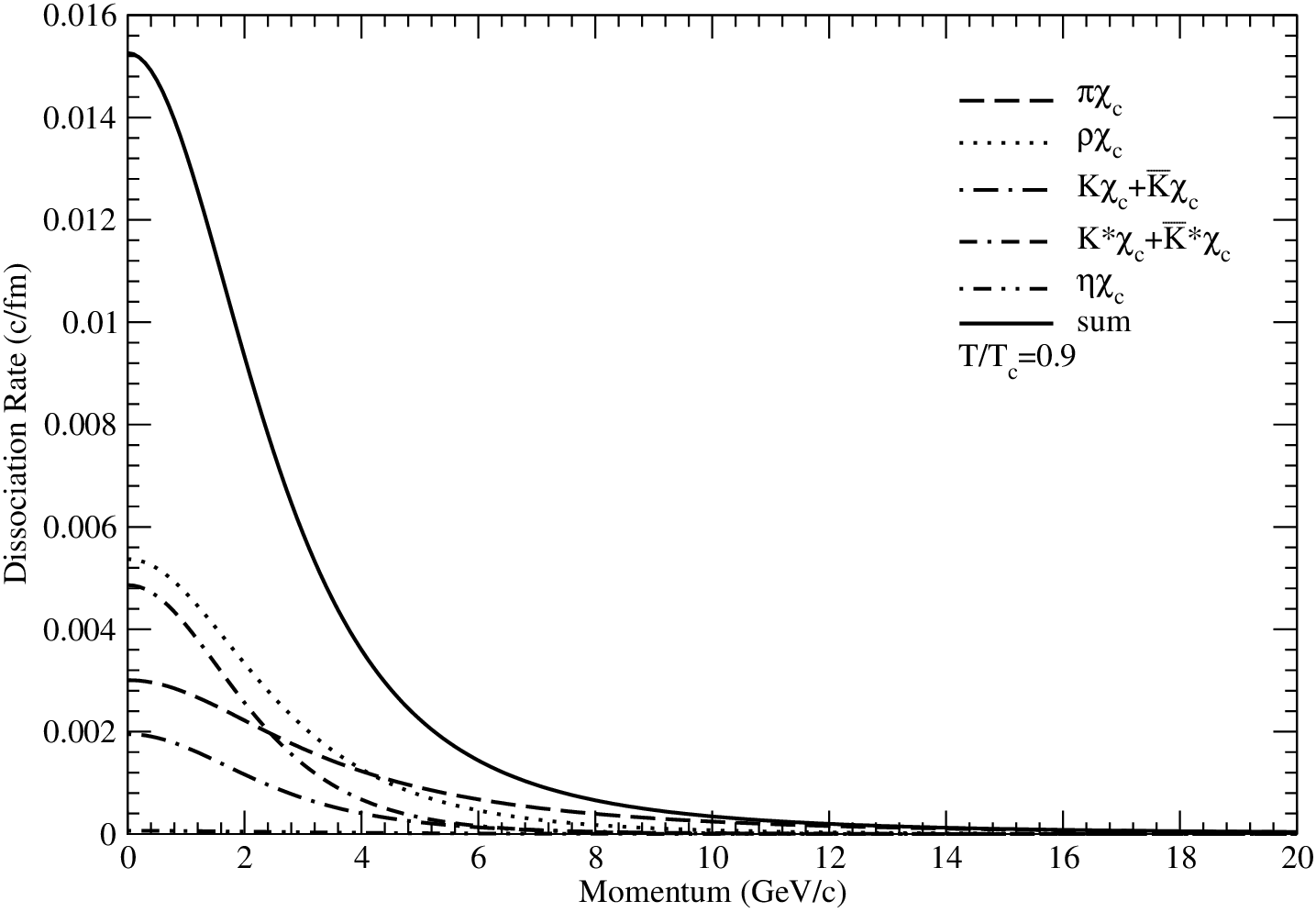}%
\caption{The same as Fig. 25 except for the temperature $0.9T_{\rm c}$.}
\label{fig28}
\end{figure}

\newpage
\begin{figure}[htbp]
  \centering
    \includegraphics[scale=0.6]{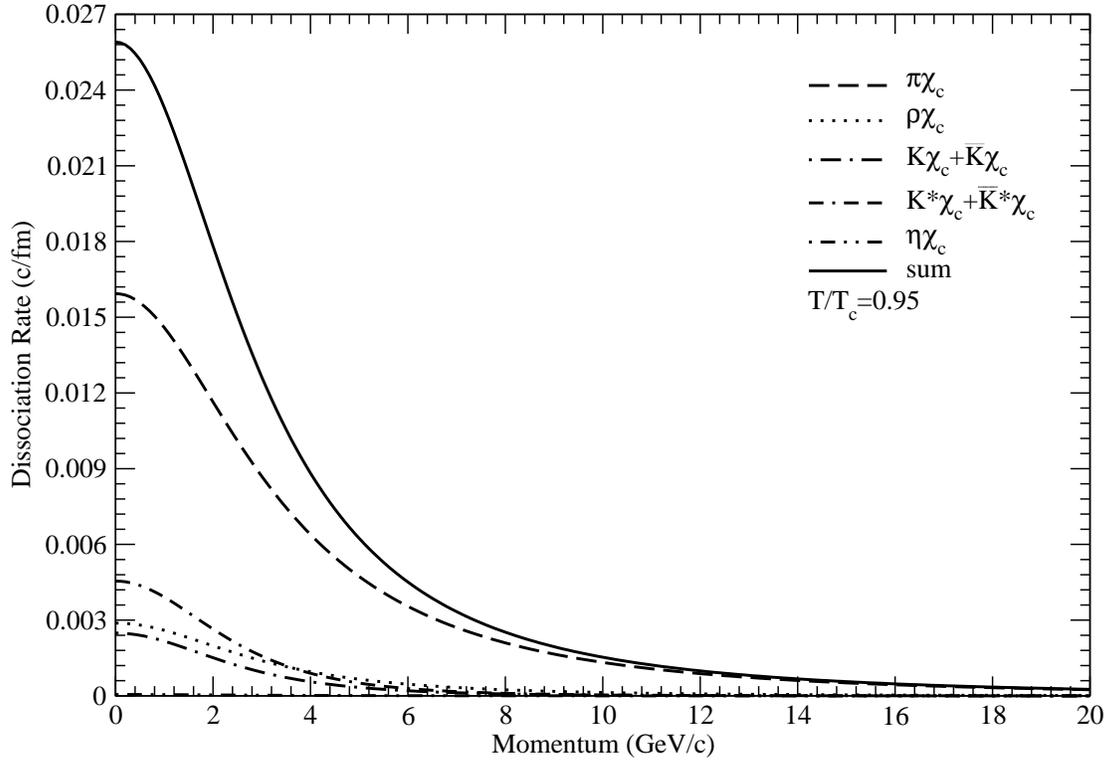}%
\caption{The same as Fig. 25 except for the temperature $0.95T_{\rm c}$.}
\label{fig29}
\end{figure}

\newpage
\begin{figure}[htbp]
  \centering
    \includegraphics[scale=0.65]{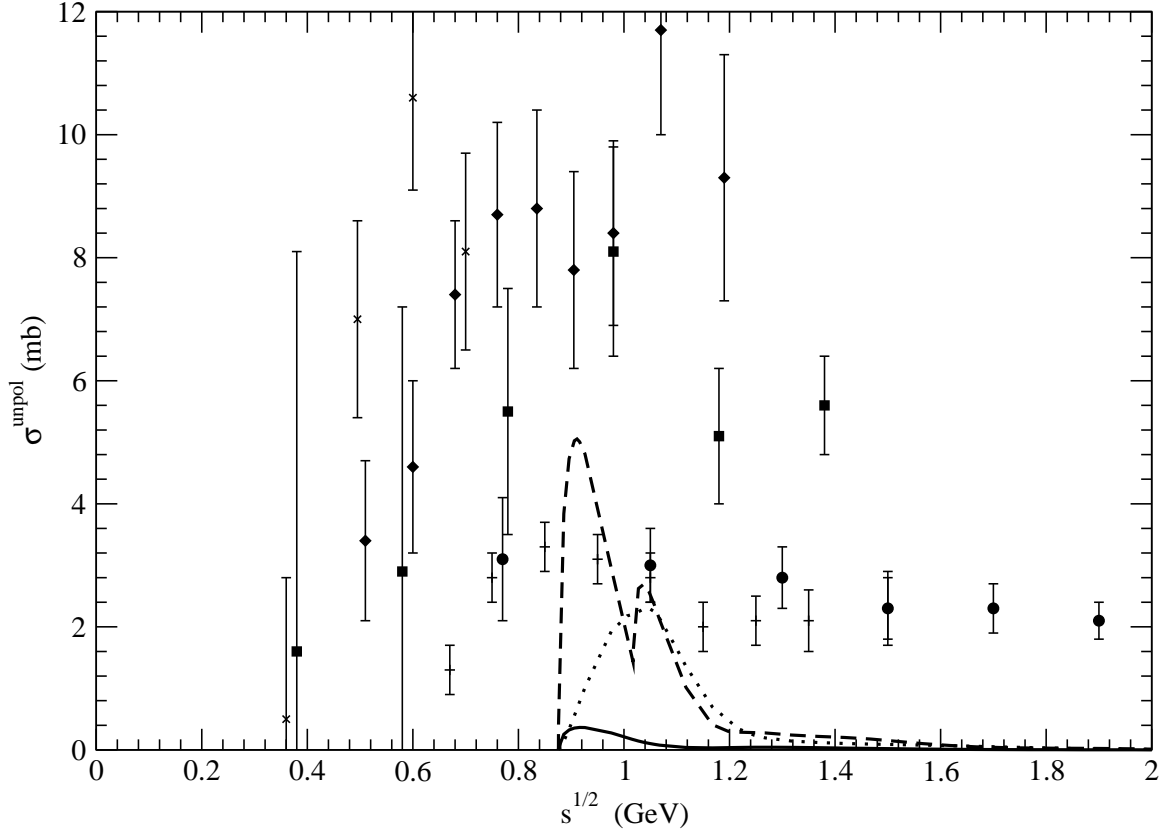}%
\caption{Cross sections for $\pi$-induced reactions that are governed by
quark interchange. Our theoretical results are shown by the solid, dashed, and
dotted curves which stand for unpolarized cross sections for
$\pi J/\psi \to \bar{D}^*D+\bar{D}D^*+\bar{D}^*D^*$, 
for $\pi \psi^\prime \to \bar{D}^*D+\bar{D}D^*+\bar{D}^*D^*$,
and for $\pi \chi_c \to \bar{D}^*D+\bar{D}D^*+\bar{D}^*D^*$, respectively. The
curves translate by -3 GeV in $\sqrt s$. Experimental data of
elastic $\pi^- \pi^-$ cross section:
$\times$, Ref. \cite{Colton}; $\Box$, Ref. \cite{CFSW}; $\Diamond$,
Ref. \cite{Losty}. Experimental data of elastic $\pi^- K^-$ cross section: 
+, Ref. \cite{JMTVH}; $\bigcirc$, Ref. \cite{Linglin}.}
\label{fig30}
\end{figure}

\begin{table*}
\caption{\label{table:I}Quantities relevant to the cross sections for the 
$K^*J/\psi$
dissociation reactions. $a_1$and $a_2$ are in units of millibarns; $b_1$, 
$b_2$, $d_0$,
and $\sqrt{s_{\rm z}}$ are in units of GeV; $c_1$ and $c_2$ are dimensionless.}
\tabcolsep=5.2pt
\begin{tabular}{cccccccccc}
\hline
Reactions & $T/T_{\rm c} $ & $a_1$ & $b_1$ & $c_1$ & $a_2$ & $b_2$ & $c_2$ &
$d_0$ & $\sqrt{s_{\rm z}}$ \\
\hline
 $K^*J/\psi\to\bar{D}D^+_s$
&  0     & 0.0466 & 0.021  & 0.51  & 0.0587 & 0.26   & 5.26  & 0.25   &4.88 \\
&  0.65  & 0.145  & 0.02   & 0.51  & 0.039  & 0.04   & 1.72  & 0.03   &4.59 \\
&  0.75  & 0.13   & 0.008  & 0.50  & 0.16   & 0.039  & 1.23  & 0.022  &4.37 \\
&  0.85  & 0.21   & 0.009  & 0.46  & 0.33   & 0.021  & 0.56  & 0.015  &3.73 \\
&  0.9   & 0.04   & 0.008  & 0.82  & 0.42   & 0.013  & 0.48  & 0.012  &3.59 \\
&  0.95  & 0.23   & 0.007  & 0.59  & 0.52   & 0.010  & 0.46  & 0.0081 &3.37 \\
  \hline
 $K^*J/\psi\to\bar{D}^*D^+_s$
&  0     & 0.407  & 0.024  & 0.59  & 0.162  & 0.288  & 44.8  & 0.03   &4.98 \\
&  0.65  & 1.016  & 0.017  & 0.494 & 0.205  & 0.033  & 0.79  & 0.02   &4.47 \\
&  0.75  & 0.860  & 0.017  & 0.52  & 0.026  & 0.023  & 0.48  & 0.017  &3.92 \\
&  0.85  & 0.355  & 0.006  & 0.49  & 0.333  & 0.025  & 1.29  & 0.015  &3.71 \\
&  0.9   & 0.36   & 0.007  & 0.44  & 0.29   & 0.017  & 0.78  & 0.01   &3.57 \\
&  0.95  & 0.74   & 0.006  & 0.45  & 0.49   & 0.013  & 0.71  & 0.0081 &3.36 \\
  \hline
 $K^*J/\psi\to\bar{D}D^{*+}_s$
&  0     & 0.486  & 0.0228 & 0.560 & 0.231  & 0.270  & 52.4  & 0.02   &4.84 \\
&  0.65  & 0.539  & 0.0186 & 0.53  & 0.127  & 0.27   & 31.5  & 0.02   &4.63 \\
&  0.75  & 0.18   & 0.017  & 0.45  & 0.13   & 0.019  & 0.65  & 0.017  &4.51 \\
&  0.85  & 0.055  & 0.010  & 0.51  & 0.038  & 0.017  & 0.49  & 0.012  &4.26 \\
&  0.9   & 0.043  & 0.005  & 0.52  & 0.065  & 0.009  & 0.50  & 0.0071 &3.87 \\
&  0.95  & 0.39   & 0.004  & 0.41  & 0.52   & 0.009  & 0.68  & 0.0071 &3.35 \\
  \hline
  $K^*J/\psi\to\bar{D}^*D^{*+}_s$
&  0     & 0.743  & 0.022  & 0.54  & 0.458  & 0.26   & 5.78  & 0.025  &5.04 \\
&  0.65  & 0.327  & 0.018  & 0.49  & 0.180  & 0.228  & 5.03  & 0.018  &4.77 \\
&  0.75  & 0.199  & 0.0149 & 0.49  & 0.081  & 0.225  & 4.78  & 0.015  &4.61 \\
&  0.85  & 0.0430 & 0.0097 & 0.50  & 0.0129 & 0.1648 & 3.05  & 0.01   &4.28 \\
&  0.9   & 0.040  & 0.003  & 0.43  & 0.029  & 0.008  & 0.90  & 0.0046 &3.93 \\
&  0.95  & 1.39   & 0.003  & 0.45  & 1.33   & 0.011  & 1.13  & 0.0061 &3.34 \\
  \hline
\end{tabular}
\end{table*}

\newpage
\begin{table*}
\caption{\label{table:II}The same as Table 1 except for the $K^*\psi'$ 
dissociation.}
\tabcolsep=4.8pt
\begin{tabular}{cccccccccc}
\hline
Reactions & $T/T_{\rm c} $ & $a_1$ & $b_1$ & $c_1$ & $a_2$ & $b_2$ & $c_2$ &
$d_0$ & $\sqrt{s_{\rm z}}$ \\
\hline
 $K^*\psi'\to\bar{D}D^+_s$
&  0   & 0.012  & 0.06   & 0.41  & 0.017  & 0.11   & 0.70   & 0.1      &5.5 \\
&  0.65  & 0.0030 & 0.08   & 0.44  & 0.0006 & 0.17   & 2.27   & 0.1    &4.98 \\
&  0.75  & 0.00063& 0.010  & 0.51  & 0.00101& 0.168  & 1.44   & 0.15   &4.81 \\
&  0.85  & 0.0025 & 0.006  & 0.58  & 0.0021 & 0.023  & 0.39   & 0.007  &4.22 \\
&  0.9   & 0.022  & 0.018  & 0.389 & 0.099  & 0.025  & 3.558  & 0.022  &3.58 \\
&  0.95  & 2.91   & 0.001  & 0.52  & 3.20   & 0.005  & 1.52   & 0.0026 &3.33 \\
  \hline
 $K^*\psi'\to\bar{D}^*D^+_s$
&  0     & 0.040  & 0.03   & 0.47  & 0.057  & 0.16   & 1.44   & 0.1    &5.74 \\
&  0.65  & 0.0061 & 0.015  & 0.48  & 0.0071 & 0.182  & 1.89   & 0.168  &5.13 \\
&  0.75  & 0.0059 & 0.010  & 0.53  & 0.0066 & 0.039  & 2.09   & 0.026  &4.71 \\
&  0.85  & 0.080  & 0.007  & 0.44  & 0.041  & 0.010  & 1.04   & 0.0092 &3.78 \\
&  0.9   & 0.75   & 0.001  & 0.45  & 0.76   & 0.004  & 1.15   & 0.002  &3.54 \\
&  0.95  & 3.32   & 0.002  & 0.46  & 1.87   & 0.005  & 0.92   & 0.0031 &3.34 \\
  \hline
 $K^*\psi'\to\bar{D}D^{*+}_s$
&  0     & 0.066  & 0.07   & 0.47  & 0.024  & 0.19   & 2.11   & 0.1    &5.49 \\
&  0.65  & 0.01349& 0.018  & 0.51  & 0.00949& 0.201  & 2.56   & 0.022  &4.94 \\
&  0.75  & 0.0158 & 0.018  & 0.48  & 0.0038 & 0.040  & 1.49   & 0.023  &4.67 \\
&  0.85  & 0.637  & 0.0020 & 0.49  & 0.279  & 0.036  & 5.28   & 0.003  &3.77 \\
&  0.9   & 4.15   & 0.0024 & 0.38  & 2.91   & 0.0036 & 0.93   & 0.0031 &3.55 \\
&  0.95  & 0.93   & 0.0033 & 0.78  & 0.64   & 0.0040 & 0.27   & 0.0031 &3.34 \\
  \hline
 $K^*\psi'\to\bar{D}^*D^{*+}_s$
&  0     & 0.179  & 0.03   & 0.502 & 0.214  & 0.19   & 2.523  & 0.15   &5.48 \\
&  0.65  & 0.150  & 0.019  & 0.44  & 0.225  & 0.023  & 0.62   & 0.022  &4.67 \\
&  0.75  & 0.50   & 0.017  & 0.53  & 0.35   & 0.022  & 1.79   & 0.022  &4.11 \\
&  0.85  & 17.96  & 0.002  & 0.43  & 13.51  & 0.005  & 0.89   & 0.0031 &3.74 \\
&  0.9   & 2.06   & 0.003  & 0.65  & 1.28   & 0.006  & 0.33   & 0.0031 &3.58 \\
&  0.95  & 1.28   & 0.003  & 0.45  & 0.98   & 0.010  & 0.82   & 0.0051 &3.34 \\
  \hline
\end{tabular}
\end{table*}

\newpage
\begin{table*}
\caption{\label{table:III}The same as Table 1 except for the $K^*\chi_c$ 
dissociation.}
\tabcolsep=4.8pt
\begin{tabular}{cccccccccc}
\hline
Reactions & $T/T_{\rm c} $ & $a_1$ & $b_1$ & $c_1$ & $a_2$ & $b_2$ & $c_2$ &
$d_0$ & $\sqrt{s_{\rm z}}$ \\
\hline
 $K^*\chi_c\to\bar{D}D^+_s$
&  0     & 0.001345& 0.218  & 5.57  & 0.02402 & 0.0801  & 0.484 & 0.1  &5.31 \\
&  0.65  & 0.00377 & 0.0314 & 0.454 & 0.0074  & 0.1381  & 1.232 & 0.1  &4.93 \\
&  0.75  & 0.00013 & 0.004  & 0.39  & 0.0032  & 0.14    & 1.5   & 0.1  &4.78 \\
&  0.85  & 0.00025 & 0.0058 & 0.44  & 0.0003  & 0.13   & 1.67  & 0.15  &4.46 \\
&  0.9   & 0.1     & 0.0011 & 0.47  & 0.18    & 0.0063 & 1.22  & 0.005 &3.54 \\
&  0.95  & 1.52    & 0.014  & 2.2   & 0.81    & 0.007   & 1.2  & 0.011 &3.3  \\
  \hline
 $K^*\chi_c\to\bar{D}^*D^+_s$
&  0     & 0.013   & 0.194  & 4.03  & 0.077   & 0.079   & 0.473 & 0.1  &5.3  \\
&  0.65  & 0.00116 & 0.0047 & 0.47  & 0.029   & 0.168   & 2.3   & 0.15 &4.91 \\
&  0.75  & 0.0028  & 0.007  & 0.57  & 0.0088  & 0.195   & 3.29  & 0.15 &4.74 \\
&  0.85  & 0.00173 & 0.0017 & 0.51  & 0.00441 & 0.028   & 2.9  & 0.025 &4.19 \\
&  0.9   & 1.23    & 0.003  & 0.5   & 0.79    & 0.01    & 1.6  & 0.005 &3.51 \\
&  0.95  & 0.77    & 0.008  & 0.88  & 1.65    & 0.013   & 2.05 & 0.012 &3.32 \\
  \hline
 $K^*\chi_c\to\bar{D}D^{*+}_s$
&  0     & 0.0178  & 0.188  & 3.69  & 0.0844  & 0.079   & 0.473 & 0.1  &5.29 \\
&  0.65  & 0.0051  & 0.0069 & 0.56  & 0.048   & 0.176   & 3.62  & 0.15 &4.9  \\
&  0.75  & 0.0074  & 0.01   & 0.53  & 0.0162  & 0.197   & 5.1   & 0.15 &4.74 \\
&  0.85  & 0.5     & 0.0103 & 1.83  & 1.13    & 0.003   & 0.52 & 0.005 &3.7  \\
&  0.9   & 1.57    & 0.012  &1.89   & 1.32    & 0.01    & 1.38 & 0.011 &3.57 \\
&  0.95  & 0.05    & 0.003  & 1.52  & 0.46    & 0.012   & 1.88 & 0.012 &3.4  \\
  \hline
 $K^*\chi_c\to\bar{D}^*D^{*+}_s$
&  0     & 0.155  & 0.0306  & 0.45  & 0.319   & 0.178   & 2.11  & 0.15 &5.27 \\
&  0.65  & 0.56   & 0.0117  & 0.5   & 0.163   & 0.22    & 7.58  & 0.01 &4.76 \\
&  0.75  & 0.6    & 0.0038  & 0.44  & 0.6     & 0.011   & 0.78 & 0.008 &4.37 \\
&  0.85  & 4.3    & 0.009   & 0.9   & 8.8     & 0.012   & 1.9  & 0.011 &3.77 \\
&  0.9   & 0.84   & 0.012   & 1.74  & 0.29    & 0.007   & 0.65 & 0.011 &3.66 \\
&  0.95  & 0.079  & 0.023   & 0.5   & 0.151   & 0.01    & 2    & 0.011 &3.47 \\
  \hline
\end{tabular}
\end{table*}

\newpage
\begin{table*}
\caption{\label{table:IV}Reactions that contribute to the dissociation rate of 
charmonium and meson. $\sigma^{\rm unpol}$ in Eqs. (30) and (31)
is the sum of the unpolarized cross sections for the reactions listed 
in each row of the third column.}
\begin{tabular}{ccc}
\hline
charmonium  &  meson  &  reactions      \\
\hline
$J/\psi$ & $\pi$ & $\pi J/\psi \to \bar{D}^* D, \bar{D} D^*, \bar{D}^* D^*$  \\
\hline
$J/\psi$ & $\rho$ & $\rho J/\psi \to \bar{D} D, \bar{D}^* D,
\bar{D} D^*, \bar{D}^* D^*$   \\
\hline
$J/\psi$ & $K$ & $K J/\psi \to \bar{D}^* D^+_s, \bar{D} D^{*+}_s, \bar{D}^* 
D^{*+}_s$         \\
\hline
$J/\psi$ & $\bar K$ & $\bar{K} J/\psi \to D_s^{*-} D, 
D_s^- D^*, D_s^{*-} D^*$      \\
\hline
$J/\psi$ & $K^*$ & $K^* J/\psi \to \bar{D} D^+_s, \bar{D}^* D^+_s, 
\bar{D} D^{*+}_s, \bar{D}^* D^{*+}_s$         \\
\hline
$J/\psi$ & $\bar {K}^*$ & $\bar{K}^* J/\psi \to D_s^- D, D_s^{*-} D, 
D_s^- D^*, D_s^{*-} D^*$      \\
\hline
$J/\psi$ & $\eta$ & $\eta J/\psi \to \bar{D}^{*} D, \bar{D} D^{*}, 
\bar{D}^{*} D^{*}, D^{*-}_{s} D_s^+, D_s^- D^{*+}_s, D^{*-}_{s} D^{*+}_{s}$  \\
\hline
$\psi'$ & $\pi$ & $\pi \psi' \to \bar{D}^* D, \bar{D} D^*, \bar{D}^* D^*$    \\
\hline
$\psi'$ & $\rho$ & $\rho \psi' \to \bar{D} D, \bar{D}^* D,
\bar{D} D^*, \bar{D}^* D^*$       \\
\hline
$\psi'$ & $K$ & $K \psi' \to \bar{D}^* D^+_s, \bar{D} D^{*+}_s,
\bar{D}^* D^{*+}_s$       \\
\hline
$\psi'$ & $\bar K$ & $\bar{K} \psi' \to D_s^{*-} D, D_s^- D^*, D_s^{*-} D^*$ \\
\hline
$\psi'$ & $K^*$ & $K^* \psi' \to \bar{D} D^+_s, \bar{D}^* D^+_s, 
\bar{D} D^{*+}_s, \bar{D}^* D^{*+}_s$       \\
\hline
$\psi'$ & $\bar {K}^*$ & $\bar{K}^* \psi' \to D_s^- D, D_s^{*-} D, D_s^- D^*, 
D_s^{*-} D^*$ \\
\hline
$\psi'$ & $\eta$ & $\eta \psi' \to \bar{D}^{*} D, \bar{D} D^{*}, 
\bar{D}^{*} D^{*}, D^{*-}_s D_s^+, D_s^- D^{*+}_s, D^{*-}_{s} D^{*+}_{s}$    \\
\hline
$\chi_c$ & $\pi$ & $\pi \chi_c \to \bar{D}^* D, \bar{D} D^*, \bar{D}^* D^*$ \\
\hline
$\chi_c$ & $\rho$ & $\rho \chi_{c} \to \bar{D} D, \bar{D}^* D, \bar{D} D^*, 
\bar{D}^* D^*$    \\
\hline
$\chi_c$ & $K$ & $K \chi_c \to \bar{D}^* D^+_s, \bar{D} D^{*+}_s,
\bar{D}^* D^{*+}_s$    \\
\hline
$\chi_c$ & $\bar K$ & $\bar{K} \chi_c \to D_s^{*-} D, D_s^- D^*, 
D_s^{*-} D^*$      \\
\hline
$\chi_c$ & $K^*$ & $K^* \chi_c \to \bar{D} D^+_s, \bar{D}^* D^+_s, 
\bar{D} D^{*+}_s, \bar{D}^* D^{*+}_s$    \\
\hline
$\chi_c$ & $\bar {K}^*$ & $\bar{K}^* \chi_c \to D_s^- D, D_s^{*-} D, D_s^- D^*,
D_s^{*-} D^*$      \\
\hline
$\chi_c$ & $\eta$ & $\eta \chi_{c} \to \bar{D}^{*} D, \bar{D} D^{*}, 
\bar{D}^{*} D^{*}, D^{*-}_s D_s^+, D_s^- D^{*+}_s, D^{*-}_{s} D^{*+}_{s}$  \\
\hline
\end{tabular}
\end{table*}

\end{document}